\documentclass[12pt]{article}
\usepackage{amsfonts}
\usepackage{amssymb}
\usepackage{amsmath,amsthm}
\usepackage{graphicx}
\usepackage[sort]{natbib}
\usepackage{url}
\usepackage{authblk}
\usepackage{tikz}
\usepackage{booktabs}
\usepackage{bbm}
\usepackage{mathrsfs}
\usepackage{enumitem}
\usepackage{caption}
\usepackage{subcaption}
\usepackage{mathtools}
\usepackage{centernot}
\usepackage{xcolor}
\usepackage{etoolbox}
\usepackage{filecontents}
\usepackage{multirow}
\usepackage{multicol}
\usepackage{rotating}
\usepackage{xr}

\newtheorem{theorem}{Theorem}

\newtheorem{lemma}[theorem]{Lemma}
\theoremstyle{definition}

\newtheorem{example}[theorem]{Example}

\newtheorem{remark}[theorem]{Remark}

\newtheorem{assumption}{Assumption}
\newenvironment{assumptionbis}[1]
  {\renewcommand{\theassumption}{\ref{#1}$'$}%
   \addtocounter{assumption}{-1}%
   \begin{assumption}}
  {\end{assumption}}
\newcommand{\ol}[1]{\overline{#1}}

\newcommand{\EE}{\mathbb{E}}
\newcommand{\PP}{\mathbb{P}}
\newcommand{\cind}{\perp \!\!\! \perp}
\DeclareMathOperator{\Var}{Var}

\DeclareMathOperator{\tr}{tr}
\DeclareMathOperator{\plim}{plim}
\DeclareMathOperator{\an}{an}
\newtoggle{commenttoggle}
\togglefalse{commenttoggle}
\newcommand{\comment}[1]{
  \iftoggle{commenttoggle}{
  {\normalsize{\color{red}{ #1}}\normalsize}
}
{}
}
\newcommand{\blind}{1}

\begin{document}

\def\spacingset#1{\renewcommand{\baselinestretch}%
{#1}\small\normalsize} \spacingset{1}

%%%%%%%%%%%%%%%%%%%%%%%%%%%%%%%%%%%%%%%%%%%%%%%%%%%%%%%%%%%%%%%%%%%%%%%%%%%%%%

\if1\blind
{
  \title{\bf Instrumental Variable Estimation of Marginal Structural Mean Models for Time-Varying Treatment}
  \author{Haben Michael$^1$, Yifan Cui$^2$, Scott A. Lorch$^{3}$ and Eric J. Tchetgen Tchetgen$^{2}$\hspace{2px}}
  \date{$^1$Department of Mathematics and Statistics, University of Massachusetts\\
    $^2$Department of Statistics, University of Pennsylvania\\
  $^3$Children's Hospital of Philadelphia}
  \maketitle
  % \title{\bf Instrumental Variable Estimation of Marginal Structural Mean Models for Time-Varying Treatment}
  % \author{Haben Michael$^1$, Yifan Cui$^2$, Scott A. Lorch$^{3,}$\thanks{} and Eric J. Tchetgen Tchetgen$^{2,}$\thanks{}\hspace{2px}}
  % \date{$^1$Department of Mathematics and Statistics, University of Massachusetts\\
  %   $^2$Department of Statistics, University of Pennsylvania\\
  % $^3$Children's Hospital of Philadelphia}
  % \maketitle
} \fi

\if0\blind
{
  \bigskip
  \bigskip
  \bigskip
  \begin{center}
    {\Large\bf A Simple Weighted Approach for Instrumental Variable Estimation of Marginal Structural Mean Models}
\end{center}
  \medskip
} \fi

\bigskip
\begin{abstract}
  \cite{robins1997} introduced marginal structural models (MSMs),\ a
  general class of counterfactual models for the joint effects of
  time-varying treatment regimes in complex longitudinal studies
  subject to time-varying confounding.  In his work, identification of MSM
  parameters is established under a sequential randomization
  assumption (SRA), which rules out unmeasured confounding of
  treatment assignment over time. We consider sufficient conditions
  for identification of the parameters of a subclass, Marginal
  Structural Mean Models (MSMMs), when sequential randomization fails
  to hold due to unmeasured confounding, using instead a time-varying
  instrumental variable. Our identification conditions require that no
  unobserved confounder predicts compliance type for the time-varying
  treatment. We describe a simple weighted
  estimator and examine its finite-sample properties in a simulation
  study. We apply the proposed estimator to examine the effect of delivery
  hospital on neonatal survival probability.
\end{abstract}

\noindent%
\vfill

\newpage
\spacingset{1.5} % DON'T change the spacing!

\section{Introduction}

\cite{robins1997,robins2000} introduced marginal
structural models (MSMs), a class of counterfactual models that encode
the joint causal effects of time-varying treatment in the presence of
time-varying confounding. For identification, Robins relied on a
sequential randomization assumption (SRA), which rules out unmeasured
confounding of the time-varying treatment. MSMs have since become the
standard analytic approach to evaluate causal effects in time-varying
epidemiological studies \citep{hernan2002,morrison2010,cerda2010,cook2002,vanderweele2011}. However,
SRA may be hard to justify in many such settings, and unmeasured
confounding bias may invalidate causal claims inferred by the
approach. In the case of a point treatment, a large literature in
causal inference has developed over the years on the instrumental
variable method aiming to address unmeasured confounding
\cite{angrist1996,baker1994,angrist1995,robins1994,heckman2010}. Instead of assuming that
there is no unmeasured confounding, the IV approach relies on the key
assumption that one has observed a pretreatment variable that can
affect the outcome only through its effects on the treatment.  Many
commonly used IVs, such as treatment compliance and tax rates, vary
with time. Nevertheless, IV methods in longitudinal settings are far
less developed.  In this paper, we consider sufficient conditions for
identification of the parameters of a Marginal Structural Mean Model
(MSMM) with the aid of a time-varying instrumental variable when
sequential randomization fails to hold due to time-varying unmeasured
confounding. In doing so, we firmly establish the IV approach in the
context of MSMMs for complex longitudinal settings, an extension
previously believed out of reach \cite{robins2000}. Our identification
conditions require longitudinal generalizations of standard IV
assumptions, together with a key assumption that no unobserved
confounder predicts compliance type for the time-varying treatment, a
longitudinal generalization of the identification condition of
\cite{wang2018}. Under these assumptions, we establish identification
of the MSMM  and propose a simple estimation procedure
analogous to inverse-probability weighted (IPW) estimation, the most
common approach for estimating MSMs under SRA.
% The set of influence functions for MSM
% parameters under IV identification is derived for a semiparametric model with
% sole restriction on the observed data distribution given by the MSM, and is
% shown to provide a rich class of multiply robust estimators, including a
% locally semiparametric efficient estimator.

Prior to the current work, \cite{robins1994} developed a general
framework for identification and estimation of causal effects of
time-varying endogenous treatments using a time-varying instrumental
variable under a structural nested mean model (SNMM). As described in
\cite{robins2000}, the parameters of an SNMM can under certain
homogeneity conditions be interpreted as MSMM parameters, in which
case \cite{robins1994} provides alternative estimators to ours. In
contrast, the proposed methodology is more general as it directly
targets MSMM parameters irrespective of whether or not they can be
interpreted as parameters of an equivalent SNMM.  \cite{robins1998}
left open the question whether MSMs, like SNMMs, were identifiable by
IVs, a question that the current work therefore answers in the affirmative.

The remainder of the paper is organized as follows. In Section
\ref{section:background} we provide context by describing
identification and estimation of MSMM parameters under SRA. In Section
\ref{section:identification} we present an alternative set of
identification conditions to SRA, making use of a time-varying
instrumental variable. In Section \ref{section:converse} we establish the necessity of a variation of our key identification condition. In Section \ref{section:estimation} we describe
a simple weighted estimator for MSMM parameters using our instrumental
variable approach.  In Section \ref{section:simulation} we present a
simulation study to examine the finite-sample performance of our
proposed estimator. In Section \ref{section:application}, we apply the
proposed estimator to examine the effect of delivery hospital on
neonatal survival probability. We conclude in Section
\ref{section:discussion} with a brief discussion and description of
future work.  % - describe
 % -- causal models for longitudinal data (phrase in terms oflatent variables)

\section{Background}
\label{section:background} We consider i.i.d. discrete-time processes
and adopt the ``potential outcomes'' framework.  The data observed on
a process consists of $T+2$ random vectors
$\ol{L}=(L_0,\ldots,L_{T+1})$ and $T+1$ random variables
$\ol{A}=(A_0,\ldots,A_{T})$. The common state space of
$A_t,t=1,\ldots,T,$ is denoted $\mathcal{A}$. Variables at time $0$
are defined to be constant, e.g., $A_0=0$. The vectors $L_t$ and
variables $A_t$ carry the interpretation of a subject's time-varying
covariates and a time-varying treatment, respectively. % \comment{mention
  % treatment is discrete-valued? The pf given below allows for
  % non-discrete
  % treatments}
% The treatment $A_t$ is usually interpreted
% as being  administered subsequent to the observation of $L_t$ but
% before that of  $L_{t+1}$.
A variable $Y\in L_{T+1}$  is singled out as
an outcome of interest% ((, and a vector $V\in L_0$ as baseline
% covariates[if so, cant have L0 be constant))
. The number of time points $T$ is non-random. The statistical
significance of these temporal relations are conditional independence
relationships formalized in assumptions given below. We use script
fonts to refer to state spaces, $f$ to refer to densities, and $\mu$
for measures relative to which densities are given, using subscripts
to indicate the law. We use overbars to indicate the history of an RV, e.g.,
$\ol{L}_t=(L_0,\ldots,L_t)$. 
Besides the
observed data, we assume the existence of $|\mathcal{A}|^{T}$
variables $Y_{\ol{a}}, \ol{a}\in\mathcal{A}^{T}$.  These ``potential
outcomes'' or ``counterfactuals'' are not in general observed. They
are related to the observed data by the ``consistency'' assumption,
\begin{assumption}[Consistency]\hspace{1.px}
 $ Y = Y_{\ol{A}} \hspace{10px}\text{ a.s.}$
\hspace{30px}\label{assumption:consistency}
\end{assumption}
% of $2^{J+1}$ processes $L_{\ol{a}}(\cdot)$ indexed by
% $\ol{a}\in\{0,1\}^{J+1}$. The observed and partially observed data are
% related by the ``consistency'' assumption \cite{robins1997},
% \begin{align}
    %     L(\cdot) = L_{\ol{A}(J)}(\cdot)\equiv
    %     L_{\ol{a}}(\cdot)\big\rvert_{\ol{a}=\ol{A}(J)} \text{ a.s.}.\label{assumption:consistency}
    %   \end{align}
\noindent In case the treatments $\ol{A}$ are discrete, the assumption
may be written as
$Y=\sum_{\ol{a}\in\mathcal{A}^T}Y_{\ol{a}}\{\ol{A}=\ol{a}\}$, using
braces to denote the event indicator.  Thus $\ol{a}$ may be interpreted as a particular treatment
regime, and the potential outcome $Y_{\ol{a}}$ as the distribution of
$Y$ were everyone in the observed population to follow treatment
regime $\ol{a}$, i.e., if $\{\ol{A}=\ol{a}\}\equiv 1$.

% ((the following para is from my job talk. happy to omit if too prolix. the entire background section is currently about 3.5 pages.)) Outside of the potential
% outcomes framework, a treatment is usually regarded as 
% driving changes in a responsive outcome. In the potential outcome
% framework, by contrast, the potential outcomes may be thought of as
% existing prior to
% treatment, similar to baseline characteristics, and the role of a treatment
% is to expose the potential outcome of a subpopulation to
% observation. By introducing outcomes that exist prior to treatment, the potential
% outcomes framework allows us to reason about causality, i.e.,
% relationships unconfounded by treatment. At the same time, models are then required to extrapolate population
% characteristics of the potential outcomes based on the observed subpopulation.

A marginal structural mean model (``MSMM'') is a model on the marginal
means of potential outcomes
\cite{robins1997}% , possibly conditional on baseline
% covariates
.  For example, the effect of
treatment may be modeled as linear in the cumulative treatment taken,
\begin{align}
\EE(Y_{\ol{a}}) = \beta_0 +
  \beta_1\sum_{t=0}^{T}a_t.\label{eqn:linear msm}
\end{align}
In this example, $\beta\in\mathbb{R}^{2}$ parameterizes the model and encodes the
incremental effect of a unit of treatment. A link function
can be introduced to accommodate binary or count outcome variables,
e.g., for binary $Y$, $\EE(Y_{\ol{a}}) = (1+\exp(\beta_0 +
\beta_1\sum_{t=0}^{T}a_t))^{-1}.$ In general we write 
\begin{align}
  \EE(Y_{\ol{a}})=m_\beta(\ol{a})\label{eqn:msmm}
\end{align}
to describe an MSMM, where $m_\beta:\mathcal{A}^T\to\mathbbm{R}$ belongs to a family
of functions parameterized by finite-dimensional $\beta$. The model parameter $\beta$ is
the target of  inference. % Examples of models falling outside this class are
% models on the joint distribution of potential outcomes, such as
% their covariance structure, or models on other functions of the
% marginal distributions besides the mean, such as the hazard
% function.

An MSMM is defined using the unobserved quantities $Y_{\ol{a}},\ol{a}\neq\ol{A}$, and the
model parameter is not in general
identified by the observed data. \cite{robins1997} provides sufficient conditions for identification and estimation, the
sequential randomization assumption and positivity:
\begin{align}
Y_{\ol{a}}\cind A_t \mid
  \ol{L}_t,\ol{A}_{t-1},
  \quad &1\le t\le T
  \qquad\text{(SRA)}\label{assumption:sra}\\
  % 0<\PP(A_t=a_t\mid\ol{A}_{t-1},\ol{L}_t)
  % \qquad(\ol{A}_{t-1},\ol{L}_t)\text{---a.s.},\comment{clear?}  \quad
  %        &a_t\in\mathcal{A}, 1\le t\le T
  % \qquad\text{(positivity)}\label{assumption:positivity}\\
  0<f_{A_t\mid\ol{A}_{t-1},\ol{L}_t}(a_t\mid\ol{a}_{t-1},\ol{l}_t)
  \text{ when }f_{\ol{A}_{t-1},\ol{L}_t}(\ol{a}_{t-1},\ol{l}_t)>0,  \quad
         &a_t\in\mathcal{A}, 1\le t\le T,
  \qquad\text{(positivity)}\label{assumption:positivity}
\end{align}
% -- SRA, allows for identification
using $\cind$ to denote statistical independence.
 In the treatment setting, SRA will hold if the cumulative observed data at
each time point captures all systematic
associations between the treatment and outcome of interest.  Positivity will hold when, among
all
subpopulations defined by covariates $\ol{L}_t$ and a treatment regime $\ol{A}_{t-1},t\le T$, there are further subpopulations
 at each possible treatment level $a_t\in\mathcal{A}$. These conditional
independence relationships are implied by the causal directed acyclic graph
 given in Fig. \ref{fig:sra dag}, in which a node is independent of non-descendants
conditional on its parent nodes; see \cite{richardson2013} for
details.

\cite{robins1997} uses assumptions
\ref{assumption:sra} and \ref{assumption:positivity} to relate the law
of a potential outcome $Y_{\ol{a}}$ to the law of the observed data
$(Y,\ol{L},\ol{A})$. Specifically, given measurable
$g:(\mathcal{Y},\mathcal{A}^T)\to\mathbbm{R}^d$, he established that
\begin{align}
  % \sum_{\ol{a}}\EE\left(g(Y_{\ol{a}},\ol{a})\right)=\EE\left(g(Y,\ol{A})/\ol{W}^{(SRA)}\right),\label{thrm:robins}
  \int_{\mathcal{A}^T}\EE\left(g(Y_{\ol{a}},\ol{a})\right)\mu_{\mathcal{A}^T}(\ol{a})=\EE\left(g(Y,\ol{A})/\ol{W}^{(SRA)}\right),\label{thrm:robins}
\end{align}
where the observation weights $1/W^{(SRA)}$ are defined by
\begin{align}
  W^{(SRA)}_t&= f_{A_t\mid \ol{A}_{t-1},\ol{L}_{t}}(A_{t}\mid \ol{A}_{t-1},\ol{L}_{t}),\qquad \ol{W}_t^{(SRA)}=\prod_{\tau=1}^tW^{(SRA)}_\tau,
  \qquad t=1,\ldots,T,\\
\ol{W}^{(SRA)}&=\ol{W}_T^{(SRA)}=\prod_{\tau=1}^TW^{(SRA)}_\tau.\label{defn:sra weights}
\end{align}
This use of overbars to represent the running product of weights departs from our usual use of overbars to denote the history of a time-varying quantity collected in a vector. The case $T=1, g(y,a)=g_0(y)\times\mathbbm{1}_a,$ gives the inverse-probability-weighted
estimator for $g_0(Y_a)$ often used in propensity score analysis.

Besides identifying the parameter of an MSMM using fully observed
data, relation (\ref{thrm:robins}) also suggests an estimator. Let
$g(y,\ol{a})=h(\ol{a})(y-m_\beta(\ol{a}))$, where $h$ is a function on
$\mathcal{A}^T$ of the same dimension as $\beta$. Then the MSMM model
(\ref{eqn:msmm}) implies
$\EE\left(h(\ol{A})\frac{Y-\mu_{\beta}(\ol{A})}{W^{(SRA)}}\right)=\EE(h(\ol{a})\left(Y_{\ol{a}}-m_\beta(\ol{a})\right))=0$,
giving rise to estimating equations for $\beta$,
$$
0=\PP_n\left(h(\ol{A})\frac{Y-\mu_{\hat{\beta}}(\ol{A})}{\ol{W}^{(SRA)}}\right),
$$
using $\PP_n$ to denote the empirical distribution on a sample of size
$n$. In practice, $\ol{W}^{(SRA)}$ may not be known and an estimate is
substituted. Under standard regularity conditions for M-estimation, the
empirical solution $\hat{\beta}$ is asymptotically normal as the
number of observations $n$ grows, with a variance that can be
approximated by its influence function.

Furthermore, a suitable choice of $h$ in (\ref{thrm:robins}) can in some situations provide a means to stabilize  the weights (\ref{defn:sra weights}), which may become
unstable as $T$ increases. Stabilized weights are defined as % formed as the ratio of the
% weights (\ref{defn:sra weights}) with an approximations
% depending only on $\ol{A}$. A common choice for this approximation is $\prod_tf(A_t\mid\ol{A}_{t-1})$, giving
$$
\ol{W}^{(SRA,stabilized)}\coloneqq\prod_{t=1}^Tf(A_t\mid\ol{A}_{t-1})/\ol{W}^{(SRA)}= \prod_{t=1}^Tf(A_t\mid\ol{A}_{t-1})/f(A_t\mid\ol{A}_{t-1},\ol{L}_t).
$$
The quality of the approximation depends on the strength of the dependence of the density
of $A_t$  and the covariates $\ol{L}_t$ given
$\ol{A}_{t-1}$, i.e., to the extent that treatment is unconfounded. 
%  The qualitative result is
% that this estimation approach, similar to the
% one adopted below, requires for practical application that either $T$ not be too large or confounding
% not be too great.

% In Lemma \ref{theorem:1} below we establish an analogous result using IV
% assumptions in lieu of SRA.
We pursue parallel results using instrumental variables to relax SRA.
We give an analogue of the identifying relation (\ref{thrm:robins}) using IVs in Section
\ref{section:identification}, and similar estimation techniques in Section
\ref{section:estimation}.

% Although we focus on marginal structural mean models, the
% marginal structural model (``MSM'')
% theory of \cite{robins1997} is more general, allowing for models  on
% many other functionals of the marginal distributions of the potential outcomes, as well
% as providing efficient estimators. See 
% \cite{tchetgen2018} for a more closely parallel development. 

\section{Identification of causal model parameters using IVs}
\label{section:identification}

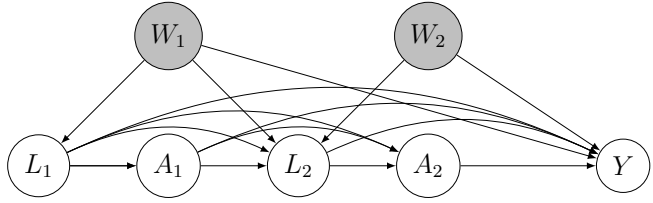
\begin{figure}
\centering
  \resizebox{250pt}{!}{%
    \begin{tikzpicture}
  \def\Ax{0}
  \def\Ay{0}
  \def\offset{2}
  \def\Bx{\Ax+4}
  \def\By{\Ay}
  \node[shape=circle,draw=black] (A0) at (\Ax,\Ay) {$A_1$};
  \node[shape=circle,draw=black] (L0) at (\Ax-\offset,\Ay) {$L_1$};
  \node[shape=circle,draw=black] (A1) at (\Bx,\By) {$A_2$};
  \node[shape=circle,draw=black] (L1) at
  (\Bx-\offset,\By) {$L_2$};

  \draw [-latex] (L0) to [bend left=0] (A0);
  \draw [-latex] (L0) to [bend left=25] (L1);
  \draw [-latex] (L0) to [bend left=25] (A1);
  \draw [-latex] (A0) to [bend left=0] (L1);
  \draw [-latex] (A0) to [bend left=25] (A1);
  \draw [-latex] (L1) to [bend left=0] (A1);
  \draw [-latex] (L0) to [bend left=0] (A0);

  \node[shape=circle,draw=black] (Y) at (\Bx+3,0) {$Y$};

  \draw [-latex] (L0) to [bend left=25] (Y);
  \draw [-latex] (A0) to [bend left=25] (Y);
  \draw [-latex] (L1) to [bend left=25] (Y);
  \draw [-latex] (A1) to [bend left=0] (Y);

  \node[shape=circle,draw=black,fill=lightgray] (U0) at (\Ax,\Ay+2) {$W_1$};
  \node[shape=circle,draw=black,fill=lightgray] (U1) at (\Bx,\Ay+2) {$W_2$};
  \draw [-latex] (U0) to (L0);
  \draw [-latex] (U0) to (L1);
  \draw [-latex] (U0) to (Y);
  \draw [-latex] (U1) to (L1);
  \draw [-latex] (U1) to (Y);

\end{tikzpicture}
% \begin{tikzpicture}
%   \def\Ax{0}
%   \def\Ay{0}
%   \def\offset{2}
%   \def\Bx{\Ax+4}
%   \def\By{\Ay}
%   \node[shape=circle,draw=black] (A0) at (\Ax,\Ay) {$A_1$};
%   \node[shape=circle,draw=black] (L0) at (\Ax-\offset,\Ay) {$L_1$};
%   \node[shape=circle,draw=black] (A1) at (\Bx,\By) {$A_2$};
%   \node[shape=circle,draw=black] (L1) at
%   (\Bx-\offset,\By) {$L_2$};

%   \draw [-latex] (L0) to [bend left=0] (A0);
%   \draw [-latex] (L0) to [bend left=25] (L1);
%   \draw [-latex] (L0) to [bend left=25] (A1);
%   \draw [-latex] (A0) to [bend left=0] (L1);
%   \draw [-latex] (A0) to [bend left=25] (A1);
%   \draw [-latex] (L1) to [bend left=0] (A1);
%   \draw [-latex] (L0) to [bend left=0] (A0);

%   \node[shape=circle,draw=black] (Y) at (\Bx+3,0) {$Y$};

%   \draw [-latex] (L0) to [bend left=25] (Y);
%   \draw [-latex] (A0) to [bend left=25] (Y);
%   \draw [-latex] (L1) to [bend left=25] (Y);
%   \draw [-latex] (A1) to [bend left=0] (Y);

% \end{tikzpicture}
  }
  \captionof{figure}{Causal DAG describing longitudinal confounding with $T=2$ time points, satisfying SRA. % The causal effect of the treatment process $\{A_t\}$ is identifiable. The causal of effect of the covariates $\{L_t\}$ may not be identifiable due to unobserved confounders, represented by the shaded nodes.
  }
  \label{fig:sra dag}
\end{figure}
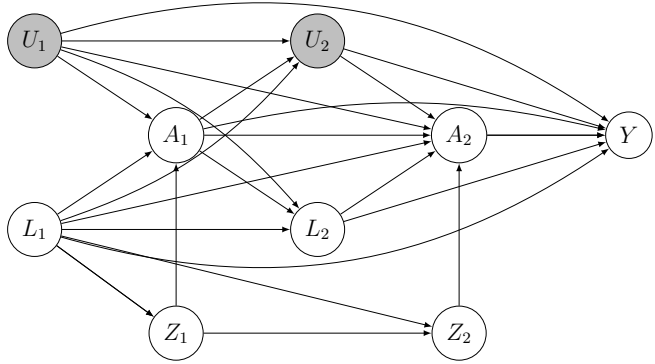
\begin{figure}
  \centering
  \vfill
  \resizebox{250pt}{!}{%
    \begin{tikzpicture}
    \def\Ax{2}
    \def\Ay{0}
    \def\offset{2.5}
    \def\Bx{\Ax+5}
    \def\By{\Ay}
    \node[shape=circle,draw=black] (A1) at (\Ax,\Ay) {$A_1$};
    \node[shape=circle,draw=black,fill=lightgray] (U1) at (\Ax-\offset,\Ay+\offset/1.5) {$U_1$};
    \node[shape=circle,draw=black] (L1) at (\Ax-\offset,\Ay-\offset/1.5) {$L_1$};
    \node[shape=circle,draw=black] (Z1) at (\Ax,\Ay-1.4*\offset) {$Z_1$};
    \node[shape=circle,draw=black] (A2) at (\Bx,\By) {$A_2$};
    \node[shape=circle,draw=black,fill=lightgray] (U2) at (\Bx-\offset,\By+\offset/1.5) {$U_2$};
    \node[shape=circle,draw=black] (Z2) at (\Bx,\By-1.4*\offset) {$Z_2$};
    \node[shape=circle,draw=black] (L2) at (\Bx-\offset,\By-\offset/1.5) {$L_2$};
    \node[shape=circle,draw=black] (Y) at (\Bx+3,\By) {$Y$};

    \draw [-latex] (U1) to (A1);
    \draw [-latex] (L1) to (A1);
    \draw [-latex] (U2) to (A2);
    \draw [-latex] (L2) to (A2);
    \draw [-latex] (Z1) to (A1);
    \draw [-latex] (Z2) to (A2);
    \draw [-latex] (Z1) to (Z2);
    \draw [-latex] (A1) to (A2);

    % \draw [-latex] (L1) to (U1);
    % \draw [-latex] (U1) to (L1);

    \draw [-latex] (L1) to (Z1);
    \draw [-latex] (L1) to (Z1);
    \draw [-latex] (L1) to (Z2);
    \draw [-latex] (L1) to [bend left=0] (L2);
    \draw [-latex] (L1) to [bend right=15] (U2);
    \draw [-latex] (L1) to [bend left=0] (A2);
    \draw [-latex] (U1) to [bend left=0] (U2);
    \draw [-latex] (U1) to [bend left=15] (L2);
    \draw [-latex] (U1) to [bend left=0] (A2);

    \draw [-latex] (L1) to [bend right=25] (Y);
    \draw [-latex] (L2) to [bend right=0] (Y);
    \draw [-latex] (U1) to [bend left=25] (Y);
    \draw [-latex] (U2) to [bend left=0] (Y);
    \draw [-latex] (A2) to [bend left=0] (Y);
    \draw [-latex] (A2) to [bend left=0] (Y);
    \draw [-latex] (A1) to [bend left=13] (Y);
    \draw [-latex] (A1) to (L2);
    \draw [-latex] (A1) to (U2);

    % \draw [-latex] (L1) to [bend left=0    ] (U2);
    % \draw [-latex] (L1) to [bend left=0] (L2);
    % \draw [-latex] (L1) to [bend right=10] (A2);
    % % \draw [-latex] (L1) to [bend left=10] (L2);
    % \draw [-latex] (A1) to [bend left=10] (U2);
    % \draw [-latex] (A1) to [bend left=0] (Z2);
    % \draw [-latex] (A1) to [bend right=10] (L2);
    % \draw [-latex] (A1) to [bend right=10] (A2);
    % \draw [-latex] (A1) to [bend right=10] (L2);
    % \draw [-latex] (Z1) to [bend left=0] (A1);
    % \draw [-latex] (Z2) to [bend left=0] (A2);
    % % \draw [-latex] (Z1) to [bend right=10] (L2);
    % \draw [-latex] (Z1) to [bend right=0] (A2);
    % % \draw [-latex] (Z0) to [bend right=10] (L2);
    % \draw [-latex] (Z0) to [bend right=0] (Z2);

    % \draw [-latex] (L0) to [bend left=25] (Y);
    % \draw [-latex] (L2) to [bend left=30] (Y);
    % \draw [-latex] (U0) to [bend left=35] (Y);
    % \draw [-latex] (U2) to [bend left=0] (Y);

    % \draw [-latex] (\Bx+2.8,\By) to [bend left=0] (Y);
    % \draw [-latex] (\Bx+2.8,\By+.4) to [bend left=0] (Y);
    % \draw [-latex] (\Bx+2.8,\By-.4) to [bend left=0] (Y);
    % \draw [-latex] (A2) to (\Bx+1.3,\By);

  \end{tikzpicture}
  }
  \vfill
  \captionof{figure}{An example of a causal DAG describing confounding with unobserved
    confounders and IV, with $T=2$ time points, satisfying the assumptions of Theorem \ref{theorem:1}. Edges emerging from the IV $Z$ only
    end at the treatment $A$. The roles of $U$ and
    $L$ are symmetric except in respect of the IV, with which $L$ but not
    $U$ is dependent.\comment{add causal links between $U_0,L_0$? this doesn't cause an identification problem so long as Z is conditionally independent of U given L. but then which direction? I dont want to suggest that only U can cause L, or vice versa. But I have to pick one or the other so as to not create a cycle.}}
  \label{fig:full dag}
% \caption{the}
\end{figure}

We now allow for the possibility of unmeasured confounders in the form
of an additional unobserved stochastic process associated with both the treatment
and outcome. The sequential randomization assumption
(\ref{assumption:sra}) is not warranted in this situation. We propose
to use ``instrumental variables'' to identify an MSMM parameter in the
absence of SRA. Informally, an IV is a random variable associated with
the treatment of interest that only affects the outcome of interest
through its effect on the treatment. % Informally, an IV is random variable associated with
% the covariate of interest, allowing it to be used as a surrogate, but
% orthogonal to the unobserved covariates. The IV induces variation in
% the covariate of interest that is independent of confounding, allowing
% for an unconfounded analysis of the relationship between the covariate
% of interest and outcome.
    %     On the other hand, suppose that one has observed a binary process $Z$ associated with $A$ that, given
% observed data, is
% independent of the potential outcomes $Y_{\ol{a}}$.

To this end, in addition to the data described in Section
\ref{section:background}, let $\ol{U}=(U_1,\ldots,U_T)$ be an
unobserved process, possibly multivariate, which may be associated
with both $\ol{A}$ and $\ol{L}$, including $Y$. We assume that
$\ol{U}$ captures all further confounding between $\ol{A}$ and $Y$
beyond $\ol{L}$, so that
SRA would hold were $\ol{U}$ observed:
\begin{assumption}[Latent SRA]
  $Y_{\overline{a}}\cind A_t\mid\overline{A}_{t-1}=\overline
{a}_{t-1},\overline{L}_t,\overline{U}_t\qquad t=1,\ldots,T,\quad\ol{a}\in\mathcal{A}^T$\label{latent ignorability}
  \label{assumption:latent sra}
\end{assumption}
That is, there are no unobserved confounders at time $t$ other than
$\ol{U}_t$. The assumptions given below impose restrictions on $\ol{U}$.

Suppose further that a binary-valued process
$\ol{Z}=(Z_1,\ldots,Z_T)$ is observed, satisfying the following IV assumptions: % Corresponding to $U$ are potential outcomes $U_{\ol{a}}$ indexed
% by treatment regimes, satisfying $U=U_{\ol{A}(j)}$ for
% all times $j$. For any time $j$, $Z(j)$ is assumed to occur
% before $A(j)$ but after $L(j)$ and $U(j)$, \comment{this ordering carrying the same
% conditional independence implications described above.}
% See Fig. \ref{fig:full dag}. % Although
% unobserved, we suppose that $U$ and $U_a$ are related as through []
% [is this actually necessary? can we just assume $U_a$?]. 
For all $1\le t\le T$ and $\ol{a}\in\mathcal{A}^T,\ol{z}\in\{0,1\}^{T}$,
% \begin{align}
\begin{assumption}[IV relevance] 
  % $Z_t\centernot{\cind} A_t\mid\overline{A}_{t-1},\overline{L}_t,\overline{Z}_{t-1} $
  $\EE(A_t\mid \overline{A}_{t-1},\overline{L}_t,\overline{Z}_{t}) \neq \EE(A_t\mid \overline{A}_{t-1},\overline{L}_t,\overline{Z}_{t-1})$
  \label{assumption:iv relevance}
\end{assumption}
\begin{assumption}[Exclusion restriction] 
$   Y_{\overline{a}\overline{z}}  = Y_{\overline{a}} $\comment{the great debate.}
\label{Exclusion Restriction}
\end{assumption}
\begin{assumption}[IV--outcome independence]
  $\ol{Z}_t \cind (Y_{\ol{a}},L_{t+1},U_{t+1}) \mid
    \overline{A}_{t}=\overline{a}_{t}
    ,\overline{L}_t,\overline{U}_t$
    \label{assumption:iv--outcome}
  \end{assumption}
  \begin{assumption}[IV--unmeasured confounder independence]
$    Z_t \cind \overline{U} \mid \overline{A}_{t-1},\overline{L}_t,\overline
{Z}_{t-1}$
\label{assumption:iv--confounder}
\end{assumption}
%   \quad
%   \begin{enumerate}
% \item $    Z_t \cind \overline{U} \mid \overline{A}_{t-1},\overline{L}_t,\overline
%   {Z}_{t-1}$,
%   \item $\EE(Y_{\ol{a}}\mid \ol{A}_t=\ol{a}_t,\ol{Z}_t,\ol{L}_t,\ol{U}_t)=\EE(Y_{\ol{a}}\mid \ol{A}_t=\ol{a}_t,\ol{Z}_{t-1},\ol{L}_t,\ol{U}_t)$ or \comment{I use this as a
%       replacement for our previous version $Z_t \cind Y_{\ol{a}} \mid
%     \overline{A}_{t-1}=\overline{a}_{t-1}
%     ,\overline{L}_t,\overline{U}_t,\ol{Z}_{t-1}$.  The difference being
%   that now the entire vector of $Z$ apperas on the LHS, and on the RHS
% the current treatment appears. (We previously had the two cases
% combined as one $Z\cind (U,Y)\mid \ldots $.) This lets me use the simpler version of
% "latent SRA" which doesn't have $Z$ in the conditioning event, and
% also gives us the analogue of the g-formula. Also I think it matches
% the literature better, since in the usual axiom  asserting that an IV
% has no non-treatment paths to the potential outcome, current treatment is conditioned
% on. Depending on the proofs we choose to go with, we may only require $\EE(Y_{\ol{a}}\mid \ol{A}_t=\ol{a}_t,\ol{ZLU}_t)=\EE(Y_{\ol{a}}\mid \ol{A}_t=\ol{a}_t,\ol{Z}_{t-1},\ol{LU}_t)$.} % \comment{consider replacing IV assumption with:}
% % \item \ol{Z}_t \cind (Y_{\ol{a}},\ol{U}_t,\ol{L}_t) | \ol{A}_t,...
%   \end{enumerate}
%   \label{assumption:iv independence}
% \end{assumption}
\begin{assumption}[IV Positivity]
$  0<\PP(  Z_t=1|\overline{A}_{t-1},\overline{L}_t,\overline{Z}_{t-1} )  <1\text{ a.s.}$
\label{assumption:iv positivity}
\end{assumption}
% \comment{Actually only need mean independence}
% \comment{We don't need the usual positivity assumption do we?} Assumption
% \ref{assumption:latent sra} is a variant of SRA, requiring SRA 
% hold were $U$
%  included among the observables.
Assumptions \ref{assumption:iv relevance}--\ref{assumption:iv
  positivity} are longitudinal generalizations of standard IV
assumptions. As in the SRA case discussed in Section
\ref{section:background}, the conditional independence relationships
described by the key assumptions \ref{assumption:latent sra},
\ref{assumption:iv--outcome}, and \ref{assumption:iv--confounder},
formalize the temporal relationships among the data $A_t,Y,L_t$, etc.,
that we use informally. A graph that provides a model of these
assumptions is given in Fig. \ref{fig:full dag}. The methods given in
\cite{richardson2013} may be used to establish that the graph in
Fig. \ref{fig:full dag}, properly interpreted, does in fact entail a
model for the conditional independence relations given in Assumptions
\ref{assumption:latent sra}, \ref{assumption:iv--outcome}, and
\ref{assumption:iv--confounder}. The DAG in Fig. \ref{fig:full dag} is
illustrative and is not meant to preclude other models compatible with these
assumptions, e.g., unmeasured confounding among the measured covariates $\{L_t\}$
or between $L_t$ and the outcome $Y$, analogous to the shaded nodes in Fig. \ref{fig:sra dag}. As a shorthand we use the
notation ``$\an$'' to refer to an ancestor set in the DAG in
Fig. \ref{fig:full dag}, e.g., $\an(Z_t)$ is
$\an(A_{t-1})\cup \an(Z_{t-1})\cup\an(L_t)$.

    %     Assumption \ref{assumption:latent sra} is more easily satisfied with a
% larger set $U$, Assumption \ref{assumption:iv independence} is more easily satisfied with smaller $U$. Assumption
% \ref{assumption:ict}, interpreted as asserting that compliance type
% \cite{frangakis2002} is conditionally independent of the
% unknown confounder. It is
% a longitudinal generalization of an assumption used in the point
% exposure setting of our problem \cite{wang2018}.
% % -- define IV, give list of axioms
% -- give identification theorem

Finally, we make an additional orthogonality assumption. For $t=1,\ldots,T.$
\begin{assumption}[Independent Compliance Type]
  \begin{align*}
    % \mathbb{E}\left[  A_t|\overline{U}_t,\overline{L}_t,\overline{A}\left(
    % t-1\right)  ,\overline{Z}_{t-1},Z_t=1\right]&-\mathbb{E}\left[
    %                                               A_t|\overline{U}_t,\overline{L}_t,\overline{A}_{t-1}
    %                                               ,\overline{Z}_{t-1},Z_t=0\right] \\
    %                                             &\eqqcolon\Delta_{t}\left(  \overline{L}_t,\overline{A}_{t-1}
    %                                               ,\overline{Z}_{t-1}  \right)
    f\left(  a_t|\overline{A}_{t-1},\overline{Z}_{t-1},Z_t=1,\overline{L}_t,\overline{U}_t\right)&-f\left(
                                                  a_t|\overline{A}_{t-1},\overline{Z}_{t-1},Z_t=0,\overline{L}_t,\overline{U}_t\right) \cind \ol{U}_t \mid \ol{A}_{t-1},\ol{Z}_{t-1},\ol{L}_t.
                                                % &\eqqcolon\Delta_{t}\left(  \overline{L}_t,\overline{A}_{t-1}
                                                %   ,\overline{Z}_{t-1}  \right),\qquad a_t\in\mathcal{A}
  \end{align*} \label{assumption:ict}
Defining
$$
\Delta_t(a_t,\overline{A}_{t-1},\overline{Z}_{t-1},\overline{L}_t,\overline{U}_t)
= f\left(  a_t|\overline{A}_{t-1},\overline{Z}_{t-1},Z_t=1,\overline{L}_t,\overline{U}_t\right)-f\left(
  a_t|\overline{A}_{t-1},\overline{Z}_{t-1},Z_t=0,\overline{L}_t,\overline{U}_t\right),
$$
the assumption is that
$\Delta_t(a_t,\overline{A}_{t-1},\overline{Z}_{t-1},\overline{L}_t,\overline{U}_t)$
does not depend on $\ol{U}_t$, and so may be written as
$\Delta_t(a_t,\overline{A}_{t-1},\overline{Z}_{t-1},\overline{L}_t)$.
The function $\Delta_t$ may be expressed using
the observed data by the relation
\begin{align*}
\Delta_t(a_t,\overline{A}_{t-1},\overline{Z}_{t-1},\overline{L}_t)
&= f\left(  a_t|\overline{A}_{t-1},\overline{Z}_{t-1},Z_t=1,\overline{L}_t,\overline{U}_t\right)-f\left(
  a_t|\overline{A}_{t-1},\overline{Z}_{t-1},Z_t=0,\overline{L}_t,\overline{U}_t\right)\\
&= f\left(  a_t|\overline{A}_{t-1},\overline{Z}_{t-1},Z_t=1,\overline{L}_t\right)-f\left(
  a_t|\overline{A}_{t-1},\overline{Z}_{t-1},Z_t=0,\overline{L}_t\right).
\end{align*}
% To do so, we
% make use of a relation on $\Delta_t_t, t=1,\ldots,T$,
% \begin{align*}
%   \Delta_t_{t}\left(  \overline{L}_t,\overline{A}_{t-1}  ,\overline{Z}_{t-1}  \right)
%   &\equiv\mathbb{E}\left(  A_t|\overline{U}_t,\overline{L}_t,\overline{A}_{t-1}  ,\overline{Z}_{t-1},Z_t=1\right)-\EE\left(  A_t|\overline{U}_t,\overline{L}_t,\overline{A}_{t-1}  ,\overline{Z}_{t-1},Z_t=0\right)\\
%   &=\mathbb{E}\left(  A_t|\overline{L}_t,\overline{A}_{t-1}  ,\overline{Z}_{t-1},Z_t=1\right)-\EE\left(  A_t|\overline{L}_t,\overline{A}_{t-1}  ,\overline{Z}_{t-1},Z_t=0\right).
% \end{align*}
    %     \comment{I cannot at all remember why I included
  % the above relation. We put it in the tech report too. but it seems like the proof
  % goes through fine without it.}
The relation follows by integrating both sides of the first line with respect
to the conditional density of $\ol{U}_t$ given
$(\ol{A}_t,\ol{Z}_{t-1},Z_t=1,\ol{L}_t)$, which is the same as the
density given $(\ol{A}_t,\ol{Z}_{t-1},Z_t=0,\ol{L}_t)$, by Assumption
\ref{assumption:iv--confounder}. %; see \cite{tchetgen2018} for details.

\end{assumption}
Assumption \ref{assumption:ict} states that while $\overline{U}_t$ may confound the causal
effects of $\overline{A}_t,$ no component of $\overline{U}_t$ interacts with
$Z_t$ in its additive effects on $A_t.$ %  A causal interpretation of the
% assumption is available if $A_t$ is binary-valued and the IV is
% conditionally independent of potential treatment outcomes under a hypothetical IV intervention, $Z_t \cind A_{z(t)}(t)
% \mid \overline{U}_t,\overline{L}_t,\overline{A}_{t-1}  ,\overline
% {Z}_{t-1}$. \comment{The time index notation is inconsistent here. I
%   switched the time index from parens $A(t)$ to subscript $A_t$
%   because we mostly don't deal with time-dependent potential outcomes
%   $L_a(t),U_a(t)$ in this manuscript. But I forgot about this one
%   place where we do.}In this case Assumption \ref{assumption:ict}
% implies that $\overline{U}_t$ is conditionally independent of the
% ``compliance type'' \cite{angrist1996} at time
% $t,$
% \begin{align}
% &  \mathbb{E}\left[  A_{z(t)=1}(t)-A_{z(t)=0}(t)|\overline{U}_t,\overline
% {L}_t,\overline{A}_{t-1}  ,\overline{Z}_{t-1}\right]  \\
% &  =\Delta_{t}\left(  \overline{L}_t,\overline{A}_{t-1}
% ,\overline{Z}_{t-1} \right)  \text{, }t=0,...,T-1.\nonumber
% \end{align}
 This assumption is a longitudinal generalization of a similar assumption
% made by \cite{wang2018,wang2018a} in the point exposure setting. 
made by \cite{wang2018} in the point exposure setting. 

Let $f_{Z_t}$ denote the density of $Z_t$ conditional on the prior observed
history $(\ol{A}_{t-1},\ol{Z}_{t-1},\ol{L}_t)$, which, by Assumption
\ref{assumption:iv--confounder}, has the same effect as conditioning on the full prior history $(\ol{A}_{t-1},\ol{Z}_{t-1},\ol{L}_t,\ol{U}_t)$. %  and, as before, let $f^*$ be any given density of the treatment
% process $A(\cdot)$ \comment{this is already defined, but that was in the SRA case}.
We define subject-specific weights $1/\ol{W}$ through:
\begin{align}
\ol{W}=\prod_{t=1}^TW_t, \qquad W_t=(-1)^{1-Z_t}f_{Z_t}\left(  Z_t\mid\overline{L}_t,\overline
{A}_{t-1},\overline{Z}_{t-1}\right)  \Delta_t\left(  \overline{L}_t,\overline{A}_{t-1}  ,\overline{Z}_{t-1}  \right).
\label{defn:weights}
\end{align}
Assumptions \ref{assumption:iv relevance} and \ref{assumption:iv positivity} ensure that the weights are nonzero.\comment{This looks pretty different from our tech report weights but I thought it was cleaner to define them
  without the stabilization terms while we're talking about identification, then describe the stabilization in
  the estimation section. The stabilization is just another function of
  treatment $A$, so it falls under the general identification theorem.}
% and
% \[
% {W}_{j,2}^{}=\frac{1}{\left(  -1\right)  ^{1-A(j)}f^{\ast}\left(
% A(j)|V,\overline{A}(j-1)\right)  }%
% \]
\begin{theorem}
Suppose that together with consistency (\ref{assumption:consistency}), Assumptions
(\ref{assumption:latent sra}--\ref{assumption:ict}) hold. %  Let
% $\EE(Y_{\ol{a}})=m_{\beta}(\ol{a})$ be an MSMM.
For measurable $g:(\mathcal{Y},\mathcal{A}^T)\to\mathbbm{R}^d$,
\begin{align}
  \EE\left(g(Y,\ol{A})/\ol{W}\right)&=\int_{\mathcal{A}^T}g(Y_{\ol{a}},\ol{a})\mu_{\ol{A}^T}(\ol{a})% \prod_{j=0}^{J-1}
                                      %               f^{\ast}\left(  a_j|\overline{a}_{j-1}\right)
                                                         ,\label{thrm:u}
\end{align}
when the expectation exists. 
\label{theorem:1}
\end{theorem}
Proofs are given in the appendix.
\comment{this is the old discrete treatment version of the proof. it's much cleaner. is it ok
  to include this and just say the continuous case is mostly the same?
\begin{proof}
Using the consistency assumption (\ref{assumption:consistency}), we
write
$$
\EE\left(g(Y,\ol{A})/\ol{W}\right)=\sum_{\ol{a}}\EE\left(\{\ol{A}=\ol{a}\}g(Y_{\ol{a}},\ol{a})/\ol{W}\right)
$$
and proceed by induction. Suppose we have established, for some $1\le t\le T$,
$$
\EE\left(g(Y,\ol{A})/\ol{W}\right)=\sum_{\ol{a}}\EE\left(\{\ol{A_t}=\ol{a_t}\}g(Y_{\ol{a}},\ol{a})/\ol{W}_t\right).
$$
By Assumptions \ref{assumption:latent sra} and \ref{assumption:iv--outcome},

\begin{align*}
  \sum_{\ol{a}}\EE\left(\{\ol{A_t}=\ol{a_t}\}g(Y_{\ol{a}},\ol{a})/\ol{W}_t\right)
  &=\sum_{\ol{a}}\EE\left(\{\ol{A_t}=\ol{a_t}\}\EE\left(g(Y_{\ol{a}},\ol{a})\mid
    \ol{A}_t=\ol{a}_t,\ol{Z}_t,\ol{L}_t,\ol{U}_t \right)/\ol{W}_t\right)\\
  &=\sum_{\ol{a}}\EE\left(\{\ol{A_t}=\ol{a_t}\}\EE\left(g(Y_{\ol{a}},\ol{a})\mid
    \ol{A}_{t-1}=\ol{a}_{t-1},\ol{Z}_{t-1},\ol{L}_t,\ol{U}_t \right)/\ol{W}_t\right).
\end{align*}
Continuing with an application of Assumption \ref{assumption:iv--confounder},
\begin{align*}
  &\sum_{\ol{a}}\EE\left(\{\ol{A_t}=\ol{a_t}\}\EE\left(g(Y_{\ol{a}},\ol{a})\mid
    \ol{a}_{t-1},\ol{Z}_{t-1},\ol{L}_t,\ol{U}_t
  \right)/\ol{W}_t\right)\\
  &=  \sum_{\ol{a}}\EE\left(\frac{\{\ol{A}_{t-1}=\ol{a}_{t-1}\}}{\ol{W}_{t-1}}\EE\left(g(Y_{\ol{a}},\ol{a})\mid
    \ol{a}_{t-1},\ol{Z}_{t-1},\ol{L}_t,\ol{U}_t  \right)
  \frac{\PP(A_t=a_t\mid \ol{a}_{t-1},\ol{Z}_{t},\ol{L}_t,\ol{U}_t)}{\ol{W}_t}\right)\\
  &=  \sum_{\ol{a}}\EE\left(\frac{\{\ol{A}_{t-1}=\ol{a}_{t-1}\}}{\ol{W}_{t-1}}\EE\left(g(Y_{\ol{a}},\ol{a})\mid
    \ol{a}_{t-1},\ol{Z}_{t-1},\ol{L}_t,\ol{U}_t  \right)
    \Delta_t^{-1}\times\right.\\
  &\hspace{.5in}\left.\sum_{z_t\in\{0,1\}}\frac{(-1)^{1-z_t}\PP(A_t=a_t\mid
    \ol{a}_{t-1},\ol{Z}_{t-1},z_t,\ol{L}_t,\ol{U}_t)f_{Z_t}(z_t\mid \ol{A}_{t-1},\ol{Z}_{t-1},\ol{L}_t,\ol{U}_t)}{f_{Z_t}(z_t\mid \ol{A}_{t-1},\ol{Z}_{t-1},\ol{L}_t)}\right)\\
  &=  \sum_{\ol{a}}\EE\left(\frac{\{\ol{A}_{t-1}=\ol{a}_{t-1}\}}{\ol{W}_{t-1}}\EE\left(g(Y_{\ol{a}},\ol{a})\mid
    \ol{a}_{t-1},\ol{Z}_{t-1},\ol{L}_t,\ol{U}_t  \right)
  \Delta_t^{-1}\sum_{z_t\in\{0,1\}}(-1)^{1-z_t}\PP(A_t=a_t\mid
    \ol{a}_{t-1},\ol{Z}_{t-1},z_t,\ol{L}_t,\ol{U}_t)\right).
\end{align*}

Continuing with an application of Assumption \ref{assumption:ict},
\begin{align*}
&\sum_{\ol{a}}\EE\left(\frac{\{\ol{A}_{t-1}=\ol{a}_{t-1}\}}{\ol{W}_{t-1}}\EE\left(g(Y_{\ol{a}},\ol{a})\mid
    \ol{a}_{t-1},\ol{Z}_{t-1},\ol{L}_t,\ol{U}_t  \right)
  \Delta_t^{-1}\sum_{z_t\in\{0,1\}}(-1)^{1-z_t}\PP(A_t=a_t\mid
  \ol{a}_{t-1},\ol{Z}_{t-1},z_t,\ol{L}_t,\ol{U}_t)\right)\\
  &=\sum_{\ol{a}}\EE\left(\frac{\{\ol{A}_{t-1}=\ol{a}_{t-1}\}}{\ol{W}_{t-1}}\EE\left(g(Y_{\ol{a}},\ol{a})\mid
    \ol{a}_{t-1},\ol{Z}_{t-1},\ol{L}_t,\ol{U}_t  \right)\right)\\
  &=\sum_{\ol{a}}\EE\left(\frac{\{\ol{A}_{t-1}=\ol{a}_{t-1}\}}{\ol{W}_{t-1}}\EE\left(g(Y_{\ol{a}},\ol{a})\mid
    \ol{a}_{t-1},\ol{Z}_{t-1},\ol{L}_{t-1},\ol{U}_{t-1}  \right)\right)\\
  % &=\sum_{\ol{a}}\EE\left(\frac{\{\ol{A}_{t-1}=\ol{a}_{t-1}\}}{\ol{W}_{t-1}}\EE\left(g(Y_{\ol{a}},\ol{a})\mid
  %   \ol{a}_{t-1},\ol{L}_{t-1},\ol{U}_{t-1},\ol{Z}_{t-1}  \right)\right)\\
  &=\sum_{\ol{a}}\EE\left(\frac{\{\ol{A}_{t-1}=\ol{a}_{t-1}\}}{\ol{W}_{t-1}}g(Y_{\ol{a}},\ol{a})\right),
\end{align*}
as required.

\end{proof}
}
\begin{remark}
The conclusion of the theorem for the particular choice $g(y,\ol{a}) = h(\ol{a})(y-m(a)),h\in L_1(\ol{A})$, i.e.,
\begin{align}
\EE\left(h(\ol{A})(Y-m(\ol{A}))/\ol{W}\right)=\int_{\mathcal{A}^T}h(\ol{A})(Y_{\ol{A}}-m(\ol{A}))\mu_{\ol{A}^T}(\ol{a}),     
\label{eqn:weakened estimating eqn}
\end{align}
may be established under weaker forms of Assumptions
\ref{assumption:consistency} and \ref{assumption:latent
  sra}. % From line \comment{...} of the proof,
Assumption \ref{assumption:consistency} may be replaced with

\begin{assumptionbis}{assumption:consistency}\hspace{.1px}
  $\EE(Y\mid \ol{A}_t=\ol{a}_t,\ol{Z}_t,\ol{L}_t,\ol{U}_t)=  \EE(Y_{\ol{a}}\mid \ol{A}_t=\ol{a}_t,\ol{Z}_t,\ol{L}_t,\ol{U}_t)$,
  \label{assumption:weakened consistency}
\end{assumptionbis}
\noindent and Assumption \ref{assumption:latent sra} may be replaced with 
\begin{assumptionbis}{assumption:latent sra}\hspace{.1px}
  $\EE(Y_{\overline{a}}\mid \ol{A}_{t}=\ol{a}_{t},\ol{Z}_{t-1},\ol{L}_t,\ol{U}_t)=
  \EE(Y_{\ol{a}}\mid \ol{A}_{t-1}=\ol{a}_{t-1},\ol{Z}_{t-1},\ol{L}_t,\ol{U}_t)$,
  \label{assumption:weakened latent sra}
\end{assumptionbis}
\noindent where $\qquad t=1,\ldots,T,\quad\ol{a}\in\mathcal{A}^T.$
\end{remark}

As the range of $\ol{W}$ includes negative values, $1/\ol{W}$ are not
weights in the usual sense, a phenomenon that also occurs in other
IV-weighted moment equations for point exposure
\cite{wang2018,abadie2003}. The weights in fact have mean zero, as
follows by taking expectations on both sides of
$$
\EE(\ol{W}_T^{-1}\mid \an(Z_T))
= \ol{W}_{T-1}^{-1}\Delta_T^{-1}\EE\left(\frac{(-1)^{1-Z_T}}{f_{Z_T}(Z_T\mid \ol{A}_{T-1},{Z}_{T-1},\ol{L}_{T})}\bigg\vert\an(Z_T)\right)
= \ol{W}_{T-1}^{-1}\Delta_T^{-1} (1 - 1),
$$
although, as mentioned previously, they are almost surely non-zero under the assumptions for identification.

\begin{example}[Binary treatment]
  
  Assumption \ref{assumption:ict} may, in some situations, be interpreted as a
  condition on the ``compliance types'' \cite{angrist1996} of the population.
  
  When the treatment is binary, $A_t\in\{0,1\},t=1,\ldots,T$, so that
  $P(A_t=0\mid \an(A_t))=1-P(A_t=1\mid \an(A_t))$, the differences
  $\Delta_t$ satisfy
  $\Delta_t(A_t=1)=-\Delta_t(A_t=0),t=1,\ldots,T$. Consider an
  application in which $Z_t$ indicates whether a subject has been
  assigned to take an experimental or control treatment at time $t$
  and $A_t$ indicates whether the assigned treatment was or was not in fact
  taken. Then
  $0<\Delta(A_t=1)=P(A_t=1\mid \an(A_t),Z_t=1)-P(A_t=1\mid
  \an(A_t),Z=0)=P(A_t=0\mid \an(A_t),Z=0)-P(A_t=0\mid \an(A_t),Z=1)$
  has the interpretation that $A_t$ must concord with the IV $Z_t$, in
  the sense that individuals at stratum $\an(A_t,Z_t)$ are more likely
  at time $t$ to take the treatment when assigned to do so than when
  assigned not to do so. Analogously, when $\Delta(A_t=1)<0$,
  individuals at stratum $\an(A_t,Z_t)$ are more likely to do the
  opposite of their assignment.
  
  An additional assumption leads to an interpretation in terms of a
  well-studied causal notion, the compliance type. As with the
  treatment-indexed potential outcomes $Y_a$ defined earlier,
  IV-indexed potential treatments may also be defined, which we denote
  as $A_{t,Z_t}$. These potential outcomes may be cross-classified by
  the four possible pairs of values of $A_{t,Z_t}$ and
  $Z_t$. Experimental subjects for whom $A_{t,Z_t=1}=1$ and $A_{t,Z_t=0}=0$
  are termed ``compliers,'' as they comply with the assignment $Z_t$,
  and similarly for ``defiers,'' $A_{t,Z_t=1}-A_{t,Z_t=0}<0$,
  ``never-takers'', $A_{t,Z_t=0}=A_{t,Z_t=1}=0$, and
  ``always-takers,'' $A_{t,Z_t=0}=A_{t,Z_t=1}=1$. Suppose that,
  analogous to SRA, these potential outcomes are conditionally
  independent of the IV,
  $$
Z_t \cind A_{t,Z_t}
 \mid \overline{U}_t,\overline{L}_t,\overline{A}_{t-1}  ,\overline
 {Z}_{t-1}.
 $$
  Then Assumption \ref{assumption:ict} asserts that, at each stratum
  $\an(A_t,Z_t)$, the compliance type is mean-independent of unknown confounders,
  \begin{align*}
  \mathbb{E}\left(  A_{t,z_t=1}-A_{t,z_t=0}|\overline{U}_t,\overline
{L}_t,\overline{A}_{t-1}  ,\overline{Z}_{t-1}\right)
  =\Delta_{t}\left(  \overline{L}_t,\overline{A}_{t-1}
,\overline{Z}_{t-1} \right)  \text{, }t=0,...,T-1.\nonumber
  \end{align*}
  Under this interpretation, the inequalities $\Delta(A_t=1)>0$ or
  $\Delta(A_t=1)<0$ assert that a given stratum consists only of
  compliers or defiers. In any event, whether or not this
  interpretation is available, a population stratum cannot consist of
  never-takers or always-takers due to Assumption \ref{assumption:iv
    relevance}.   In this application, therefore, Assumption \ref{assumption:ict} is warranted when enough data on the
  patients are obtained to account for any systematic differences in
  compliance type.
  
  % In this application, Assumption \ref{assumption:ict} is warranted when enough data on the
  % patients are obtained to account for any systematic differences in
  % compliance type. However, less is necessary. Assumption \ref{assumption:ict}
  % requires only that compliance type be independent of unmeasured
  % treatment-outcome confounders. Viewed this way, a benefit of Theorem \ref{theorem:1} over
  % the SRA theorem is to allow an MSMM parameter to be estimated upon collecting
  % covariates that account for systematic differences both between treatment
  % and outcome and also compliance types, rather than all
  % treatment-outcome confounders.    
\end{example}

\begin{example}[Continuous treatment]
  \label{example:continuous treatment}
  We consider the implications of Assumption \ref{assumption:ict} for
  continuous treatment densities. First, because
  $f_{A_t\mid \an(A_t)}(a_t,Z_t=1)$ and
  $f_{A_t\mid \an(A_t)}(a_t,Z_t=0)$ both integrate to 1
  with respect to $\mu_{A_t}$, their difference $\Delta_t(a_t)$ must
  integrate to 0.  As the densities vanish at infinity, so must
  $\Delta_t$. As discussed in Section \ref{section:estimation}, IV
  estimators are typically unstable when the magnitude of $\Delta_t$
  is small, and therefore the tails must decay quickly for good
  performance. Second, $\Delta_t$ must be nonzero almost surely with
  respect to $\mu_{A_t}$, by Assumption \ref{assumption:iv
    positivity}.  Third, the nonnegativity of
  $f_{A_t\mid \an(A_t)}(a_t,Z_t=1)$ requires, for all
  $U_t$, that
  $|\Delta_t(a_t)|\le f_{A_t\mid \an(A_t)}(a_t,Z_t=0)$
  for $a$ such that $\Delta_t(a_t)<0$. The first two requirements hold
  for the difference of any two densities that are unequal
  a.s.$-\mu_{A_t}$, but the last is not as easily satisfied. It
  requires that for a range of densities obtained by varying $U_t$,
  adding $\Delta_t$ doesn't lead to a function that has negative
  values.

  An example is a location-scale parametrization for the treatment
  density. Let the baseline density $f_{A_t\mid \an(A_t)}(A_t=a,L_t=(l_1,l_2),U_t=u,Z_t=0)$ be
  normal $\phi((a-l_1)/u)/u$. The first component
  of the observed confounder $L_t$ controls the location and the
  unobserved confounder $U_t$ controls the spread. Let
  $\Delta_t(a\mid L_t=(l_1,l_2))$ be a difference between normal densities
  that does not depend on $U_t$, say, $\phi(a)-\phi(a/l_2)/l_2$. If the spread
  $u$ of the baseline density lies within an appropriate range, then $f_{A_t\mid \an(A_t)}(A_t=a,L_t=(l_1,l_2),U_t=u,Z_t=0) +
  \Delta_t(a)$ is a valid density for $A_t$. % (Controlling this
  % baseline spread is analogous to how previously we chose $\Delta$ using
  % $L_t$ only, and next chose the baseline $\PP(A_t\mid L_t,U_t)$ so that
  % $\PP(A_t\mid L_t,U_t)+\Delta$ fit in $(0,1)$.)
  In particular,
  given $L_t=(l_1,l_2)$ with $l_2\in (0,1)$, suppose $l_2<u<\min(1,l_2/(1-l)2))$ for the standard
  deviation $u$ of the baseline. Let $l_1=0$ since the location
  is irrelevant to the argument. Then, as shown in Appendix \ref{appendix:continuous treatment},
      $$f_{A_t\mid \an(A_t)}(A_t=a,L_t=(l_1,l_2),U_t=u,Z_t=1)>0.$$
      This method can be extended to other location-scale
  families but the restriction on the scale given in this example
  needs to be obtained anew, depending on the form of the densities. A
  small simulation is given in Appendix \ref{appendix:continuous treatment}.
\end{example}

\section{Partial converse}
\label{section:converse}
  Let data $(A_1,Z_1,L_1,U_1),(A_2,Z_2,L_2,U_2),\ldots,$ be given. Suppose there exists a process $\omega_1,\omega_2,\ldots$ adapted
  to the observed data such that for any $T,h,m$, and $Y$ compatible with $m$,
  $$
  \EE(h(\ol{A})\omega_T(\ol{A}_T,\ol{L}_T,\ol{Z}_T)(Y-m(A))=0.
  $$
  For example, under the assumptions of Theorem \ref{theorem:1},
  $\omega_t=1/W_t$, with $W_t$ given in (\ref{defn:weights}), is an
  example of such a process. In this section, we consider whether
  there are other processes in the class $\omega_1,\omega_2,\ldots,$
  that require less than Assumption \ref{assumption:ict}. % how much
  % Assumption \ref{assumption:ict} restricts the class of processes
  % $\omega_1,\omega_2,\ldots$.
  We relax the
  assumption that the time-varying instrument process $\{Z_t\}$ is
  binary. We do restrict the treatment process $\{A_t\}$ and the
  instrument process $\{Z_t\}$ to be discrete-valued.

  Given an MSMM $m(\ol{A})$, the residual $Y-m(\ol{A})$ may be
decomposed as the sum of two noise terms $\epsilon$ and $\eta$,
$$
\epsilon + \eta = Y - \EE(Y\mid \ol{A},\ol{Z},\ol{L},\ol{U}) +  \EE(Y\mid \ol{A},\ol{Z},\ol{L},\ol{U}) - m(\ol{A}).
$$
The first difference, $\epsilon$, is orthogonal to the vectors $(\ol{A},\ol{Z},\ol{L},\ol{U})$, whereas the
second, $\eta$, need not be.

Under Assumptions \ref{assumption:weakened latent sra}, \ref{assumption:iv--outcome}, and an MSMM $\EE(Y_{\ol{a}})=m(\ol{a})$, $\eta$ may be written as a sum of martingales restricted to the treatment levels $\ol{a}\in\mathcal{A}^T$,
\begin{align*}
  \eta &= \EE(Y\mid \ol{A},\ol{Z},\ol{L},\ol{U}) - m(\ol{A}) = \\
  % &=\sum_{\ol{a}\in\mathcal{A}^T}\{\ol{A}=\ol{a}\}\left(\EE(Y_{\ol{a}}\mid \ol{a},\ol{Z},\ol{L},\ol{U}) - m(\ol{a})\right)\\
  % &=\sum_{\ol{a}\in\mathcal{A}^T}\{\ol{A}=\ol{a}\}\left(\sum_{t=1}^T\left(\EE(Y_{\ol{a}}\mid \ol{a}_t,\ol{Z}_t,\ol{L}_t,\ol{U}_t)-\EE(Y_{\ol{a}}\mid \ol{a}_{t-1},\ol{Z}_{t-1},\ol{L}_{t-1},\ol{U}_{t-1})\right) +\EE(Y_{\ol{a}}) - m(\ol{a})\right)\\
       &= \sum_{\ol{a}\in\mathcal{A}^T}\{\ol{A}=\ol{a}\}\sum_{t=1}^T\left(\EE(Y_{\ol{a}}\mid\ol{a}_{t-1},\ol{Z}_{t-1},\ol{L}_t,\ol{U}_t)
         - \EE(\EE(Y_{\ol{a}}\mid\ol{a}_{t-1},\ol{Z}_{t-1},\ol{L}_t,\ol{U}_t) \mid \ol{a}_{t-1},\ol{Z}_{t-1},\ol{L}_{t-1},\ol{U}_{t-1}) \right) .
\end{align*}

For $1\le t\le T,\ol{a}\in\mathcal{A}^T$, let
$\eta_t(\ol{a},\ol{a}'_t,\ol{z}_t,\ol{l}_t,\ol{u}_t)=\EE(Y_{\ol{a}}\mid\ol{a'}_{t-1},\ol{z}_{t-1},\ol{l}_t,\ol{u}_t)
- \EE(\EE(Y_{\ol{a}}\mid\ol{a'}_{t-1},\ol{Z}_{t-1},\ol{L}_t,\ol{U}_t) \mid
\ol{a'}_{t-1},\ol{Z}_{t-1},\ol{L}_{t-1},\ol{U}_{t-1})$. Then for any $\ol{a}\in\mathcal{A}^T$,
         \begin{align*}
           \EE(Y\mid \ol{a},\ol{Z},\ol{L},\ol{U}) - m(\ol{a}) = \sum_{t=1}^T\eta_t(\ol{a},\ol{a}_t,\ol{Z}_t,\ol{L}_t,\ol{U}_t)
         \end{align*}
         and $\EE\left(\eta_t(\ol{a},\ol{a}_t,\ol{Z}_t,\ol{L}_t,\ol{U}_t)\mid \ol{a}_{t-1},\ol{Z}_{t-1},\ol{L}_{t-1},\ol{U}_{t-1}\right)=0$ for all $t$.

         Conversely,
         \begin{lemma}
           Let
           $\eta_t(\ol{a},\ol{a}'_{t-1},\ol{z}_{t-1},\ol{l}_t,\ol{u}_t),1\le
           t\le T,$ be measurable functions
           $\mathcal{A}^T\times\mathcal{A}^{t-1}\times\mathcal{Z}^{t-1}\times\mathcal{L}^t\times\mathcal{U}^t\to
           \mathbb{R}$ such that for all $\ol{a}\in\mathcal{A}^T$,
           $\eta_t(\ol{a},\ol{a}_{t-1},\ol{Z}_{t-1},\ol{L}_t,\ol{U}_t)$
           is integrable and
           $$\EE\left(\eta_t(\ol{a},\ol{a}_{t-1},\ol{Z}_{t-1},\ol{L}_t,\ol{U}_t)\mid
             \ol{a}_{t-1},\ol{Z}_{t-1},\ol{L}_{t-1},\ol{U}_{t-1}\right)=0.$$ Suppose for all
           $\ol{a}\in\mathcal{A}^T$, variables $Y_{\ol{a}}$ satisfy
         \begin{align}
           \EE(Y_{\ol{a}}\mid \ol{a'},\ol{z},\ol{l},\ol{u}) - m(\ol{a}) = \eta = \sum_{t=1}^T\eta_t(\ol{a},\ol{a'}_{t-1},\ol{z}_{t-1},\ol{l}_t,\ol{u}_t)
           \label{defn:converse:eta}
         \end{align}
          and a variable $Y$ satisfies
         \begin{align}
           \EE(Y\mid \ol{a},\ol{z},\ol{l},\ol{u})=\EE(Y_{\ol{a}}\mid\ol{a},\ol{z},\ol{l},\ol{u}),
           \label{defn:converse:Y}
         \end{align}
         with $\EE(Y_{\ol{a}}\mid\ol{a},\ol{z},\ol{l},\ol{u})$ as in (\ref{defn:converse:eta}). Then the data $(Y,\ol{A},\ol{Z},\ol{L},\ol{U})$ satisfy the MSMM $\EE(Y_{\ol{a}})=m(\ol{a})$ and Assumptions \ref{assumption:weakened consistency}, \ref{assumption:weakened latent sra}, and \ref{assumption:iv--outcome}.
         \label{lemma:1}
         \end{lemma}
       Lemma \ref{lemma:1} gives a class of distributions for outcomes $Y$ compatible with the previously described identification results. That is, if the remaining data $(\ol{A},\ol{Z},\ol{L},\ol{U})$ satisfy Assumptions \ref{assumption:iv--confounder} and \ref{assumption:ict}, then (\ref{eqn:weakened estimating eqn}) holds. This class of distributions for outcomes $Y$ are described by the endogenous noise $\eta$ in (\ref{defn:converse:eta}). As an example, $\eta_t(\ol{a},\ol{a}'_{t-1},\ol{z}_{t-1},\ol{l}_t,\ol{u}_t)=\zeta(\ol{a}'_{t-1},\ol{z}_{t-1},\ol{l}_t,\ol{u}_t)$ for some $1\le t\le T$ and $\eta_{t'}(\ol{a},\ol{a}'_{t'-1},\ol{z}_{t'-1},\ol{l}_{t'},\ol{u}_{t'})=0$ for $t'\neq t$, satisfies the requirements for (\ref{defn:converse:eta}) whenever $\zeta\in L_1(\ol{A}_{t-1},\ol{Z}_{t-1},\ol{L}_t,\ol{U}_t)$ and $\EE(\zeta \mid \ol{A}_{t-1},\ol{Z}_{t-1},\ol{L}_{t-1},\ol{U}_{t-1})=0$.

\begin{theorem}
  \label{theorem:2}
  Let data $A,Z,L,U,\ldots$ be given. Suppose there exists a process $\omega_t$ adapted
  to the observed data such that for any $T,h,m$,
  \begin{align}
  \EE(h(\ol{A}_T)\omega_T(\ol{A},\ol{L},\ol{Z}_T)(Y-m(\ol{A}_T))=0\label{eqn:theorem:converse}
  \end{align}
  whenever the data satisfy Assumptions \ref{assumption:weakened latent sra} and \ref{assumption:iv--outcome}
  and the MSMM
  $\EE(Y_{\ol{a}})=m(\ol{a}), \ol{a}\in\mathcal{A}^T$, holds. Then for
  any $t,a_t\in\mathcal{A},$ 
  \begin{align}
    \sum_{z_t\in\mathcal{Z}}f_{A_t\mid A_{t-1},Z_{t},L_t,U_t}(a_t\mid \ol{A}_{t-1},\ol{Z}_{t-1},z_t,\ol{LU}_t)\omega_t'(\ol{A}_{t-1},a_t,\ol{Z}_{t-1},z_t,\ol{L}_t)
    \label{conclusion:thrm:converse}
  \end{align}
  does not depend on $(L_t,U_t)$, where the process $\omega_t'$ also satisfies $\ref{eqn:theorem:converse}$.
\end{theorem}

The condition (\ref{conclusion:thrm:converse}) on the data imposed by the conclusion of Theorem
\ref{theorem:2} is similar to Assumption \ref{assumption:ict} insofar
as it requires a linear combination of the levels of the treatment
density given by the IV to be mean-independent of $U_t$ for each
$t$. Assumption \ref{assumption:ict} corresponds to the particular
linear combination given by the difference of the two levels of the
IV, assumed binary. Assumption \ref{assumption:ict} is, however,
stronger than the necessary condition (\ref{conclusion:thrm:converse})
since it requires mean-independence of the entire vector $\ol{U}$, not
just $U_t$.  Condition (\ref{conclusion:thrm:converse}) requires
additionally mean-independence of $L_t$, but that too is implied by
Assumption \ref{assumption:ict} by the choice of the weights
$\omega$. For example, if
$\alpha=\sum_{z_t\in\mathcal{Z}}f_{A_t\mid
  A_{t-1},Z_{t},L_t,U_t}(a_t\mid
\ol{A}_{t-1},\ol{Z}_{t-1},z_t,\ol{L}_t,\ol{U}_t)\omega_t(\ol{A}_{t-1},a_t,\ol{Z}_{t-1},z_t,\ol{L}_t)$
does not depend on $\ol{U}_t$ one may take
$\omega'_t=\omega_t/\alpha$. The difficulty in meeting the condition
is mean-independence of $U_t$, since the weights $\omega$ can only
depend on the observed data.
  
  On the other hand, Theorem \ref{theorem:2} allows the data
  $(A_1,Z_1,L_1,U_1),(A_2,Z_2,L_2,U_2),\ldots,$ to be given a priori, i.e.,
  (\ref{conclusion:thrm:converse}) is a necessary condition even if
  the weights are chosen based on the data process so long as the
  weights satisfy (\ref{eqn:theorem:converse}). Theorem \ref{theorem:2} therefore makes weaker assumptions about the weights than Theorem \ref{theorem:1}, % The assumptions
  % of the theorem are therefore weaker than what we demand of the
  % weights in Theorem \ref{theorem:1}, 
  where the same weights must hold for
  any data that satisfy Assumptions \ref{assumption:latent sra}--\ref{assumption:ict}. % Stronger
  % conditions naturally follow from imposing this stronger requirement.

\begin{example}[Point exposure, binary treatment and IV] When $T=1,
  A\in\{0,1\}$, and $Z\in\{0,1\}$, the conclusion of Theorem \ref{theorem:2} is that 
  % $$
  % \EE(\{A=0\}h(0)\omega(A=0,ZL)\mid LU) +
  % \EE(\{A=1\}h(1)\omega(A=1,ZL)\mid LU) = c_h, \text{ for any }h:\{0,1\}\to\mathbb{R},
  % $$
  \begin{gather}
  \begin{aligned}
  \omega(0,0,L)\PP(A=0\mid Z=0,L,U) + \omega(0,1,L)\PP(A=0\mid
    Z=1,L,U) &= c_0\\
  \omega(1,0,L)\PP(A=1\mid Z=0,L,U) + \omega(1,1,L)\PP(A=1\mid
  Z=1,L,U) &= c_1
  \label{conclusion:example:point exposure}
  \end{aligned}
\end{gather}
for constants $c_0,c_1\in\mathbb{R}$.
  % where $c_h$ is a constant that may depend on $h$. Taking
  % $h(0)=1,h(1)=0$, and then $h(0)=0,h(0)=1$, it follows that
  % \begin{align*}
  % \omega(0,0,L)\PP(A=0\mid Z=0,LU)f_Z(0\mid L) + \omega(0,1,L)\PP(A=0\mid
  %   Z=1,LU)f_Z(1\mid L) = c_0
  % \end{align*}
  Since $P(A=1\mid A,Z,L)=1-P(A=0\mid A,Z,L)$, (\ref{conclusion:example:point exposure}) is
  \begin{align*}
  \omega(0,0,L)\PP(A=0\mid Z=0,L,U) + \omega(0,1,L)\PP(A=0\mid
    Z=1,L,U) &= c_0\\
  \omega(1,0,L)\PP(A=0\mid Z=0,L,U) + \omega(1,1,L)\PP(A=0\mid
    Z=1,L,U) &= \omega(1,0,L)+ \omega(0,1,L) - c_1.
  \end{align*}
  Fixing $L$ and letting $U$ vary leads to an overdetermined system of linear equations, implying
  \begin{gather}
    \begin{aligned}
      \omega(0,1,L)/\omega(0,0,L) &= \omega(1,1,L)/\omega(1,0,L)\\%\eqqcolon R(L)\\
      \omega(1,0,L)/\omega(0,0,L) &= \omega(1,1,L) + \omega(1,0,L) - 1.
      \label{eg:point exposure:2}
    \end{aligned}
  \end{gather}
The latter must hold for any $L$ unless $P(A=1\mid Z,L,U)$ is constant
  with respect to $U$. Conversely, let the data $A,Z,L,U,$ and $\PP(A=0\mid Z=0,L,U)<1,\omega(0,0,L),\omega(0,1,L)$ be given. Then (\ref{eg:point exposure:2}) determine
  $\PP(A=0\mid Z=1,L,U),\omega(1,0,L),\omega(1,1,L)$.

  Suppose the common value of the ratios in the first line of (\ref{eg:point exposure:2}) is -1. Then for $a\in\{0,1\}$
  \begin{align*}
  c_a&=\omega(a,0,L)\PP(A=a\mid Z=0,L,U) + \omega(a,1,L)\PP(A=a\mid
    Z=1,L,U)\\
  &=\omega(a,0,L)\PP(A=a\mid Z=0,L,U) - \omega(a,0,L)\PP(A=a\mid
    Z=1,L,U)
  \end{align*}
  or,
  $$
  \PP(A=a\mid Z=0,L,U) - \PP(A=a\mid
    Z=1,L,U)% \in m(L), a\in\{0,1\}    
    $$
  is a function of $L$ only, for $a\in\{0,1\}$. In \cite{wang2018} the authors establish that
  this condition is, in fact, sufficient to identify the MSMM
  parameter in the binary IV, binary exposure, $T=1$, setting
  considered here, when that parameter is  the ATE (defined in Example \ref{example:wald}).
  % Suppose we insist that the same weights solve ((ref est eqn)) for
  % any data $AZLU$. That is, the functional form of
  % $W:\{0,1\}\times\{0,1\}\times\mathbb{R}$ is fixed, whatever the law
  % of the data. Then taking $\PP(A=0\mid Z=1,L,U)=\PP(A=0\mid
  % Z=0,L,U)+\delta(L)$ for some real-valued function$\delta$ and choice
  % of $\PP(A=0\mid
  % 0,L,U)$, ((ref above eqn)) implies
  % $$
  % (W(0,0,L)+W(0,1,L))\PP(A=0\mid 0,L,U) = c_0-W(0,1,L)\delta(L)
  % $$
  % implying either $\PP(A=0\mid Z=0,L,U)$ is a function only of $L$ or
  % that $W(0,0,L)+W(0,1,L)$ a.s. By choosing $\PP(A=0\mid Z=0,L,U)$ to vary
  % with $U$ and the support of $L$ to include $\mathbb{R}$, it follows
  % that $W(0,0,\cdot)=-W(0,1,\cdot)$. Then ((ref above W conditions))
  % implies $W(\cdot,0,\cdot)=-W(\cdot,1,\cdot)$. In ((linbo paper)) the authors establish that
  % this condition is, conversely, sufficient to identify the MSMM
  % parameter in the binary IV, binary exposure, $T=1$, setting
  % considered here, when that parameter is  the ATE ((see example
  % with ATE for defn)).
  
\end{example}

\section{Estimation and Inference}
\label{section:estimation}
Let $\EE(Y_{\ol{a}})=m_\beta(\ol{a})$ be an MSMM, and suppose that 
the assumptions of Theorem \ref{theorem:1} hold. Then
$\EE\left(h(\ol{A})(Y-m_\beta(\ol{A}))/\ol{W})\right)=0$, and
\begin{align}
\PP_ns_\beta = \PP_n\left(
  h(\ol{A})(Y-m_\beta(\ol{A}))/\ol{W}\right)\label{eqn:estimating eqn}
\end{align}
may serve as estimating equations for $\beta$, where $s_\beta=h(\ol{A})(Y-m_\beta(\ol{A}))/\ol{W}$. When the MSMM is
linear, $m_\beta(\ol{A})=\beta^T\ol{A}$,
the solution to $\PP_ns_\beta=0$ is a weighted least squares estimator,
\begin{align}
\hat{\beta}=(\PP_n(h(\ol{A})\ol{A}^T/\ol{W}))^{-1}\PP_nh(\ol{A})Y/\ol{W}.
\label{eqn:estimation:closed form}
\end{align}
In practice, $\ol{W}$ may not be known, and a $\sqrt{n}$-consistent estimate, say $\widehat{\ol{W}}$, may be substituted,\begin{align}
\hat{\beta}=(\PP_n(h(\ol{A})\ol{A}^T/\widehat{\ol{W}}))^{-1}\PP_nh(\ol{A})Y/\widehat{\ol{W}}.                                                                                                                                                             \label{eqn:estimation:closed form observed}
                                                                                                                                                           \end{align}                                                                                                                                                           To describe the practical behavior of the estimator $\hat{\beta}$ given by (\ref{eqn:estimation:closed form observed}),
suppose nuisance parameters include $\alpha$, parameterizing
$\Delta_t$; $\gamma$, parameterizing $f_{Z_t}, t=1,\ldots,T$; and
$\nu$, containing any additional nuisance parameters. In the
parametrization described in Section \ref{section:simulation} below,
for example, $\nu$ parametrizes the ``baseline'' probability
$\PP(A_t=1\mid \ol{A}_{t-1},\ol{L}_t,\ol{Z}_{t-1},Z_t=0)$.  Besides
$s_\beta$, let $s_{\alpha}, s_{\gamma},$ and $s_{\nu}$ be estimating
equations for $\alpha_t, \gamma_t,$ and $\nu_t$, collected as
$s=(s_\beta,s_\alpha,s_\gamma,s_\nu)$. That is, they are functions of
the observed data $\ol{O}=(\ol{A},\ol{L},\ol{Z})$ and parameters
$(\beta,\alpha,\gamma,\nu)$ such that, if the data is generated under
parametrization $(\beta_0,\alpha_0,\gamma_0,\nu_0)$, then
$\EE(s_\beta(O;\beta_0))=\EE(s_\alpha(O;\alpha_0))=\EE(s_\gamma(O;\gamma_0))=\EE(s_\nu(O;\nu_0))=0$. In the parametrization described in
Section \ref{section:simulation} below, for example, we use maximum
likelihood to estimate $\alpha,\gamma,$ and $\nu$, and the
estimating equations are scores for the model. By a standard
expansion, the influence function for the estimator
$(\hat{\beta},\hat{\alpha},\hat{\gamma},\hat{\nu})$ is
\begin{align}
-\left(\EE\left(\frac{\partial s}{\partial
  (\beta^T,\alpha^T,\gamma^T,\nu^T)}\right)\right)^{-1}s.\label{eqn:influence function}
\end{align}
Provided the usual regularity conditions for M-estimation hold, the
solution $\hat{\beta}$ to (\ref{eqn:influence function}) is asymptotically normal with
influence function given by the first $p$ components of (\ref{eqn:influence function}),
where $p$ is the dimension of $\beta$. Inference may be carried  out
with the nonparametric bootstrap or the ``sandwich'' asymptotic
variance estimator; we compare both in Section \ref{section:simulation}.

If many observations $n$ are available relative to $T$, separate
models may be imposed and estimated at different time points; if $T$
is small relative to $n$, the data may be pooled to estimate a single
model common to all time points. In the latter case, 

\begin{align}
\frac{\partial
  s_\beta}{\partial (\beta^T,\alpha^T,\gamma^T,\nu^T)} =
-h(\ol{A})\left(\widehat{\ol{W}}^{-1}\frac{\partial}{\partial\beta^T}m_\beta,
  \frac{Y-m_\beta}{\widehat{\ol{W}}}\sum_t\frac{\frac{\partial}{\partial\alpha^T}\Delta_t(\alpha)}{\Delta_t(\alpha)},\frac{Y-m_\beta}{\widehat{\ol{W}}}\sum_t\frac{\frac{\partial}{\partial\gamma^T}f_{Z_t}(\gamma)}{f_{Z_t}(\gamma)},0\right).
  \label{eqn:information}
\end{align}
% \comment{did I take logs for the Delta component...doesnt work if Delta can be negative. also seems like I forgot h function at least in last 2 components, should be part of score for beta.}
The form of the remaining components of the matrix $\frac{\partial s}{\partial
  \beta,\alpha,\gamma,\nu}$ will depend on the parametrization chosen;
see Section \ref{section:simulation} for an example.

\begin{example}[Linear omitted variables model, comparing biases]
  \label{example:linear}
  % We compare the bias of estimators obtained by varying the choice of
  % weights.
  Given a linear MSMM, suppose an estimator is obtained as the root of a weighted
  estimating equation
  $
  \PP_n\left(\omega h(\ol{A})\left(Y-\beta^T\ol{A}\right)\right)=0,
  $
  where the weight $\omega$ is an integrable function of the observed data
  $(\ol{A},\ol{Z},\ol{L})$. This root is a weighted least squares
  estimator
  \begin{align}
    \hat{\beta}=\left(\PP_n(\omega h(\ol{A})\ol{A}^T)\right)^{-1}\PP_n\left(\omega h(\ol{A})Y\right).
    \label{eqn:example:bias}
  \end{align}
   Suppose the data satisfy the assumptions of Theorem \ref{theorem:1} and the observed outcome is
  $$
  % \EE(Y\mid \ol{A},\ol{Z},\ol{L},\ol{U})
  Y = \sum_t
(\beta_{L_t}g_{L_t}(L_t) + \beta_{U_t}g_{U_t}(U_t)) + \beta^T\ol{A} +
\epsilon,
  $$
  with $\epsilon$ exogenous, $\beta_{L_t},\beta_{U_t}\in\mathbbm{R}^p$,
  and
  $\EE(g_{L_t}(L_t)\mid\ol{L}_{t-1},\ol{A}_{t-1})=\EE(g_{U_t}(U_t)\mid\ol{U}_{t-1},\ol{A}_{t-1})=0$.  As
  discussed in the passage following Lemma \ref{lemma:1}, this
  outcome model is consistent with the MSMM
  $$
  \EE(Y_{\ol{a}})=\beta^T\ol{a}.
  $$
  %  The estimator for the MSMM parameter $\beta$ using
  % empirical estimating equations is a weighted least squares
  % The estimator (\ref{eqn:example:bias}) is 
  % $$
  % \hat{\beta} = \left(\PP_n\left(h(\ol{A})\ol{A}^T/\omega\right)\right)^{-1}\PP_n\left(\omega^{-1}h(\ol{A})\left(\sum_t(\beta_{L_t}g_{L_t}(L_t) + \beta_{U_t}g_{U_t}(U_t)+\epsilon)\right)\right) + \beta.
  % $$
  We consider the asymptotic bias of estimator (\ref{eqn:example:bias}) in this setting,
  $$
  \plim \hat{\beta}-\beta=\left(\PP_n\left(h(\ol{A})\ol{A}^T/\omega\right)\right)^{-1}\PP_n\left(\omega^{-1}h(\ol{A})\left(\sum_t(\beta_{L_t}g_{L_t}(L_t) + \beta_{U_t}g_{U_t}(U_t))\right)\right),
  $$
   for various choices of weights $\omega$. Details are given in Appendix \ref{appendix:linear}.

  When $\omega=1$, the resulting estimator $\hat{\beta}$, known as the
  ``associational'' or ``crude'' estimator, ignores all confounding. The implied model is misspecified by omitting covariates $L_t,U_t$. The
  bias is
  $$
  \left(\EE\left(h(\ol{A})\ol{A}^T\right)\right)^{-1}\EE\left(h(\ol{A})\left(\sum_tg_{L_t}(L_t)+g_{U_t}(U_t)\right)\right).
  $$
  This bias is related to the strength of the dependency between the
  treatments and all confounders, known and unknown.
  The SRA estimator is given by the choice
  $\omega=1/\ol{W}^{(SRA)}=1/\prod_tf(A_t\mid \ol{L}_t,\ol{A}_{t-1})$. Its bias
  $$
  \left(\int_{\mathcal{A}}h(\ol{a})\ol{a}^t\mu_{\ol{A}}(\ol{a})\right)^{-1}\EE\left(h(\ol{A})\left(\sum_tg_{U_t}(U_t)\right)/\ol{W}^{(SRA)}\right)
  $$
  is related to the dependence between the treatment and unknown
  confounders, as expected due to the violation of SRA. In comparison with the bias of the associational estimator, the
  term corresponding to treatment and known confounder dependency
  is eliminated. % As $\sum_tg_{U_t}(U_t)$ and
  % $h(\ol{A})/\ol{W}^{(SRA)}$ are generally correlated when $U_t$ are in fact confounders, the bias is
  % nonzero.
  % Unless it happens that the treatment--known
  % confounder and treatment--unknown confounder dependencies cancel,
  % the associational estimator will have greater bias.--not true
  % because weight term could change things
  When $\omega$ is the IV weights (\ref{defn:weights}), the asymptotic bias is
  zero since we have assumed the conditions of Theorem \ref{theorem:1}.
  
  A Monte Carlo simulation comparing these three estimators is described in Section \ref{section:simulation}.
  % 2.  focus on 1 time point. suppose $Y$ is linear in the
  % confounders and treatment, $Y=\beta_LL+\beta_UU+\beta A+\epsilon$,
  % for $\epsilon$ independent of $L,U$, and Assumption ((latent SRA))
  % is satisfied. Then for $a\in\mathcal{A}$,
  % $\EE(Y_a\mid L,U)=\EE(Y_a\mid A=a,L,U)=\EE(Y\mid
  % A=a,L,U)=\beta_LL+\beta_UU+\beta a$, and the data satisfies the
  % MSMM $\EE(Y_a)=\beta a$. If an instrument $Z$ satisfying
  % Assumptions ((...)) is available, then by lemma,
  % 2. bias when sra weights are used\\
  % 3. When $W=1$,
  % $\hat{\beta}=(A^TA)^{-1}(A^TA)\beta + (A^TA)^{-1}(L+U+\epsilon)$
  % is the OLS estimator. The bias in this case is
  % $\EE(\hat{beta}-\beta)=\EE((A^TA)^{-1}(L+U))$. ((possibility of
  % known and unknown bias cancelling)).
\end{example}

\begin{example}[Wald Estimator]\label{example:wald} Suppose $T=1$, %and  the conditional
  % density of the IV and $\Delta_t$ terms in ((our weights)) are
  % functions of treatment alone,
    $\Delta_1(a_1,L_1)=\Delta_1(a_1)$, and $f_{Z_1\mid L_1,U_1 }(z_1,L_1,U_1)=1/2$.  For purposes of estimation, both $f_{Z_1}$ and $\Delta_1$
  terms in the IV weights (\ref{defn:weights}) may be canceled by taking
  $h$ in the estimating equations (\ref{eqn:estimating eqn}) to be
  $$
  h(a_1)=h_1(a_1)\Delta_1(a_1)/2,%(f_{Z_1\mid L_1,U_1 }(z_1)\Delta_1(a_1,L_1)=\Delta_t(a_t))^{-1}.
  $$
  with $h_1(a_1)$ available to be specified.  The remaining weight term is just $(-1)^{1-Z_1}$. Consider the
  regression model
  $$
  \EE(Y_a)=\beta a.
  $$
  Taking $h_1(a_1)=1$, the solution to the estimating equation (\ref{eqn:estimating eqn}) is then
  \begin{gather}
  \begin{aligned}
    \hat{\beta}&=\left(\PP_nA(-1)^{1-Z}\right)^{-1}\PP_n(-1)^{1-Z}Y\\
    &=\frac{\PP_nY\{Z=1\} - \PP_nY\{Z=0\}}{\PP_nA\{Z=1\} - \PP_nA\{Z=0\}}.
    \label{defn:example:wald:wald}
  \end{aligned}
\end{gather}
This estimator is known as the Wald estimator. If $Z$ is an IV and the
consistency assumption (\ref{assumption:consistency}) is satisfied,
the Wald estimator is consistent for the ``average treatment effect,''
the average difference in the potential outcome $Y_a$ across the two
groups defined by $a$. In \cite{wang2018}, the authors directly
establish identification of the ATE using IVs and provide further
results on estimation.
  
  The finite sample mean of the Wald estimator may be infinite. For example, when $A$ is discrete, the denominator, viewed as a random walk, is 0 with a positive probability on the order of $1/\sqrt{n}$. The variance estimator
  obtained from the influence function, suggested in Section
  \ref{section:estimation}, is asymptotic.%  % \comment{e.g. when
%   % the density of the denominator is positive and continuous at $0$--state condition in terms of A density, or drop}
% For example, suppose $A$ is discrete, and multiply the numerator and denominator of the ratio in (\ref{defn:example:wald:wald}) by $n$. The 
%   denominator, viewed as a process indexed by $n$, is a random walk on the integers that
%   goes up with probability $\PP(A=1,Z=1)$, down with probability
%   $\PP(A=1,Z=0)$% \comment{after multiplying top and bottom by n}
%   , and does not move otherwise. The probability the
%   denominator is 0 is then the probability the walk is at 0, a positive probability that % is $\mathcal{O}(4^{-n})$
%   decreases as $1/\sqrt{n}$. 
\end{example}

\begin{example}[Two-state markov chain]\label{eg:markov}
  We examine the relationship between confounding 
  and the variance of the estimator obtained from the estimating equation (\ref{eqn:estimating eqn}), using a simple model to
  compare expressions in the SRA and IV contexts. Details are given in Appendix \ref{appendix:markov}.

  SRA weights include probability densities at each time point, and IV
  weights include a difference of densities. As the number of time
  points $T$ grows and these weights are multiplied, an estimator may
  quickly become unstable. Let $\hat{\beta}$ be
  obtained as the solution to (\ref{eqn:estimating eqn}). Assuming standard
  regularity conditions, the asymptotic variance of $\hat{\beta}$ is the variance of
  the influence function,
  \begin{align}
  \Var(\sqrt{n}(\hat{\beta}-\beta_0))\to
    \left(\EE\frac{\partial}{\partial \beta}(hm_\beta/\ol{W})\right)^{-2}\EE\left((h(\ol{A})(Y-m_\beta)/\ol{W})^2\right).\label{eg:markov:IF}    
  \end{align}
  % The initial factor is, by ((main lemma)),
  % $$
  % \EE((\sum_tA_t)^2/\ol{W})=\sum_{\ol{a}}(\sum_ta_t)^2=\sum_{j=1}^Tj^2{T\choose j}=2^{T-1}T(T+1),
  % $$ and does not depend
  % on the weights. A first order approximation to the asymptotic
  % variance is
  % $$
  % (2^{T-1}T(T+1))^2\EE(\eta^2+\epsilon^2)\EE(1/\Pi_tW_t^2).
  % $$
  In this display, the weights $\ol{W}$ refer generically to either
  SRA weights (\ref{defn:sra weights}) or IV weights (\ref{defn:weights}). The term $h(\ol{A})m_\beta(\ol{A})$
  is a function of the treatments, so by (\ref{thrm:robins}), in the
  case $\ol{W}$ are SRA weights, or by Theorem \ref{theorem:1}, in the case of
  IV weights,
  $$
  \EE\frac{\partial}{\partial
    \beta}\left(hm_\beta/\ol{W}\right)\vert_{\beta=\beta_0}
  =\int \frac{\partial}{\partial
    \beta}hm_\beta d\mu_{\ol{A}}
  $$
  does not depend on the weights. A first order approximation to the asymptotic
  variance is
  \begin{align}
  \left(\int \frac{\partial}{\partial
    \beta}hm_\beta d\mu_{\ol{A}}\right)^{-2}\EE\left((h(\ol{A})(Y-m_\beta))^2\right)\EE\left(1/\Pi_tW_t^2\right).\label{eg:markov:2}
  \end{align}
  This expression appears to grow exponentially in the number of time
  points. In the SRA framework,
  various techniques have been proposed to stabilize the
  weights. Since weights are only needed to identify the the causal parameter when confounding is present, an approach to improve efficiency is, speaking loosely, to use SRA weights only to the extent required by confounding present in the data. This approach is carried out by a suitable choice of $h:\mathcal{A}^T\to\mathbb{R}$ in (\ref{eg:markov:IF}). We consider an analogous approach in the case of IV weights.

  To illustrate SRA weight stabilization, consider a simple two-state
  markov chain as a model for the treatment and confounding
  process. In this model, the covariates and treatment are binary and
  $\PP(L_t\mid\an(L_t))=\PP(L_t\mid A_{t-1})=p_{LA}$,
  $\PP(A_t\mid \an(A_t))=\PP(A_t\mid L_{t-1})=p_{AL}$. See Fig. \ref{eg:markov:DAG:SRA}.
  In Appendix \ref{appendix:markov}, it is shown that the contribution of the weight
  term
  is $\EE(1/\ol{W}^{2})=(\EE(1/W_1^2))^T=(p_{LA}(1-p_{LA}))^{-T}$. Thus,
  the contribution is minimized over $p_{LA}$ at $1/2$, when
  treatment and covariate are independent, and increases without bound
  as $|p_{LA}-1/2|\to 1/2$. On the other hand, suppose $h$ is chosen
  so that the modified weights
  $\ol{W}=\prod_t\frac{f(A_t|L_t)}{f(A_t\mid A_{t-1})}$ are used. With
  this choice of weights, $\EE(1/\ol{W}^{2})$ may in fact be
  bounded. Its value is determined not by $1/(p_{LA}(1-p_{LA}))$ as in
  the unstabilized case, but by the ratio
  $(p_{AL}(1-p_{AL}))/(p_{LA}(1-p_{LA}))$. The qualitative result is that
  stabilized weights are bounded as $t$ grows when the degree of
  treatment-covariate confounding does not grow faster than the
  treatment's predictiveness of the covariate.

  Next, consider an extension of the two-state markov model allowing
  for unknown confounding and satisfying the assumptions of Theorem
  \ref{theorem:1}. The model is a mixture of two independent chains of
  the type just described, say,
  $\ldots \to L_{t-1}\to A_{t-1}^L\to L_t\to\ldots$ with parameters
  $p_{AL},p_{LA}$, and
  $\ldots\to U_{t-1}\to A_{t-1}^U\to U_t\to\ldots$ with parameters
  $p_{AU},p_{UA}$. Suppose an exogenous IV process $\{Z_t\}$ is also
  available, and let
  $\delta_l=\PP(A_t=1\mid L_t=l,Z_t=1)-\PP(A_t=1\mid L_t=l,Z_t=0)$ for
  $l\in\{0,1\}$. See Fig. \ref{eg:markov:fig:DAG:IV}. The IV weights
  are then given by $(-1)^{1-Z_t}\delta_{L_t},t=1,\ldots,T$. As with the SRA weights, the
  second moment of the IV weights $\EE(1/\ol{W}^{2})$ is exponential
  in the number of time points. Corresponding to the confounding term
  $p_{AL}$, which determined the rate of growth in the unstabilized
  SRA case, $\omega = 1/(\delta_0\delta_1)$ and
  $\kappa = 1/\delta_0^2-1/\delta_1^2$ determine the rate of growth of
  the IV weights. The former may be interpreted as a measure of IV
  weakness, and the latter as a measure of the degree of confounding
  by the IV process and the known confounder process $\{L_t\}$.

  Analogously to SRA weights, we consider stabilizing a weight term $\delta_{L_t}$ by
  an arbitrary term depending on the treatment previous to $L_t$, say,
  $\gamma_{A_{t-1}}$, with values $\gamma_0,\gamma_1$. Upon
  computing the second moment $\EE(1/\ol{W}^{2})$, one finds that the
  contribution due to $\kappa$ may be controlled, but the variance due $\omega$
  remains. The growth remains exponential in time. Therefore, while
  stabilization is helpful, it is not as helpful as in the SRA
  setting, where the variance of the weights may be bounded.

  The difference between the SRA and IV cases seems to be the
  following. In both cases the stabilization terms may be assumed to
  integrate to 1, since multiplying $h$ by a constant does not change
  the influence function (\ref{eg:markov:IF}) or its variance. In the
  case of SRA weights, the weights themselves also satisfy this type of property, being densities. Specifically, the terms
  $\prod_tf(a_t\mid l_{t-1})$ cannot be uniformly small across all
  choices $a_t,l_{t-1},t=1,\ldots,T$. One may therefore hope to choose
  the stabilizing terms to match the magnitude of the corresponding
  weight terms. The IV weights do not satisfy this type of property,
  i.e., $\delta_0$ and $\delta_1$ may both be arbitrarily small at the
  same time, and no choice of $(\gamma_0,\gamma_1)$, which cannot both
  be small at the same time due to the mentioned scale invariance, will control
  the weights.
  \begin{figure}
    \centering
    \begin{tikzpicture}
  \def\Tx{0}
  \def\Ty{0}
  \def\offset{2.5}
  \def\Ux{\Tx+2*\offset}
  \def\Uy{\Ty}
  \node[shape=circle,draw=black] (L1) at (\Tx,\Ty) {$L_1$};
  \node[shape=circle,draw=black] (A1) at (\Tx+\offset,\Ty) {$A_1$};
  \node[shape=circle,draw=black] (L2) at (\Ux,\Uy) {$L_2$};
  \node[shape=circle,draw=black] (A2) at (\Ux+\offset,\Uy) {$A_2$};

  \draw [-latex] (L1) to [bend left=0] (A1);
  % \draw [-latex] (L1) to [bend left=25] (L2);
  % \draw [-latex] (L1) to [bend left=25] (A2);
  \draw [-latex] (A1) to [bend left=0] (L2);
  % \draw [-latex] (A1) to [bend left=25] (A2);
  \draw [-latex] (L2) to [bend left=0] (A2);
  \draw [-latex] (L1) to [bend left=0] (A1);

  \node[shape=circle,draw=black] (Y) at (\Ux+2.2*\offset,0) {$Y$};

  \draw [-latex] (L1) to [bend left=25] (Y);
  % \draw [-latex] (A1) to [bend left=25] (Y);
  \draw [-latex] (L2) to [bend left=25] (Y);
  \draw [-latex] (A1) to [bend left=25] (Y);
  \draw [-latex] (A2) to [bend left=0] (Y);

  \node[below] at ({\Tx+\offset+(\Ux-\offset)/2},\Ty-.2) {$p_{AL}$};  
  \node[below] at (\Ux+\offset/2,\Ty-.2) {$p_{LA}$};  
\end{tikzpicture}

% \begin{tikzpicture}
%   \def\Ax{0}
%   \def\Ay{0}
%   \def\offset{2}
%   \def\Bx{\Ax+5}
%   \def\By{\Ay}
%   \node[shape=circle,draw=black] (A0) at (\Ax,\Ay) {$A(j)$};
%   \node[shape=circle,draw=black] (L0) at (\Ax-\offset,\Ay) {$L(j)$};
%   \node[shape=circle,draw=black,font=\tiny] (A1) at (\Bx,\By) {$A(j+1)$};
%   \node[shape=circle,draw=black,font=\tiny] (L1) at
%   (\Bx-\offset,\By) {$L(j+1)$};

%   \draw [-latex] (L0) to [bend left=0] (A0);
%   \draw [-latex] (L0) to [bend left=30] (L1);
%   \draw [-latex] (L0) to [bend left=30] (A1);
%   \draw [-latex] (A0) to [bend left=0] (L1);
%   \draw [-latex] (A0) to [bend left=30] (A1);
%   \draw [-latex] (L1) to [bend left=0] (A1);
%   \draw [-latex] (L0) to [bend left=0] (A0);

%   % \draw[->]  (\Ax-3,\Ay) -- (L0);
  
%   \node at (\Ax-4,0) {. . . };
%   \node at (\Bx+1.8,0) {. . . };

%   \node[shape=circle,draw=black] (Y) at (\Bx+3.2,0) {$Y$};
%   \draw [-latex] (\Bx+2.5,.5) to (Y);
%   \draw [-latex] (\Bx+2.4,.4) to (Y);
%   \draw [-latex] (\Bx+2.3,.3) to (Y);
%   \draw [-latex] (\Bx+2.25,.2) to (Y);
% \end{tikzpicture}
    \caption{DAG for the two-state markov model meeting the sequential
      randomization assumption, with 2 time points.}
    \label{eg:markov:DAG:SRA}
  \end{figure}
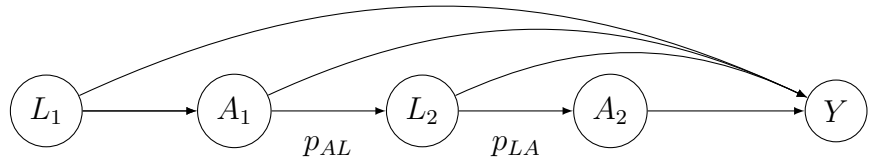

  \begin{figure}
    \centering
    \begin{tikzpicture}
    \def\Ax{2}
    \def\Ay{0}
    \def\offset{2.5}
    \def\Bx{\Ax+5}
    \def\By{\Ay}
    \node[shape=circle,draw=black] (A1) at (\Ax,\Ay) {$A_1$};
    \node[shape=circle,draw=black,fill=lightgray] (U1) at (\Ax-\offset,\Ay+\offset/1.5) {$U_1$};
    \node[shape=circle,draw=black] (L1) at (\Ax-\offset,\Ay-\offset/1.5) {$L_1$};
    \node[shape=circle,draw=black] (Z1) at (\Ax,\Ay-1.4*\offset) {$Z_1$};
    \node[shape=circle,draw=black] (A2) at (\Bx,\By) {$A_2$};
    \node[shape=circle,draw=black,fill=lightgray] (U2) at (\Bx-\offset,\By+\offset/1.5) {$U_2$};
    \node[shape=circle,draw=black] (Z2) at (\Bx,\By-1.4*\offset) {$Z_2$};
    \node[shape=circle,draw=black] (L2) at (\Bx-\offset,\By-\offset/1.5) {$L_2$};
    \node[shape=circle,draw=black] (Y) at (\Bx+3,\By) {$Y$};

    \draw [-latex] (U1) to (A1);
    \draw [-latex] (L1) to (A1);
    \draw [-latex] (U2) to (A2);
    \draw [-latex] (L2) to (A2);
    \draw [-latex] (Z1) to (A1);
    \draw [-latex] (Z2) to (A2);
    % \draw [-latex] (Z1) to (Z2);
    % \draw [-latex] (A1) to (A2);
    \draw [-latex] (A1) to (L2);
    \draw [-latex] (A1) to (U2);

    % \draw [-latex] (L1) to (Z1);
    % \draw [-latex] (L1) to (Z1);
    % \draw [-latex] (L1) to (Z2);
    % \draw [-latex] (L1) to [bend left=0] (L2);
    % \draw [-latex] (L1) to [bend right=15] (U2);
    % \draw [-latex] (L1) to [bend left=0] (A2);
    % \draw [-latex] (U1) to [bend left=0] (U2);
    % \draw [-latex] (U1) to [bend left=15] (L2);
    % \draw [-latex] (U1) to [bend left=0] (A2);

    \draw [-latex] (L1) to [bend right=25] (Y);
    \draw [-latex] (L2) to [bend right=0] (Y);
    \draw [-latex] (U1) to [bend left=25] (Y);
    \draw [-latex] (U2) to [bend left=0] (Y);
    \draw [-latex] (A2) to [bend left=0] (Y);
    \draw [-latex] (A2) to [bend left=0] (Y);

    % \draw [-latex] (L1) to [bend left=0    ] (U2);
    % \draw [-latex] (L1) to [bend left=0] (L2);
    % \draw [-latex] (L1) to [bend right=10] (A2);
    % % \draw [-latex] (L1) to [bend left=10] (L2);
    % \draw [-latex] (A1) to [bend left=10] (U2);
    % \draw [-latex] (A1) to [bend left=0] (Z2);
    % \draw [-latex] (A1) to [bend right=10] (L2);
    % \draw [-latex] (A1) to [bend right=10] (A2);
    % \draw [-latex] (A1) to [bend right=10] (L2);
    % \draw [-latex] (Z1) to [bend left=0] (A1);
    % \draw [-latex] (Z2) to [bend left=0] (A2);
    % % \draw [-latex] (Z1) to [bend right=10] (L2);
    % \draw [-latex] (Z1) to [bend right=0] (A2);
    % % \draw [-latex] (Z0) to [bend right=10] (L2);
    % \draw [-latex] (Z0) to [bend right=0] (Z2);

    % \draw [-latex] (L0) to [bend left=25] (Y);
    % \draw [-latex] (L2) to [bend left=30] (Y);
    % \draw [-latex] (U0) to [bend left=35] (Y);
    % \draw [-latex] (U2) to [bend left=0] (Y);

    % \draw [-latex] (\Bx+2.8,\By) to [bend left=0] (Y);
    % \draw [-latex] (\Bx+2.8,\By+.4) to [bend left=0] (Y);
    % \draw [-latex] (\Bx+2.8,\By-.4) to [bend left=0] (Y);
    % \draw [-latex] (A2) to (\Bx+1.3,\By);

  \end{tikzpicture}
    \caption{DAG for the two-state markov model with unknown
      confounding and IV, with 2 time points. The model may be obtained by starting with two processes of the type depicted in Fig. \ref{eg:markov:DAG:SRA}, mixing the treatment process, and adding an exogenous IV.}
    % \label{eg:markov:fig:DAG:IV:2}
    \label{eg:markov:fig:DAG:IV}
  \end{figure}
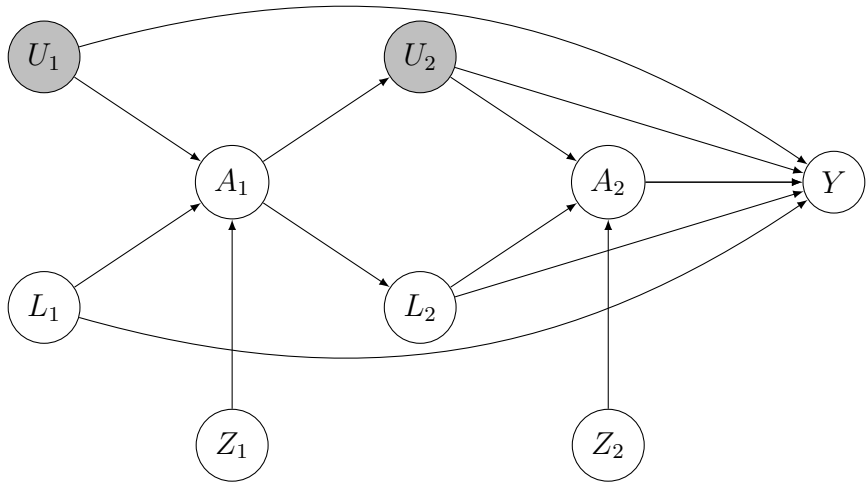
\end{example}

  \section{Simulation}
\label{section:simulation}
   
We examine the finite-sample behavior of the simple weighted estimator described in Section
\ref{section:estimation} under a data-generation process in
which SRA does not hold but the assumptions of Theorem \ref{theorem:1}, allowing for IV weights, do
hold. We consider the following linear MSMM:
$$
\EE(Y_{\overline{a}})=\beta_0 + \beta_1\sum_ta_t.
$$
Details on the data generation and estimation procedure are given in Appendix \ref{appendix:simulation details}. An additional simulation using data-generation process described in Example \ref{eg:markov} is given in Appendix \ref{appendix:markov}.
% \subsection{Results}
Besides our inverse-weighted estimator, also computed for comparison were an ``oracle'' estimator, an
SRA estimator, and
the associational or ``crude'' estimator. The oracle estimator uses inverse
probability weighting with the true propensity score $P(A_t=1\mid \ol{A}_{t-1},\ol{L}_t,\ol{Z}_t,\ol{U}_t)$, i.e.,
treating $\overline{U}$ as known and taking into account all confounders. The SRA estimator uses inverse probability weighting with
the propensity score taking into account only observed confounders,
$P(A_t=1\mid \ol{A}_{t-1},\ol{L}_t,\ol{Z}_t)$.
The associational estimator uses no weights, ignoring all confounding.

% For the first
% estimator,  
% $\Delta_j, 0\le j\le J-1,$ requires consistent estimation. The MLE is used, obtained by
% maximizing the full observed data
% likelihood for $A(j)$,
% \begin{align*}
% \PP(A(j+1)=1\mid \ol{LUZ}(j+1),\ol{A}(j)) &= \Phi\left((\alpha_0 + \alpha_1
% L(j+1))/\sqrt{1+\alpha_2^2}\right)\times
%   (1-\Delta(j+1))\\
%   &\quad +Z(j+1)\Delta(j+1),
% \end{align*}
% with 
% $$
% \Phi^{-1}(\Delta(j+1)) = \gamma_0 + \gamma_1L(j+1),
% $$
% obtained by integrating $U(j+1)$ out of (\ref{simulation:algo}), the maximization being
% taken over the coefficients $\alpha,$ viewed as nuisance parameters, and $\gamma,$ the parameters of
% interest for estimating $\Delta$. % H
% owever, as $U$ is unobserved, we
% compute the MLE using the compatible model 
% $$
% \PP(A(j+1)=1\mid \ol{LUZ}(j+1),\ol{A}(j)) = \Phi((\alpha_0 + \alpha_1
% L(j+1))/\sqrt{1+\alpha_2^2})\times
% (1-\Delta(j+1)),
% $$
% obtained by integrating out $U(j+1)$.
% We simulate data under the following laws:
% [copy from report]
% described by a causal diagram [] in [].
% -- full graph
% -- simplified graphical structure
% -- comparison with other estimators: SRA, oracle

\begin{figure}
\centering
\begin{subfigure}{.33\textwidth}
  \centering
  \includegraphics[width=\linewidth]{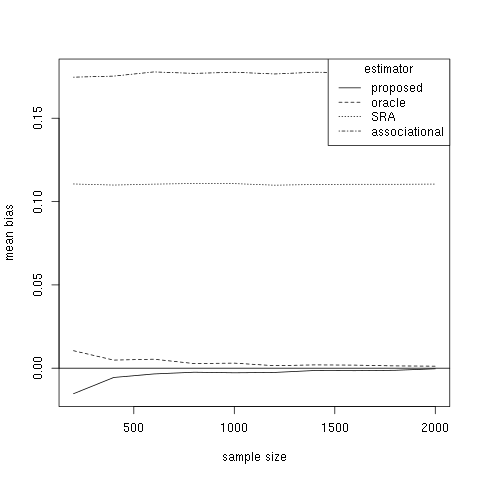}
  \caption{T=2}
  \label{fig:1a}
\end{subfigure}%
\begin{subfigure}{.33\textwidth}
  \centering
  \includegraphics[width=\linewidth]{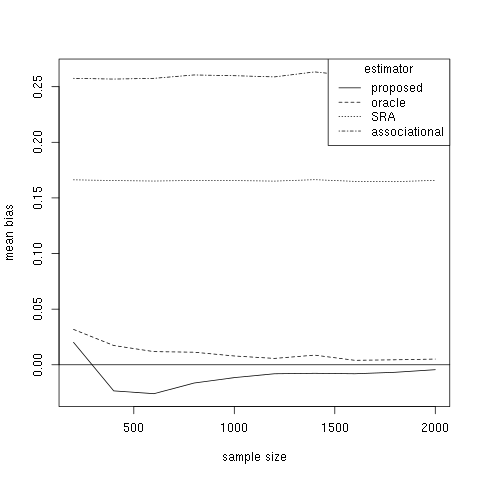}
  \caption{T=3}
  \label{fig:1b}
\end{subfigure}
\begin{subfigure}{.33\textwidth}
  \centering
  \includegraphics[width=\linewidth]{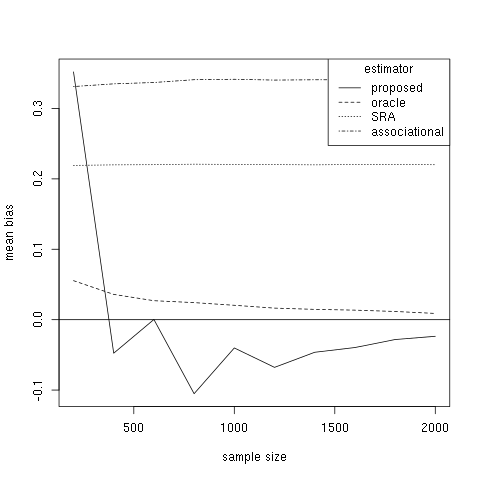}
  \caption{T=4}
  \label{fig:1c}
\end{subfigure}
\caption{Mean bias versus sample size of the proposed weighted
  estimator, for T=2, 3, and 4, time points, compared with oracle
  (weights including observed and unobserved confounders), SRA
  (weights including observed confounders), and associational (no
  weighting) estimators.}
\label{fig:1}
\end{figure}

For few time points, $2\le T\le 4$, the bias of the proposed estimator
falls off at a comparable rate to that of the oracle estimator. As
expected, the SRA and associational estimators are biased. See Figure
\ref{fig:1} for plots of the mean bias versus sample size. The
estimator is relatively noisy, however, with standard deviations on
the order of 1/10 when the bias is on the order of 1/1000. See Table
\ref{table:1} for measures of scale. A semiparametric efficient
estimator may mitigate the noisiness, although such an estimator is
beyond the scope of this paper; see \cite{tchetgen2018} for details.

For inference, we use the sandwich estimator and nonparametric
bootstrap.  Using each, we examine the empirical coverage of a nominal
95\% CI, varying the sample size $n$ and total number of time points
$T$, with the observed standard deviation of the estimators reported
for comparison. The coverage is close to the nominal level for smaller
$T$ and larger $n$, and overconservative for larger $T$ and smaller
$n$. The sample size needed for efficient coverage grows about
exponentially with the number of time points $T$, consistent with the
discussion in Example \ref{appendix:markov}. % \comment{Should there be comment on the large sample sizes needed for good performance? I could mention that these numbers are consistent with the large sample sizes of other big IV studies like the draft lottery or compulsory education studies. Or could mention that they are well under the size of our sample data from CHOP (for $T=2$).}
Table \ref{table:1}
presents the detailed results.

% \input{../sim/112418/112418.tex}
%https://tex.stackexchange.com/questions/147895/produce-2-side-by-side-tables-from-tables-as-external-files-source-xtable-in-r/147898?noredirect=1#comment335695_147898
\newcommand\insertmytabular[3][.45]%
{%
    {% extra group to make the redefinition of table local
        \renewenvironment{table}[1][]{}{}
        \begin{subtable}[b]{#1\textwidth}
            \centering
            \input{#2}
            % \caption{#3}
        \end{subtable}%
    }%
  }
  
  % \clearpage
  % \input{../sim/022219/122519/122519a.tex}{}

  \begin{table}
  \insertmytabular{122519a.txt}{}
  % \insertmytabular{../sim/112418/112418a.tex}{}\\
  % \insertmytabular{../sim/112418/112418b.tex}{}
  \caption{Empirical bias, Monte Carlo $\sigma_{mc}$, sandwich
    $\sigma_{sw}$, and bootstrap $\sigma_{bs}$ standard deviations,
    and the coverage of nominal 95\% sandwich and bootstrap CIs. The
    sandwich variance estimator appears more efficient than the
    bootstrap for this model, in contrast with the markov model
    discussed in Appendix \ref{appendix:markov}. The jumps in sample
    size, approximately exponential in time $T$, are consistent with
    the discussion in Example \ref{eg:markov} and Appendix
    \ref{appendix:markov}. For our application, $T=2$ and
    $n\approx 270,000$.}\label{table:1}
\end{table}

The code used to carry out the simulations described above is
available at \url{https://github.com/haben-michael/iv-msmm}. Also provided is an R
package to estimate the parameters of an MSMM under the models
described in Section \ref{subsection:data generation} and Example
\ref{eg:markov}.

\section{Application}
\label{section:application}
We examine the effect on neonatal mortality of delivery at a
high-volume, high-technology hospital. High-level neonatal intensive
care units (``NICUs'') have facilities for advanced care and average
50 or more premature births per year. Unadjusted analyses show a
harmful association of delivery at a high-level NICU on neonatal
mortality, but the effect is likely confounded. For example, more
complicated pregnancies are often directed to high-level NICUs.
Analyses controlling for a number of observed covariates have found a
protective effect of delivery at a high-level NICU. Possible
unmeasured confounders that remain in such analyses include unrecorded
comorbidities on which a treating physician bases the decision to
direct a mother to a high-level NICU. To account for these unmeasured confounders, \cite{lorch2012} conducted an IV analysis using the relative distance of a mother's residence to a high-level versus low-level NICU. This analysis found a protective effect. We consider the cumulative
effect over time of delivery at a high-level NICU.

We consider a repeated measurements model $Y_t\sim \beta_0 + \beta_1\sum_{\tau=1}^tA_{\tau}$, i.e.,
\begin{align*}
  Y_1&\sim \beta_0 + \beta_1 A_1\\
  Y_2&\sim \beta_0 + \beta_1 (A_1+A_2).
\end{align*}
The outcome $Y_t$ represents the occurrence of an event at time $t$ and $A_t$
represents the treatment at time $t$, delivery at a high-level NICU,
$t=1,2$. The parameters $(\beta_0,\beta_1)$ are the targets of
inference, with $\beta_1$ representing the additive effect of
cumulative treatment on neonatal mortality.

The data consists of 270,831 mothers who had exactly two births in
Pennsylvania between 1995 and 2005. This data was drawn from a larger
set consisting of all births in Pennsylvania between 1995 and 2005 for
which birth certificates, death certificates and hospital records
could be linked \cite{lorch2012}. The NICU level of the delivery
hospital, coded based on previous work in four levels from least to
most advanced facility, was dichotomized as ``low'' (levels 1, 2) and
``high'' (levels 3, 4). Delivery at a high-level NICU serves as the treatment
$A_t$, for $t=1$ or $2$ representing the birth order. Our instrument $Z_t$ is the mother's
residence's distance to the nearest low-level NICU minus nearest
high-level NICU, dichotomized at $0$. That is, $Z_t>0$ indicates that
a high-level NICU is closer than a low-level NICU to the mother's
residence. This IV is longitudinal in nature, with over 30\% of the
mothers changing residence between pregnancies, and about 10\% of
these changes in residence constituting a change in IV status.

Assumption \ref{assumption:iv relevance} is supported by previous
research suggesting that mothers tend to deliver at NICUs near their
residence. In the data, the correlation between IV and treatment
exceeds 60\%. Assumption \ref{assumption:iv positivity} requires
that no population stratum consists deterministically of individuals
living closer to a high-level NICU, nor does any stratum consist
deterministically of individuals farther from a high-level NICU, where
the population strata are determined by available covariates and the
treatment and IV history. For example, regressing the IV at $T=2$ on
available covariates and history, one finds a pseudo-$R^2$ of just 65.1\%.

The remaining assumptions involve unobservables and cannot be directly
tested using the data. Assumption \ref{assumption:iv--outcome}
requires that the relative distance to a high-level or low-level NICU
not affect neonatal outcomes except through the type of hospital at
which the delivery occurred, conditional on available data. Assumption
\ref{assumption:iv--confounder} requires that the mother's relative
distance to a high-level NICU is independent of unmeasured confounders
of the association between NICU and neonatal death, at least upon
controlling for socioeconomic data and other measured covariates. For
example, the assumption requires that the relative distance of the
mother's residence is independent of the presence of unrecorded fetal
heart tracing results, if these results indeed confound the
relationship. Further discussion of the plausibility of the IV
assumptions may be found in \cite{lorch2012}, which discusses a related
IV, and qualifications may be found in \cite{yang2014}.

Assumption \ref{assumption:ict} requires that all factors are recorded that relate
to whether a mother delivers at a high-level NICU when living closer
to one. For example, if premature births are likely to occur at a
high-level NICU irrespective of the mother's residence's distance,
then the data ought to capture whether a birth is premature or
not. Possible violations of this assumption are discussed in \cite{yang2014}.

The parameters $(\beta_0,\beta_1)$ were estimated as the solution
to the estimating equation
\begin{align*}
\PP_n\left(\begin{pmatrix}
  1 & A_1\\
  1 & A_1+A_2
\end{pmatrix}^T
\begin{pmatrix}
  W_1 & 0\\
  0 & \ol{W}_2
\end{pmatrix}
\left(\begin{pmatrix}
  Y_1\\
  Y_2
\end{pmatrix}
-
\begin{pmatrix}
  1 & A_1\\
  1 & A_1+A_2
\end{pmatrix}
\begin{pmatrix}
  \beta_0\\
  \beta_1
\end{pmatrix}\right)\right).
\end{align*}
The model for the density of the instrument conditional on the past
observed history was modeled using a logistic regression. Likewise,
$\Delta_t=\PP(A_t=1\mid Z_t=1,\ol{L}_t,\ol{A}_{t-1}) - \PP(A_t=1\mid
Z_t=0,\ol{L}_t,\ol{A}_{t-1})$ was estimated by first fitting
$\PP(A_t=1\mid Z_t,\ol{L}_t,\ol{A}_{t-1})$, again using a logistic
regression. In both regressions, the covariates used were gestational
age, mother's educational level, and month that prenatal care began,
following \cite{yang2014}. Besides this IV adjusted estimator, also
computed were estimates using no weights, i.e., an associational
estimate, and using SRA weights, using the covariates just described
to form propensity
scores. % The variable for the month of start of prenatal
% care had a missingness of about 48\%. Missing values to 0 and created
% a separate variable to indicate missingness. The number of
% observations is 270,831, where each observation includes a mother and
% her two births.

% The point estimates for $\beta_1$ are 0.0002632631 (associational),
% 0.0002905478 (SRA), 0.0001459692 (IV). This parameter represents the
% linear effect on neonatal mortality on a mother's second birth of a
% mother's first two deliveries at a high-level NICU. Bootstrap 95\% CIs
% are $(0.0001538,0.0003557)$ (associational), $(0.0002058,0.0004103)$
% (SRA), and $(-0.0001631,0.0004891)$ (IV). Thus while the associational
% and SRA analyses find a significantly harmful effect on neonatal
% mortality of delivery at a high-level NICU, the IV analysis fails to
% reject the null of no effect. The direction of these results is
% similar to the results reported in \cite{lorch2012}, though not as
% strong. There, unadjusted/associational analyses found a significantly
% harmful effect of delivery at a high-level NICU on infant death and
% other complications whereas the IV analysis found a significantly
% protective effect. Moreover, further IV analysis reported in
% \cite{yang2014} finds little effect for most infants, similar to the
% result found here, with the significantly protective effect limited
% primarily to premature infants.

The point estimates for $\beta_1$, given as the number of deaths per
$10,000$ births, are 2.63 (associational), 2.91
(SRA), 1.46 (IV). This parameter represents the linear effect on
neonatal mortality of the second of a mother's first two
deliveries at a high-level NICU. Bootstrap 95\% CIs are
$(1.54,3.56)$ (associational), $(2.06,4.10)$
(SRA), and $(-1.63,4.89)$ (IV). Thus while the associational
and SRA analyses find a significantly harmful effect on neonatal
mortality of delivery at a high-level NICU, the IV analysis fails to
reject the null of no effect. The direction of these results is
similar to the results reported in \cite{lorch2012}, though not as
strong. There, unadjusted/associational analyses found a significantly
harmful effect of delivery at a high-level NICU on infant death and
other complications whereas the IV analysis found a significantly
protective effect. Moreover, further IV analysis reported in
\cite{yang2014} finds little effect for most infants, similar to the
result found here, with the significantly protective effect limited
primarily to premature infants.

\section{Discussion}
\label{section:discussion}
We have shown how IVs may be used to identify causal parameters in
marginal structural mean models. Most of our assumptions are mainly
variations of standard IV or MSM assumptions. Our key assumption
requires that unknown confounders not interact with the IV in the latter's
additive effect on the treatment.  We further showed that the
conclusion of our identification theorem requires an assumption of a similar form.

Several extensions to these results suggest themselves. First, the
method of proof of our identification result may be generalized to
apply to other MSMs besides mean models. In \cite{cui2020}, a Cox MSM
for right-censored survival data is considered. The technical report
\cite{tchetgen2018} provides a theoretical framework for MSMs in
general, although it lacks analysis of the finite-sample behavior and
certain theoretical results for MSMMs given in the current work, such
as the extension to continuous treatments.

Second, we have required that the instrumental variable be
binary. Continuous IVs are often encountered% , including the IV
% considered in Section \ref{section:application}
, such as the difference in distances encountered in our
application. Dichotomizing such IVs to fit our framework entails a
loss of efficiency and may introduce other difficulties into the
estimation procedure. Therefore, it would be useful to extend our
identification and estimation results to allow for ordinal or
continuous IVs. The resulting estimator would generalized two-stage
least squares to the longitudinal setting in the way that the
estimator proposed here generalizes the Wald estimator (Example
\ref{example:wald}).

Third, the estimator proposed here, the solution to the estimating
equation (\ref{eqn:estimating eqn}), while convenient, does not make efficient use of all
the available data. We expect improved performance from a robust, semiparametric efficient estimator.

\bibliographystyle{Chicago}

% \bibliography{Bibliography-MM-MC}
\bibliography{causal}

\section{Appendix: Markov model estimation}
\label{appendix:markov}
The data and model are described in Example \ref{appendix:markov}. The treatment model given in (\ref{eg:markov:kernel:IV}),
\begin{align*}
  \PP(A_t=a&\mid L_t=l,U_t=u,Z_t=z)=\\
  &(1-q)p_L^{\{l=a\}}(1-p_L)^{\{l\neq a\}} + qp_U^{\{u=a\}}(1-p_U)^{\{u\neq a\}}+(-1)^{1-z}(-1)^{1-a}\delta_l/2,
  % \PP(A_t=a\mid L_t=l,U_t=u)&=
\end{align*}
implies the observed-data model
$$
\PP(A_t=a\mid L_t=l,Z_t=z)=q/2 + (1-q)p_L^{\{l=a\}}(1-p_L)^{\{l\neq a\}}.
  % &(1-q)p_L^{\{l=a\}}(1-p_L)^{\{l\neq a\}} + qp_U^{\{u=a\}}(1-p_U)^{\{u\neq a\}}+(-1)^{1-z}(-1)^{1-a}\delta_l/2
$$
In order to identify the model we
assume the mixing probability $q$ is known. For all $t$, the differences $\Delta_t$ are parametrized by $(\delta_0,\delta_1)$, and the remaining parameter for the treatment model is $p_L$. The MSMM model is
$$
\EE(Y_{\ol{a}}) = m_{\beta}(\ol{a})=\beta\sum_ta_t.
$$
As discussed in Section \ref{subsection:data generation}, outcomes consistent with this MSMM may be generated as
\begin{align*}
  Y &= \eta + \epsilon\\
  \eta &= \sum_{t=1}^T\tau_t(L_t - (1-q)p_L^{A_{t-1}}(1-p_L)^{1-A_{t-1}} - q/2) +\\
  &\sum_{t=1}^T\rho_t(U_t - qp_U^{A_{t-1}}(1-p_U)^{1-A_{t-1}} - (1-q)/2) +\EE(Y_{\ol{a}}),
\end{align*}
where $\epsilon$ is standard normal and exogenous, and $\tau_t=\rho_t=1,t=1,\ldots,T.$

Using the notation in Section \ref{section:estimation}, the parameters are the MSMM
parameter $\beta, \alpha=(\delta_0,\delta_1)$ and $\nu=p_L$. Theorem
\ref{theorem:1} is used to estimate $\beta$ and $\alpha$ and $\nu$ are
estimated by maximum likelihood. That is, the weighted residuals
$\PP_n (\sum_tA_t)(Y-m_\beta(\ol{A}))/\ol{W}$  serve as an estimating equation for
$\beta$ and the scores as estimating equations for $\alpha$ and
$\nu$. Formulas for these scores and the information for all the
parameters are obtained as in Section \ref{sec:estimation} by substituting
\begin{align*}
  \pi_{\alpha,\nu}(A,L,Z) &= q/2 + (1-q)p_L^L(1-p_L)^{1-L}\\
  \frac{\partial \pi_{\alpha,\nu}}{\partial\alpha,\nu}(A,L,Z)&=((-1)^{1-Z}(1-L)(\delta_1/\delta_0)^L,(-1)^{1-Z}(\delta_0/\delta_1)^{1-L},\\
                                                        &(1-q)L(1/p_L-1)^{1-L}-(1-q)(1-L)(1/p_L-1)^{-L}).
\end{align*}
The second derivative $\frac{\partial^2 \pi_{\alpha,\nu}}{\partial(\alpha,\nu)^2}$ is 0.

The results of a simulation are given in Table \ref{table:2}. In contrast to the model described in Section \ref{section:simulation}, the sandwich-derived CI appears more conservative than the bootstrap CI.

  \begin{table}
  \insertmytabular{010620b.tex}{}
  % \insertmytabular{../sim/112418/112418a.tex}{}\\
  % \insertmytabular{../sim/112418/112418b.tex}{}
  \caption{Empirical bias, Monte Carlo standard deviation, sandwich
    sd, bootstrap sd, and the coverage of nominal 95\% sandwich and
    bootstrap CIs, using the simple markov model discussed in Appendix
    \ref{appendix:markov}.\comment{I will prettify table once
      numbers/formatting is finalized.}}\label{table:2}
\end{table}

\section{Appendix: Continuous treatment density}\label{appendix:continuous treatment}
We first show that the treatment density given in Example
\ref{example:continuous treatment} is a valid density. As there, let the baseline density
$f_{A_t\mid \an(A_t)}(A_t=a,L_t=(l_1,l_2),U_t=u,Z_t=0)$ be normal
$\phi((a-l_1)/u)/u$. The first component of the observed confounder
$L_t$ controls the location and the unobserved confounder $U_t$
controls the spread. Let $\Delta_t(a\mid L_t=(l_1,l_2))$ be a
difference between normal densities that does not depend on $U_t$,
say, $\phi(a)-\phi(a/l_2)/l_2$. If the spread $u$ of the baseline
density lies within an appropriate range, then
$f_{A_t\mid \an(A_t)}(A_t=a,L_t=(l_1,l_2),U_t=u,Z_t=0) + \Delta_t(a)$
is a valid density for $A_t$. % (Controlling this
  % baseline spread is analogous to how previously we chose $\Delta$ using
  % $L_t$ only, and next chose the baseline $\PP(A_t\mid L_t,U_t)$ so that
  % $\PP(A_t\mid L_t,U_t)+\Delta$ fit in $(0,1)$.)
  In particular,
  given $L_t=(l_1,l_2)$ with $l_2\in (0,1)$, suppose $l_2<u<\min(1,l_2/(1-l)2))$ for the standard
  deviation $u$ of the baseline. Let $l_1=0$ since the location
  is irrelevant to the argument. Then, 
  \begin{align*}
    % &\text{baseline }+\text{ delta at }a\\
      &f_{A_t\mid \an(A_t)}(A_t=a,L_t=(l_1,l_2),U_t=u,Z_t=1)\\
      &=f_{A_t\mid \an(A_t)}(A_t=a,L_t=(l_1,l_2),U_t=u,Z_t=0) +
        \Delta_t(a)\\
      &=\phi(a/u)/u + \phi(a)-\phi(a/l_2)/l_2\\
      &=\phi(a)\left\{1+\exp(a^2(1-1/u^2)/2)/u-\exp(a^2(1-1/l_2^2)/2)/l_2\right\}\\
      &=\phi(a)\exp(a^2(1-1/l_2^2)/2)/l_2\left\{l_2\exp(a^2(1/l_2^2-1)/2)+(l_2/u)\exp(a^2(1/l_2^2-1/u^2)/2)-1\right\}\\
      &\ge \phi(a)\exp(a^2(1-1/l_2^2)/2)/l_2\left\{l_2+l_2/u-1\right\}\\
      &> 0.
  \end{align*}
  The first inequality follows from the condition that $l_2\in (0,1)$
  and $l_2<u$, and the second inequality from the requirement
  $u<l_2/(1-l_2)$. Since $\Delta_t$ integrates to 0 by construction,
  $f_{A_t\mid \an(A_t)}(A_t=a,L_t=(l_1,l_2),U_t=u,Z_t=1)$ is a
  valid density. 

\begin{figure*}[t]
  % \begin{multicols}{3}
  \centering
    \includegraphics[width=\linewidth*19/60]{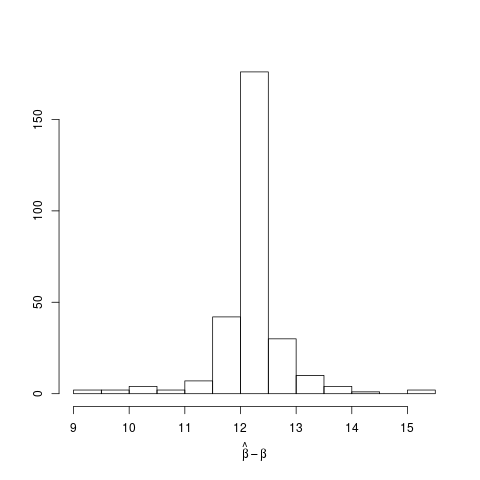}
    \includegraphics[width=\linewidth*19/60]{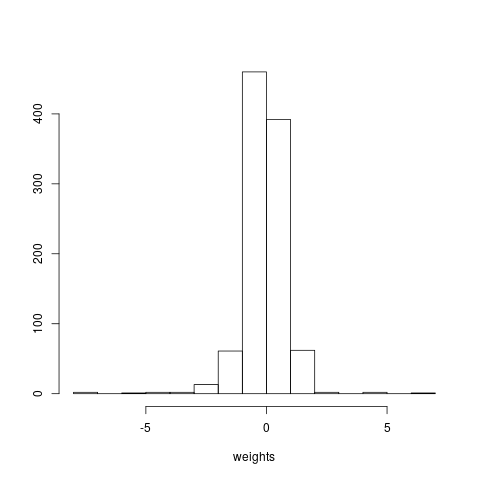}
    \includegraphics[width=\linewidth*19/60]{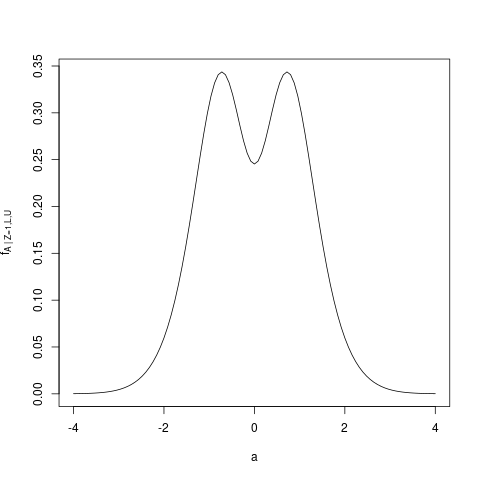}
% \end{multicols}
\caption{(Continuous treatment density.) The left panel is a histogram of the observed bias of the estimated coefficient in a simulation of size 1000. The middle panel is a histogram of the weights in one simulated data set. The right panel is a plot of the density $f_{A\mid Z=1,L,U}$, i.e., the sum of the baseline treatment density, which is normal, and the function $\Delta(a,l,u)$ for sample values of $l$ and $u$.}
\label{fig:appendix:continuous}
\end{figure*}

A small simulation using a continuous treatment density for a single time point is presented below. Following Example \ref{example:continuous treatment}, $L$ and $U$ are sampled from a uniform distribution on the unit interval, $Z$ is a standard bernoulli, and the treatment density is defined as
\begin{align*}
  % L &\sim \text{Unif[0,1]}\\
  % U &\sim \text{Unif[0,1]}\\
  \Delta(a \mid L) &= \phi(a)-\phi(a/l)/l\\
  f_{A\mid Z,L,U}(A=a,L=l,U=u) &= \phi(a/u)/u + Z\Delta(a\mid L=l),\\
\end{align*}
using $\phi$ to denote the standard normal density.
The outcome is sampled as  $ Y = (L-\EE(L)) + (U-\EE(U)) + \beta A + \epsilon$,
with $\beta=2$. The sample size is 1000. The observed bias and
standard deviation of the estimates are -.195 and 0.64, and the median
absolute error is .249. Figure \ref{fig:appendix:continuous} gives a
histogram of the observed biases, as well as histograms of the weights
and the plot of the conditional treatment density $f_{A\mid Z=1,L,U}$ for one choice of $L,U$.
% \begin{figure}
%   \includegraphics{../figs/39a}
% \end{figure}

% \begin{figure*}[t]
%   \centering
%   \subcaptionbox{}[.3\linewidth][c]{%
%     \includegraphics[width=.3\linewidth]{../figs/39a}}\quad
%   \subcaptionbox{Fig2}[.3\linewidth][c]{%
%     \includegraphics[width=.3\linewidth]{../figs/39b}}\quad
%   \subcaptionbox{Fig3}[.3\linewidth][c]{%
%     \includegraphics[width=.3\linewidth]{}}
%   \label{fig:appendix:continuous}
% \end{figure*}
\comment{if time available, extend to 2 time points. or at least do inference. right now the bootstrap standard errors are a little too wide.}

\section{Appendix: Proof of Theorem \ref{theorem:1}}
\begin{proof}\comment{Move proofs to appendix when finalized}
  To simplify notation, we use a
single overbar and single time index to indicate a history of random vectors,
e.g.,
$\ol{AZL}_t=(\ol{A}_t,\ol{Z}_t,\ol{L}_t)=(A_{\tau},Z_{\tau},L_{\tau})_{\tau=0}^{\tau=t}$.

  \comment{  As a typographical simplification, we sometimes write a vector of vector of variables with the same final time point, e.g., $(\ol{A}_t,\ol{Z}_t,\ol{L}_t)$, using a single overbar and time point, e.g., $(\ol{AZL}_t)$. I am conflicted on notation. Using the unsimplified notation $(\ol{A}_t,\ol{Z}_t,\ol{L}_t)$ throughout makes this proof too messy, I don't think it's an option. I had the idea to use the DAG notation, e.g., instead of $(\ol{A}_t,\ol{Z}_t,\ol{L}_t)$ we could write the ancestor set of $Y$, e.g., $\an(Y)$. But half the time in this proof I'm not referring to the random variables $(\ol{A}_t,\ol{Z}_t,\ol{L}_t)$ but elements of the state space $(\ol{a}_t,\ol{z}_t,\ol{l}_t)$, so an additional notation is needed to distinguish the two. Also it may  be helpful for readers to write explicitly the variables the weights depend on, e.g., $W_t(\ol{A}_t,\ol{Z}_{t-1},\ol{L}_t)$ instead of just $W_t$....but there's already a lot going.}
  \begin{align*}
    \EE(\ol{W}^{-1}g(Y,\ol{A})) &= \EE(\ol{W}^{-1}\EE(g(Y,\ol{A})\mid
                                  \ol{AZLU}))\\
                                &=
                                  \int_{\mathcal{A}^T\times\mathcal{Z}^T\times\mathcal{L}^T\times\mathcal{U}^T}
                                  \ol{W}^{-1}\EE(g(Y_{\ol{a}},\ol{a})\mid\ol{azlu})
                                  f_{\ol{AZLU}}(\ol{azlu})\mu_{\ol{AZLU}}(\ol{azlu})% \\
                                % &=
                                %   \int_{\mathcal{A}^T\times\mathcal{Z}^T\times\mathcal{L}^T\times\mathcal{U}^T}
                                %   \ol{W}^{-1}\EE(g(Y_{\ol{a}},\ol{a})\mid\ol{a},z_{T-1},\ol{lu}) f_{\ol{AZLU}}(\ol{azlu})\mu_{\ol{AZLU}}(\ol{azlu}),
  \end{align*}
  % using Assumption \ref{assumption:iv--outcome} for the last equality.
  The measure $\mu_{\ol{AZLU}}(\ol{azlu})$ is a product measure on
  $\mathcal{A}^T\times\mathcal{Z}^T\times\mathcal{L}^T\times\mathcal{U}^T$
  relative to which the density $f_{\ol{AZLU}}$ is given, and analogously for $\mu_{\ol{A},\ol{Z}_{T-1}\ol{LU}}(\ol{a}\ol{z}_{T-1}\ol{lu})$ on $\mathcal{A}^T\times\mathcal{Z}^{T-1}\times\mathcal{L}^T\times\mathcal{U}^T$ and $f_{\ol{A}\ol{Z}_{T-1}\ol{LU}}$, and so forth. We have assumed the marginal measure $\mu_{Z_t},1\le t\le T,$ is counting measure on $\mathcal{Z}_t=\{0,1\}$.

  Proceeding by induction, suppose we
  have established for some $s$, $0\le s< T$,
  \begin{align}
    \EE(\ol{W}^{-1}&g(Y,\ol{A})) =\\
    &\int_{\mathcal{A}^T\times\mathcal{Z}^{T-s}\times\mathcal{L}^{T-s}\times\mathcal{U}^{T-s}}
    \ol{W}^{-1}_{T-s}\EE(g(Y_{\ol{a}},\ol{a})\mid\ol{azlu}_{T-s}) f_{\ol{AZLU}_{T-s}}(\ol{azlu}_{T-s})\mu_{\ol{A},\ol{ZLU}_{T-s}}(\ol{a},\ol{zlu}_{T-s}).
    \label{lemma:identification:ind hyp}
  \end{align}
  By Assumption \ref{assumption:iv--outcome},  (\ref{lemma:identification:ind hyp}) is
  \begin{align*}
  \EE(\ol{W}^{-1}g(Y,\ol{A})) =  \int_{\mathcal{A}^T\times\mathcal{Z}^{T-s}\times\mathcal{L}^{T-s}\times\mathcal{U}^{T-s}}
  \ol{W}^{-1}_{T-s}\EE(g(Y_{\ol{a}},\ol{a})\mid\ol{a}_{T-s},\ol{z}_{T-s-1},\ol{lu}_{T-s})\times\\
  f_{\ol{AZLU}_{T-s}}(\ol{azlu}_{T-s})\mu_{\ol{A},\ol{ZLU}_{T-s}}(\ol{a},\ol{zlu}_{T-s}).
  \end{align*}
  By Assumption \ref{assumption:iv--confounder},
  \begin{align*}
    f_{\ol{AZLU}_{T-s}} &=
                          f_{A_{T-s}\mid \ol{A}_{T-s-1}\ol{ZLU}_{T-s}}
                          \times  f_{Z_{T-s}\mid
                          \ol{AZ}_{T-s-1}\ol{LU}_{T-s}}
                          \times
                          f_{\ol{AZ}_{T-s-1}\ol{LU}_{T-s}}\\
    &=
                          f_{A_{T-s}\mid \ol{A}_{T-s-1}\ol{ZLU}_{T-s}}
                          \times  f_{Z_{T-s}\mid
                          \ol{AZ}_{T-s-1}\ol{L}_{T-s}}
                          \times
                          f_{\ol{AZ}_{T-s-1}\ol{LU}_{T-s}},
  \end{align*}
  so (\ref{lemma:identification:ind hyp}) is the same as
  \begin{align*}
  &\int_{\mathcal{A}^T\times\mathcal{Z}^{T-s}\times\mathcal{L}^{T-s}\times\mathcal{U}^{T-s}}
  \ol{W}^{-1}_{T-s-1}\EE(g(Y_{\ol{a}},\ol{a})\mid\ol{a}_{T-s},\ol{z}_{T-s-1},\ol{lu}_{T-s})
    (-1)^{1-Z_{T-s}}\Delta_{T_s}^{-1}\times\\
    & f_{A_{T-s}\mid \ol{A}_{T-s-1}\ol{ZLU}_{T-s}}(\ol{azlu}_{T-s}) f_{\ol{AZ}_{T-s-1}\ol{LU}_{T-s}}(\ol{az}_{T-s-1}\ol{lu}_{T-s})
    \mu_{\ol{A},\ol{ZLU}_{T-s}}(\ol{a},\ol{zlu}_{T-s})\\
    &=\int_{\mathcal{A}^T\times\mathcal{Z}^{T-s-1}\times\mathcal{L}^{T-s}\times\mathcal{U}^{T-s}}
  \ol{W}^{-1}_{T-s-1}\EE(g(Y_{\ol{a}},\ol{a})\mid\ol{a}_{T-s},\ol{z}_{T-s-1},\ol{lu}_{T-s})
  \Delta_{T_s}^{-1}\times\\
  &\sum_{z\in\{0,1\}} (-1)^{z}f_{A_{T-s}\mid
    \ol{A}_{T-s-1}\ol{ZLU}_{T-s}}(\ol{alu}_{T-s},\ol{z}_{T-s-1},z_{T-s}=z)\times\\
    &f_{\ol{AZ}_{T-s-1}\ol{LU}_{T-s}}(\ol{az}_{T-s-1}\ol{lu}_{T-s})
    \mu_{\ol{A},\ol{Z}_{T-s-1},\ol{LU}_{T-s}}(\ol{a},\ol{z}_{T-s-1}\ol{lu}_{T-s})\\
    &=\int_{\mathcal{A}^T\times\mathcal{Z}^{T-s-1}\times\mathcal{L}^{T-s}\times\mathcal{U}^{T-s}}
  \ol{W}^{-1}_{T-s-1}\EE(g(Y_{\ol{a}},\ol{a})\mid\ol{a}_{T-s},\ol{z}_{T-s-1},\ol{lu}_{T-s})\times\\
    &f_{\ol{AZ}_{T-s-1}\ol{LU}_{T-s}}(\ol{az}_{T-s-1}\ol{lu}_{T-s})
    \mu_{\ol{A},\ol{Z}_{T-s-1},\ol{LU}_{T-s}}(\ol{a},\ol{z}_{T-s-1}\ol{lu}_{T-s}).
  \end{align*}
Assumption \ref{assumption:ict} was used to cancel $\Delta_{T_s}^{-1}$ in obtaining the last equality.
  Applying Assumption \ref{assumption:latent sra} and integrating out $L_{t-s},U_{T-s}$, the last expression is
  \begin{align*}
    &\int_{\mathcal{A}^T\times\mathcal{Z}^{T-s-1}\times\mathcal{L}^{T-s}\times\mathcal{U}^{T-s}}
  \ol{W}^{-1}_{T-s-1}\EE(g(Y_{\ol{a}},\ol{a})\mid\ol{az}_{T-s-1},\ol{lu}_{T-s})\times\\
    &f_{\ol{AZ}_{T-s-1}\ol{LU}_{T-s}}(\ol{az}_{T-s-1}\ol{lu}_{T-s})
      \mu_{\ol{A},\ol{Z}_{T-s-1},\ol{LU}_{T-s}}(\ol{a},\ol{z}_{T-s-1}\ol{lu}_{T-s})\\
    &=\int_{\mathcal{A}^T\times\mathcal{Z}^{T-s-1}\times\mathcal{L}^{T-s-1}\times\mathcal{U}^{T-s-1}}
  \ol{W}^{-1}_{T-s-1}\EE(g(Y_{\ol{a}},\ol{a})\mid\ol{azlu}_{T-s-1})\times\\
    &f_{\ol{AZLU}_{T-s-1}}(\ol{azlu}_{T-s-1})
    \mu_{\ol{A},\ol{ZLU}_{T-s-1}}(\ol{a},\ol{zlu}_{T-s-1}),
  \end{align*}
  completing the inductive step.
\end{proof}

\section{Appendix: Proof of Lemma \ref{lemma:1}}
         \begin{proof}
           The RHS of (\ref{defn:converse:eta}) does not depend on $z_T$ or $a'_{T-1}$, so that $\EE(Y_{\ol{a}}\mid\ol{a}_{T-1},a_T',\ol{ZLU})=\EE(Y_{\ol{a}}\mid\ol{a}_{T-1},a_T',\ol{Z}_{T-1}\ol{LU})=\EE(Y_{\ol{a}}\mid\ol{a}_{T-1},\ol{Z}_{T-1}\ol{LU}).$ Therefore,         
         \begin{align*}
           \EE(Y_{\ol{a}}\mid\ol{aZLU}_{T-1}) &= \EE(\EE(Y_{\ol{a}}\mid\ol{aZ}_{T-1},\ol{LU})\mid \ol{aZLU}_{T-1})\\
                                              &= \EE\left(m(\ol{a}) + \sum_{t=1}^T\eta_t(\ol{a},\ol{aZ}_{t-1},\ol{LU}_t) \biggm| \ol{aZLU}_{T-1}\right)\\
                                              &= m(\ol{a}) + \sum_{t=1}^{T-1}\eta_t(\ol{a},\ol{aZ}_{t-1},\ol{LU}_t),
         \end{align*}         
         and by induction, $\EE(Y_{\ol{a}}\mid\ol{azlu}_t)=m(\ol{a}) + \sum_{t'=1}^{t}\eta_{t'}(\ol{a},\ol{az}_{t'-1},\ol{lu}_{t'})$ for all $t$. Consequently, $\EE(Y_{\ol{a}}\mid\ol{a}_{t-1},a_t',\ol{ZLU})=\EE(Y_{\ol{a}}\mid\ol{a}_{t-1},a_t',\ol{Z}_{t-1}\ol{LU})=\EE(Y_{\ol{a}}\mid\ol{a}_{t-1},\ol{Z}_{t-1}\ol{LU})$ for all $t$, so that the data satisfy Assumption \ref{assumption:weakened latent sra} and \ref{assumption:iv--outcome}. Additionally, $\EE(Y_{\ol{a}})=\EE(\EE(Y_{\ol{a}}\mid L_1U_1))=\EE(m(\ol{a}) + \eta_{1}(\ol{a},\ol{LU}_{1}))=m(\ol{a}),$ so the data satisfies the MSMM given by $m(\ol{a})$. Finally, $Y$ is defined in (\ref{defn:converse:Y}) so as to satisfy Assumption \ref{assumption:weakened consistency}. Therefore, so long as $\eta$ satisfies (\ref{defn:converse:eta}), outcomes $Y$ satisfying (\ref{defn:converse:Y}) are consistent with the assumptions implying the identification result (\ref{eqn:weakened estimating eqn}).
       \end{proof}

       \section{Appendix: Proof of Theorem \ref{theorem:2}}
       \begin{proof}
  In view of Lemma \ref{lemma:1}, (\ref{eqn:theorem:converse}) is equivalent to the
  requirement that
  $$
  \EE(h(\ol{A_T})\omega_T(ALZ_T)\eta(\ol{AZLU}_T))=0
  $$
  hold for any choice of $T,h$ and $\eta$ as in
  (\ref{defn:converse:eta}). An example of such $\eta$ is any function of
  $\ol{AZ}_{T-1}\ol{LU}_T$ conditionally mean-zero given
  $\ol{AZLU}_{t-1}$, i.e.,
  $\zeta(\ol{AZ}_{T-1}\ol{LU}_T)-\EE(\zeta(\ol{AZ}_{T-1}\ol{LU}_T)\mid
  \ol{AZLU}_{T-1})$ for arbitrary $\zeta\in
  L_1(\ol{AZ}_{T-1}\ol{LU}_T)$. The condition becomes
  \begin{align*}
    \EE(h(\ol{A}_T)\omega_T(\ol{ALZ}_T)\zeta(\ol{AZ}_{T-1}\ol{LU}_T)) &= \EE(h(\ol{A}_T)\omega_T(\ol{ALZ}_T)\EE(\zeta(\ol{AZ}_{T-1}\ol{LU}_T)\mid\ol{AZLU}_{T-1}))\\
    &= \EE(\EE(h(\ol{A}_T)\omega_T(\ol{ALZ}_T)\mid\ol{AZLU}_{T-1})\zeta(\ol{AZ}_{T-1}\ol{LU}_T)),
  \end{align*}
  for all $\zeta\in L_1(\ol{AZ}_{T-1}\ol{LU}_T)$, implying that
  $\EE(h(\ol{A}_T)\omega_T(\ol{ALZ}_t)\mid\ol{AZLU}_{T-1})$ is a version of
  the conditional expectation
  $\EE(h(\ol{A}_T)\omega_T(\ol{ALZ}_T)\mid\ol{AZ}_{T-1}\ol{LU}_T)$, i.e.,
  $h(\ol{A}_T)\omega_T(\ol{ALZ}_T)$ is conditionally mean-independent of
  $L_T,U_T$ given $\ol{AZLU}_{T-1}$.

  Taking $h(\ol{A}_T)=\{\ol{A}_T=\ol{a}_T\}$ for $\ol{a}_T\in\mathcal{A}^T$,  
  \begin{align*}
    &\sum_{z_T\in\mathcal{Z}}f_{A_T\mid ...}(a_T\mid \ol{A}_{T-1},\ol{Z}_{T-1},z_T,\ol{LU}_T)\omega_T(\ol{A}_{T-1},a_T,\ol{Z}_{T-1},z_T,\ol{L}_T)f_{z_T\mid ...}(z_T\mid \ol{AZ}_{T-1}\ol{L}_{T})\\
    &\sum_{z_T\in\mathcal{Z}}f_{A_T\mid ...}(a_T\mid \ol{A}_{T-1},\ol{Z}_{T-1},z_T,\ol{LU}_T)\omega_T(\ol{A}_{T-1},a_T,\ol{Z}_{T-1},z_T,\ol{L}_T)f_{z_T\mid ...}(z_T\mid \ol{AZ}_{T-1}\ol{LU}_{T})\\
    &=\EE(f_{A_T\mid ...}(a_T\mid \ol{A}_{T-1}\ol{ZLU}_T)\omega_T(\ol{A}_{T-1},a_T,\ol{ZL}_T)\mid\ol{AZ}_{T-1}\ol{LU}_{T})\\
    % &=\EE(\{\ol{A}_T=\ol{a}_T\}\omega_T(\ol{aLZ}_T)\mid\ol{aZ}_{T-1}\ol{LU}_{T})\\
    &=\EE(\{A_T=a_T\}\omega_T(\ol{A}_{T-1},a_T,\ol{ZL}_T)\mid\ol{AZ}_{T-1}\ol{LU}_{T})\\
    &=\EE(\{A_T=a_T\}\omega_T(\ol{A}_{T-1},a_T,\ol{ZL}_T)\mid\ol{AZLU}_{T-1})
  %   &\sum_{z_T\in\mathcal{Z}}f_{A_T\mid ...}(a_T\mid \ol{a}_{T-1}\ol{LU}_T,\ol{Z}_{T-1},z_T)\omega_T(\ol{aL}_T,\ol{Z}_{T-1},z_T)f_{z_T\mid ...}(z_T\mid \ol{aZ}_{T-1}\ol{L}_{T})\\
  %   &\sum_{z_T\in\mathcal{Z}}f_{A_T\mid ...}(a_T\mid \ol{a}_{T-1}\ol{LU}_T,\ol{Z}_{T-1},z_T)\omega_T(\ol{aL}_T,\ol{Z}_{T-1},z_T)f_{z_T\mid ...}(z_T\mid \ol{aZ}_{T-1}\ol{LU}_{T})\\
  %   &=\EE(f_{A_T\mid ...}(a_T\mid \ol{a}_{T-1}\ol{ZLU}_T)\omega_T(\ol{aLZ}_T)\mid\ol{aZ}_{T-1}\ol{LU}_{T})\\
  %   &=\EE(\{\ol{A}_T=\ol{a}_T\}\omega_T(\ol{aLZ}_T)\mid\ol{aZ}_{T-1}\ol{LU}_{T})\\
  %   &=\EE(\{\ol{A}_T=\ol{a}_T\}\omega_T(\ol{aLZ}_T)\mid\ol{aZLU}_{T-1})
  % % &=\EE(\{\ol{A}=\ol{a}\}\omega_T(aLZ_T)\mid\ol{aLU}_{T-1})\text{
  % %   a.s.}
  \end{align*}
  In particular, $\sum_{z_T\in\mathcal{Z}}f_{A_T\mid ...}(a_T\mid \ol{A}_{T-1},\ol{Z}_{T-1},z_T,\ol{LU}_T)\omega_T'(\ol{A}_{T-1},a_T,\ol{Z}_{T-1},z_T,\ol{L}_T)$ is mean-independent of $(L_T,U_T)$ given $\ol{aZLU}_{T-1}$, where $$\omega_T'(\ol{A}_{T-1},a_T,\ol{Z}_{T-1},z_T,\ol{L}_T)=\omega_T(\ol{A}_{T-1},a_T,\ol{Z}_{T-1},z_T,\ol{L}_T)f_{z_T\mid ...}(z_T\mid \ol{AZ}_{T-1}\ol{L}_{T}).$$ Since $T$ was arbitrary, the result follows.
  
\end{proof}

\section{Appendix: Linear omitted variables model, comparing biases}\label{appendix:linear}
 \label{example: linear}
  % We compare the bias of estimators obtained by varying the choice of
  % weights.
  Given a linear MSMM, suppose an estimator is obtained as the root of weighted
  estimating equations
  $$
  \PP_n\left(\omega h(\ol{A})\left(Y-\beta^T\ol{A}\right)\right)=0
  $$
  where the weight $\omega$ is an integrable function of the observed data
  $(\ol{A},\ol{Z},\ol{L})$. This root is a weighted least squares
  estimator
  \begin{align}
    \hat{\beta}=\left(\PP_n(\omega h(\ol{A})\ol{A}^T)\right)^{-1}\PP_n\left(\omega h(\ol{A})Y\right).
    \label{eqn:example:bias}
  \end{align}
 
   Suppose the data satisfy the assumptions of Theorem \ref{theorem:1} and the observed outcome is
%   % $$
%   % \EE(Y\mid \ol{A},\ol{Z},\ol{L},\ol{U}) = % \sum_t
% \beta_{L_t}(g_{L_t}(L_t)-\EE(g_{L_t}(L_t)\mid %
% \ol{L}_{t-1},\ol{A}_{t-1})) + %
% \beta_{U_t}(g_{U_t}(U_t)-\EE(g_{U_t}(U_t)\mid %
% \ol{U}_{t-1},\ol{A}_{t-1})) + % \beta^TD(\ol{A}) + \epsilon,
%   % $$
  $$
  % \EE(Y\mid \ol{A},\ol{Z},\ol{L},\ol{U})
  Y = \sum_t
(\beta_{L_t}g_{L_t}(L_t) + \beta_{U_t}g_{U_t}(U_t)) + \beta^T\ol{A} +
\epsilon,
  $$
  with $\epsilon$ exogenous, $\beta_{L_t},\beta_{U_t}\in\mathbbm{R}^p$,
  and
  $\EE(g_{L_t}(L_t)\mid\ol{L}_{t-1},\ol{A}_{t-1})=\EE(g_{U_t}(U_t)\mid\ol{U}_{t-1},\ol{A}_{t-1})=0$.  As
  discussed in the passage following Lemma \ref{lemma:1}, this
  outcome model is consistent with the MSMM
  $$
  \EE(Y_{\ol{a}})=\beta^T\ol{a}.
  $$
  %  The estimator for the MSMM parameter $\beta$ using
  % empirical estimating equations is a weighted least squares
  The estimator (\ref{eqn:example:bias}) is 
  $$
  \hat{\beta} = \left(\PP_n\left(h(\ol{A})\ol{A}^T/\omega\right)\right)^{-1}\PP_n\left(\omega^{-1}h(\ol{A})\left(\sum_t(\beta_{L_t}g_{L_t}(L_t) + \beta_{U_t}g_{U_t}(U_t)+\epsilon)\right)\right) + \beta.
  $$
  We consider the asymptotic bias of this estimator,
  $$
  \plim \hat{\beta}-\beta=\left(\PP_n\left(h(\ol{A})\ol{A}^T/\omega\right)\right)^{-1}\PP_n\left(\omega^{-1}h(\ol{A})\left(\sum_t(\beta_{L_t}g_{L_t}(L_t) + \beta_{U_t}g_{U_t}(U_t))\right)\right),
  $$
   for various choices of weights $\omega$.

  When $\omega=1$, the resulting estimator $\hat{\beta}$, known as the
  ``associational'' or ``crude'' estimator, ignores all confounding. The implied model is misspecified by omitting covariates $L_t,U_t$. The
  bias is
  $$
  \left(\EE\left(h(\ol{A})\ol{A}^T\right)\right)^{-1}\EE\left(h(\ol{A})\left(\sum_tg_{L_t}(L_t)+g_{U_t}(U_t)\right)\right).
  $$
  This bias is related to the strength of the dependency between the
  treatments and all confounders, known and unknown. When $g_{L_t}$,
  $g_{U_t}$, and $h$ are linear, for example, the bias is linear in
  the covariance between the treatments and the sum of the confounders.

  The SRA estimator, given by the choice
  $\omega=1/\ol{W}^{(SRA)}=1/\prod_tf(A_t\mid \ol{L}_t,\ol{A}_{t-1})$, has bias
  \begin{align}
    \left(\EE\left(h(\ol{A})\ol{A}^T/\prod_tf(A_t\mid L_t,A_{t-1})\right)\right)^{-1}\EE\left(\frac{h(\ol{A})\sum_t(\beta_{L_t}g_{L_t}(L_t) + \beta_{U_t}g_{U_t}(U_t))}{\prod_tf(A_t\mid L_t,A_{t-1})}\right).
    \label{eqn:example:bias:sra}
  \end{align}
  Since it is assumed $\EE(g_{L_T}(L_T)\mid A_{T-1},L_{T-1})=0$,
  \begin{align*}
    \EE\left(\frac{h(\ol{A})g_{L_T}(L_T)}{\prod_{t=1}^Tf(A_t\mid L_t,A_{t-1})}\right) &=
                                                                             \EE\left(\EE\left(\frac{h(\ol{A})g_{L_T}(L_T)}{\prod_{t=1}^Tf(A_t\mid L_t,A_{t-1})}\biggm| L_T,A_{T-1}\right)\right)\\
                                                                           &=\EE\left(\int_{\mathcal{A}}\frac{h(\ol{A}_{T-1},a_T)g_{L_T}(L_T)}{\prod_{t=1}^{T-1}f(A_t\mid L_t,A_{t-1})}\mu_{A_T}(a_T)\right)\\
    &=\EE\left(\int_{\mathcal{A}}\frac{h(\ol{A}_{T-1},a_T)}{\prod_{t=1}^{T-1}f(A_t\mid L_t,A_{t-1})}\mu_{A_T}(a_T)\times\EE\left(g_{L_T}(L_T)\mid A_{T-1},L_{T-1}\right)\right)=0,
  \end{align*}
  and similarly for $t<T$.
  The inverted factor in (\ref{eqn:example:bias:sra}) is, by (\ref{thrm:robins}),
  $\int_{\mathcal{A}}h(\ol{a})\ol{a}^t\mu_{\ol{A}}(\ol{a})$.
  The resulting expression
  $$
  \left(\int_{\mathcal{A}}h(\ol{a})\ol{a}^t\mu_{\ol{A}}(\ol{a})\right)^{-1}\EE\left(h(\ol{A})\left(\sum_tg_{U_t}(U_t)\right)/\ol{W}^{(SRA)}\right),
  $$
  shows that the bias is a quantity related to the dependence between the treatment and unknown
  confounders, as expected due to the violation of SRA. In comparison with the bias of the associational estimator, the
  term corresponding to treatment and known confounder dependency
  is eliminated. As $\sum_tg_{U_t}(U_t)$ and
  $h(\ol{A})/\ol{W}^{(SRA)}$ are generally correlated when $U_t$ are in fact confounders, the bias is
  nonzero. %  Unless it happens that the treatment--known
  % confounder and treatment--unknown confounder dependencies cancel,
  % the associational estimator will have greater bias.--not true
  % because weight term could change things

  When $\omega$ are the IV weights (\ref{defn:weights}), the asymptotic bias is
  zero since we have assumed the conditions of Theorem \ref{theorem:1}, which entails
  $$
  \EE\left(\ol{W}^{-1}h(\ol{A})\sum_t\left(\beta_{L_t}g_{L_t}(L_t) +
  \beta_{U_t}g_{U_t}(U_t)\right)\right)
  =  \EE\left((\ol{W}^{-1}h(\ol{A})(Y - m_{\beta}(\ol{A}))\right) -
  \EE\left(\ol{W}^{-1}h\cdot\epsilon\right) = 0.
  $$

  A Monte Carlo simulation comparing these three estimators is described in Section \ref{section:simulation}.
  % 2.  focus on 1 time point. suppose $Y$ is linear in the
  % confounders and treatment, $Y=\beta_LL+\beta_UU+\beta A+\epsilon$,
  % for $\epsilon$ independent of $L,U$, and Assumption ((latent SRA))
  % is satisfied. Then for $a\in\mathcal{A}$,
  % $\EE(Y_a\mid L,U)=\EE(Y_a\mid A=a,L,U)=\EE(Y\mid
  % A=a,L,U)=\beta_LL+\beta_UU+\beta a$, and the data satisfies the
  % MSMM $\EE(Y_a)=\beta a$. If an instrument $Z$ satisfying
  % Assumptions ((...)) is available, then by lemma,
  % 2. bias when sra weights are used\\
  % 3. When $W=1$,
  % $\hat{\beta}=(A^TA)^{-1}(A^TA)\beta + (A^TA)^{-1}(L+U+\epsilon)$
  % is the OLS estimator. The bias in this case is
  % $\EE(\hat{beta}-\beta)=\EE((A^TA)^{-1}(L+U))$. ((possibility of
  % known and unknown bias cancelling)).

\section{Appendix: Two-state markov chain}\label{appendix:markov}
  We examine the relationship between confounding 
  and the variance of the estimator obtained from the estimating equation (\ref{eqn:estimating eqn}), using a simple model to
  compare expressions in the SRA and IV contexts.

  SRA weights include probability densities at each time point, and IV
  weights include a difference of densities. As the number of time
  points $T$ grows and these weights are multiplied, an estimator may
  quickly become unstable. Let $\hat{\beta}$ be
  obtained as the solution to (\ref{eqn:estimating eqn}). Assuming standard
  regularity conditions, the asymptotic variance of $\hat{\beta}$ is the variance of
  the influence function,
  \begin{align}
  \Var(\sqrt{n}(\hat{\beta}-\beta_0))\to
    \left(\EE\frac{\partial}{\partial \beta}(hm_\beta/\ol{W})\right)^{-2}\EE\left((h(\ol{A})(Y-m_\beta)/\ol{W})^2\right).\label{eg:markov:IF}    
  \end{align}
  % The initial factor is, by ((main lemma)),
  % $$
  % \EE((\sum_tA_t)^2/\ol{W})=\sum_{\ol{a}}(\sum_ta_t)^2=\sum_{j=1}^Tj^2{T\choose j}=2^{T-1}T(T+1),
  % $$ and does not depend
  % on the weights. A first order approximation to the asymptotic
  % variance is
  % $$
  % (2^{T-1}T(T+1))^2\EE(\eta^2+\epsilon^2)\EE(1/\Pi_tW_t^2).
  % $$
  In this display, the weights $\ol{W}$ refer generically to either
  SRA weights (\ref{defn:sra weights}) or IV weights (\ref{defn:weights}). The term $h(\ol{A})m_\beta(\ol{A})$
  is a function of the treatments, so by (\ref{thrm:robins}), in the
  case $\ol{W}$ are SRA weights, or by Theorem \ref{theorem:1}, in the case of
  IV weights,
  $$
  \EE\frac{\partial}{\partial
    \beta}\left(hm_\beta/\ol{W}\right)\vert_{\beta=\beta_0}
  =\int \frac{\partial}{\partial
    \beta}hm_\beta d\mu_{\ol{A}}
  $$
  does not depend on the weights. A first order approximation to the asymptotic
  variance is
  \begin{align}
  \left(\int \frac{\partial}{\partial
    \beta}hm_\beta d\mu_{\ol{A}}\right)^{-2}\EE\left((h(\ol{A})(Y-m_\beta))^2\right)\EE\left(1/\Pi_tW_t^2\right).\label{eg:markov:2}
  \end{align}
  This expression appears to grow exponentially in the number of time
  points. In the SRA framework,
  various techniques have been proposed to stabilize the
  weights. These involve 
  using a function of the treatments to cancel out the
  weights, functions of both treatment and confounders. The stability of the weights therefore depends on the
  strength of the dependence between treatment and confounder, a
  relationship that can be quantified in simple models. We consider
  analogous stabilization for the IV estimator.

  \emph{SRA weights.} Suppose treatment and covariates are binary, and%transitions from a to l and l to a are given as \\
  \begin{gather}
  \begin{aligned}
  \PP(L_t\mid\an(L_t))&=\PP(L_t\mid A_{t-1})=p_{LA},\\
    \PP(A_t\mid \an(A_t))&=\PP(A_t\mid L_{t-1})=p_{AL},\\
    \PP(L_1=0)&=\PP(L_1=1)=1/2.\label{eg:markov:kernel}
  \end{aligned}
\end{gather}
The data is a two-state markov chain with alternating doubly stochastic
  transition matrices,
  \begin{align*}
    \begin{pmatrix} p_{LA} & 1-p_{LA}\\
      1-p_{LA} & p_{LA}
    \end{pmatrix},
                 \begin{pmatrix} p_{AL} & 1-p_{AL}\\
                   1-p_{AL} & p_{AL}
                 \end{pmatrix}.
  \end{align*}

  \begin{figure}
    \centering
    \begin{tikzpicture}
  \def\Tx{0}
  \def\Ty{0}
  \def\offset{2.5}
  \def\Ux{\Tx+2*\offset}
  \def\Uy{\Ty}
  \node[shape=circle,draw=black] (L1) at (\Tx,\Ty) {$L_1$};
  \node[shape=circle,draw=black] (A1) at (\Tx+\offset,\Ty) {$A_1$};
  \node[shape=circle,draw=black] (L2) at (\Ux,\Uy) {$L_2$};
  \node[shape=circle,draw=black] (A2) at (\Ux+\offset,\Uy) {$A_2$};

  \draw [-latex] (L1) to [bend left=0] (A1);
  % \draw [-latex] (L1) to [bend left=25] (L2);
  % \draw [-latex] (L1) to [bend left=25] (A2);
  \draw [-latex] (A1) to [bend left=0] (L2);
  % \draw [-latex] (A1) to [bend left=25] (A2);
  \draw [-latex] (L2) to [bend left=0] (A2);
  \draw [-latex] (L1) to [bend left=0] (A1);

  \node[shape=circle,draw=black] (Y) at (\Ux+2.2*\offset,0) {$Y$};

  \draw [-latex] (L1) to [bend left=25] (Y);
  % \draw [-latex] (A1) to [bend left=25] (Y);
  \draw [-latex] (L2) to [bend left=25] (Y);
  \draw [-latex] (A1) to [bend left=25] (Y);
  \draw [-latex] (A2) to [bend left=0] (Y);

  \node[below] at ({\Tx+\offset+(\Ux-\offset)/2},\Ty-.2) {$p_{AL}$};  
  \node[below] at (\Ux+\offset/2,\Ty-.2) {$p_{LA}$};  
\end{tikzpicture}

% \begin{tikzpicture}
%   \def\Ax{0}
%   \def\Ay{0}
%   \def\offset{2}
%   \def\Bx{\Ax+5}
%   \def\By{\Ay}
%   \node[shape=circle,draw=black] (A0) at (\Ax,\Ay) {$A(j)$};
%   \node[shape=circle,draw=black] (L0) at (\Ax-\offset,\Ay) {$L(j)$};
%   \node[shape=circle,draw=black,font=\tiny] (A1) at (\Bx,\By) {$A(j+1)$};
%   \node[shape=circle,draw=black,font=\tiny] (L1) at
%   (\Bx-\offset,\By) {$L(j+1)$};

%   \draw [-latex] (L0) to [bend left=0] (A0);
%   \draw [-latex] (L0) to [bend left=30] (L1);
%   \draw [-latex] (L0) to [bend left=30] (A1);
%   \draw [-latex] (A0) to [bend left=0] (L1);
%   \draw [-latex] (A0) to [bend left=30] (A1);
%   \draw [-latex] (L1) to [bend left=0] (A1);
%   \draw [-latex] (L0) to [bend left=0] (A0);

%   % \draw[->]  (\Ax-3,\Ay) -- (L0);
  
%   \node at (\Ax-4,0) {. . . };
%   \node at (\Bx+1.8,0) {. . . };

%   \node[shape=circle,draw=black] (Y) at (\Bx+3.2,0) {$Y$};
%   \draw [-latex] (\Bx+2.5,.5) to (Y);
%   \draw [-latex] (\Bx+2.4,.4) to (Y);
%   \draw [-latex] (\Bx+2.3,.3) to (Y);
%   \draw [-latex] (\Bx+2.25,.2) to (Y);
% \end{tikzpicture}
    \caption{DAG for the two-state markov model meeting the sequential
      randomization assumption, with 2 time points.}
    \label{eg:markov:DAG:SRA}
  \end{figure}
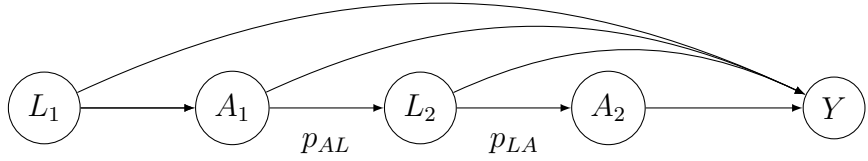
  
  The parameter $p_{LA}$ is the probability that the state of
  $A_{t+1}$ is the same as $L_t$. It may be interpreted as the
  strength of the dependence of $A$ on $L$, a type of confounding,
  with the strongest confounding occurring as $p_{LA}$ nears the
  border of $[0,1]$, and the weakest at $p_{LA}=.5$. The situation is
  analogous for $p_{AL}$. The marginal distributions of both $L_t$ and
  $A_t, t=1,\ldots,T,$ are bernoulli with success probability 1/2.
  % The parameter $p_{LA}$ describes one part of confounding, the
  % dependence of treatment on the confounding covariate.
  % which is
  % constant across time in this example
  % The parameter $\lambda$
  % describes the other part of confounding, the dependence of the outcome on the
  % covariate.
  
  The SRA weights are $\prod_tf(A_t\mid
  L_{t-1})=\prod_tp_{LA}^{\{A_t=L_{t-1}\}}(1-p_{LA})^{\{A_t\neq
    L_{t-1}\}}=(1-p_{LA})^T\prod_t(p_{LA}/(1-p_{LA}))^{\{A_t=L_{t-1}\}}$. Because
  $\EE(\{A_t=L_{t-1}\}\mid A_{t-1})=p_{LA}$ and the states are binary, the factors that make up
  the weights are 
  independent and identically distributed, and
  \begin{align}
  \EE(1/\ol{W}^{2})=(\EE(1/W_1^2))^T=(p_{LA}(1-p_{LA}))^{-T}.\label{eg:markov:1}
  \end{align}
  The variance is
 polynomial in the inverse of $p_{LA}(1-p_{LA})$, a
  measure of
  treatment--covariate dependence, with order given by the number
  of time points $T$. The parameter $p_{AL}$ determining $A_{t-1}\to
  L_t$ transitions does not play a role, although it plays the main role
  in weight stabilization discussed below. The dependence on $p_{LA}$ is through 
  $p_{LA}(1-p_{LA})=1/4-(p_{LA}-1/2)^2$, so that (\ref{eg:markov:1}) is
  minimized over $p_{LA}$ at $1/2$, when treatment and covariate are
  independent, and increases without bound as $|p_{LA}-1/2|\to 1/2$.

\begin{figure}
    \centering
    \includegraphics[scale=.5]{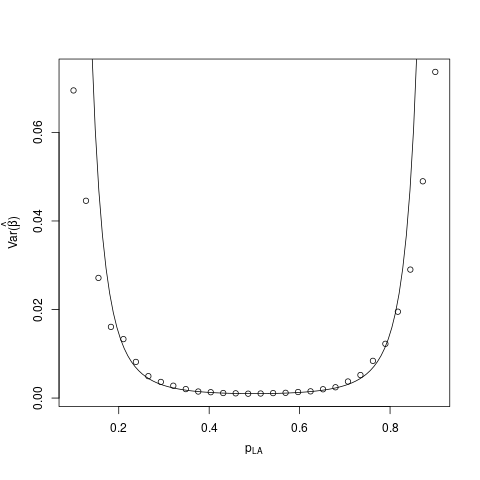}
    \caption{The variance of the unstabilized SRA estimator $\hat{\beta}$ in the two-state markov model. The solid line is the first-order approximation (\ref{eg:markov:3}) and the plotted characters come from a Monte Carlo simulation. For the simulation the number of time points $T$ is $7$ and sample size is $500$.}\label{eg:markov:fig:1}
  \end{figure}

  Although the focus on this example is the behavior of the weights,
  an estimate of the variance of the full estimator is straightforward
  once an outcome model is specified. Suppose the observed outcome $Y$ satisfies
  $$
  \EE(Y\mid \ol{A},\ol{Z},\ol{L},\ol{U}) = \lambda \sum_t
  (L_t-\EE(L_t\mid \ol{L}_{t-1},\ol{A}_{t-1})) + \beta\sum_tA_t +
  \epsilon,
  $$
  where $\epsilon$ has mean zero and variance $\sigma^2$. While parameter $p_{LA}$ describes one part of confounding, the dependence of treatment on the confounding covariate, the parameter $\lambda$
  describes the other part of confounding, the dependence of the outcome on the
  covariate. By Lemma \ref{lemma:1},  this model for $Y$ is consistent with the MSMM 
  $$
  \EE(Y_{\ol{a}})=\beta \sum_t a_t.
  $$
  It follows that the first order approximation (\ref{eg:markov:2}) is
  \begin{align}
  \frac{\lambda^2(1+\sigma^2/(Tp_{LA}(1-p_{LA})))}{(T+1)(4p_{LA}(1-p_{LA}))^{T-1}}\label{eg:markov:3}
  \end{align}
  The principal difference from the second moment of the weights (\ref{eg:markov:1}) is that the
  exponent is $T-1$ rather than $T$, and quadratic dependence on
  $\lambda$. A plot of the dependence on $p_{LA}$, along with the empirical variance
  from a small simulation to indicate the quality of the approximation, is given in Fig. \ref{eg:markov:fig:1}.
  \begin{figure}
    \centering
    \includegraphics[scale=.5]{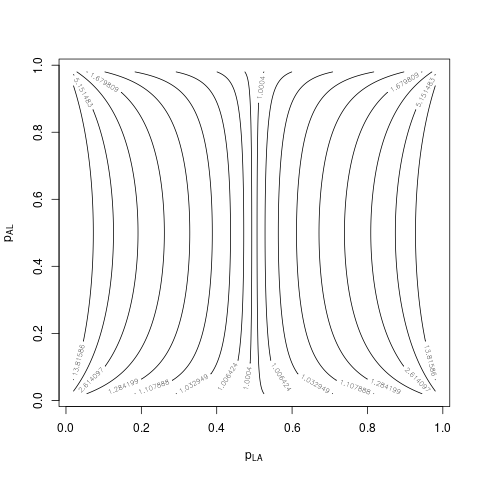}
    \caption{Variance of stabilized SRA weights in the two-state markov model. The variance blows up as $p_{LA}$ approaches 0 or 1, as in the unstabilized case, but remains bounded if $p_{AL}$ approaches 0 or 1 with at least the same rate.}\label{eg:markov:fig:2}
  \end{figure}
  % blowing up as $p_{LA}$ is farther from .5, i.e., the
  % treatment-covariate confounding increasing.
  
  Modified weights are often used to mitigate the instability of the
  SRA estimator.
  A factor $h'$ in the function $h(\ol{A})$ is chosen to approximate
  $f(A_t|\ol{L}_t,\ol{A}_{t-1})$, with a view to minimizing the
  mean square of the influence function
  (\ref{eg:markov:IF}). In the trivial case that $L$ is not in
  fact a confounder,  $h'$ may be taken to be
  $f(A_t|\ol{L}_t,\ol{A}_{t-1})=f(A_t|\ol{A}_{t-1})$. The
  weights are cancelled out and the estimator is no longer exponential in
  $T$. In general, the quality of an approximation of $f(A|L)$ using a
  function of $A$ depends on how well $A$ predicts $L$, 
  controlled in this example by $p_{AL}$. The variance of the influence function (\ref{eg:markov:IF}) does not change on multiplying $h$
  by a constant, so the minimization is well-posed, and there is no loss of generality to assume $\int
  h'=1$. 
  
  A common choice of stabilized weights, which we consider, uses
  the density $f(A_t|\ol{A}_{t-1})$ as the approximation to
  $f(A_t|\ol{L}_t,\ol{A}_{t-1})$%  ((though this minimizes the
  % difference with $f(A|L)$, not (ratio - 1)))
  , that is, $h$
  contains as a factor the
  joint density of $\ol{A}$. For the two-state markov model, the
  markov property gives as stabilized weights,
  $$
  \ol{W}=\prod_t\frac{f(A_t|L_t)}{f(A_t\mid A_{t-1})}.
  $$
  The factors are again i.i.d., and the second moment of the
  inverted weights is computed to be
  $$
  \EE(1/\ol{W}^2)=\left(\EE(\frac{f(A_2\mid A_1)}{f(A_2\mid L_1)})\right)^T=\left(1+4\frac{p_{AL}(1-p_{AL})}{p_{LA}(1-p_{LA})}(p_{LA}-1/2)^2\right)^T.
  $$
  % \comment{express all in terms of $p_{LA}(1-p_{LA})$ or $(p_{LA}-1/2)^2$}
  
  Holding $T$ fixed, consider the behavior of the variance as the
  parameters $p_{LA},p_{AL}$ vary. When $p_{LA}=p_{AL}=p$, this
  expression is $(1+4(p-1/2))^T$, and the blowup at the boundary
  points present in the case of unstabilized weights is eliminated
  (Fig. \ref{eg:markov:fig:2}).  For $p\in[0,1]$ let
  $\rho(p)=p(1-p)=1/4-(p-1/2)^2$, a measure of the distance of $p$ to
  the boundary of $[0,1]$. With this notation,
  \begin{align}
  \EE(1/\ol{W}^2)=\left(1+4\frac{\rho(p_{AL})}{\rho(p_{LA})}(p_{LA}-1/2)^2\right)^T\le \left(1+\frac{\rho(p_{AL})}{\rho(p_{LA})}\right)^T.\label{eg:markov:4}
  \end{align}
  The behavior of stabilized weights as $p_{LA}$ nears the boundary of $[0,1]$ is governed not by $1/\rho(p_{LA})$, as in the unstabilized case, but the ratio $\rho(p_{AL})/\rho(p_{LA})$, and will be bounded when $\rho(p_{AL})=\mathcal{O}(\rho(p_{LA}))$.
  % when $|p_{AL}-1/2|$ is equal to or greater than $|p_{LA}-1/2|$, the
  % weights are polynomial not in $1/p_{LA}$ but
  % $1+p_{LA}^2$.
  Qualitatively, this situation occurs when the
  degree of treatment-covariate confounding does not grow faster than the
  treatment's predictiveness of the covariate. % ((lrage p2 helpful
  % for stabilized variance, large p1 harmful))

  Next, let $T$ grow. It follows from (\ref{eg:markov:4}) that the variance can be stabilized by controlling the decay of $\rho(p_{AL})/\rho(p_{LA})$. By comparison with $t\mapsto(1+1/t)^t$ it follows $\rho(p_{AL})/\rho(p_{LA})=\mathcal{O}(1/T)$ is sufficient. This possibility is not available with unstabilized weights. Since $p_{AL}(1-p_{AL})\le 1/4$, the unstabilized weight moment (\ref{eg:markov:1}) $(p_{AL}(1-p_{AL}))^{-T}\ge 4^T$ always diverges with $T$.

  \emph{IV weights.}

  We next consider an extension of the two-state markov model (\ref{eg:markov:kernel}) in order to examine the behavior of the IV
  weights in relation to confounding. The states $L_t$ are augmented
  with additional binary variables $U_t$ and $Z_t$, giving rise to a process $\ldots (L_{t-1},U_{t-1},Z_{t-1})\to A_{t-1}\to (L_t,U_t,Z_t)\to A_t\ldots$ . For $q,p_L,p_U\in
  (0,1)$ and suitable $\delta_0,\delta_1$, as discussed below, %satisfying $|\delta_l|<1/2-\max(|p_{l0}-1/2|,|p_{l1}-1/2|),l\in\{0,1\}$,
  define transition probilities through
  \begin{gather}
  \begin{aligned}
    \PP(A_t=a\mid \an(A_t))&=\PP(A_t=a\mid L_t=l,U_t=u,Z_t=z)\\
    &=\PP(A_t=a\mid L_t=l,U_t=u)+(-1)^{1-z}(-1)^{1-a}\delta_l/2\\
    \PP(A_t=a\mid L_t=l,U_t=u)&=(1-q)p_L^{\{l=a\}}(1-p_L)^{\{l\neq
                                a\}} + qp_U^{\{u=a\}}(1-p_U)^{\{u\neq a\}}\\
    % \PP(L_{t+1}=l,U_{t+1}=u\mid A_t=a)&=(1/2)\PP(A_t=a\mid L_t=l,U_t=u)\\
    % \PP(L_{t+1}=l,U_{t+1}=u,Z_{t+1}=z\mid A_t=a,\an(L_{t+1},U_{t+1},Z_{t+1}))&=\PP(Z_{t+1}=z)\PP(L_{t+1}=l,U_{t+1}=u\mid A_t=a)\\
                              \PP(L_{t+1}=l,U_{t+1}=u,Z_{t+1}=z&\mid A_t=a,\an(L_{t+1},U_{t+1},Z_{t+1}))=\PP(Z_{t+1}=z)\PP(L_{t+1}=l,U_{t+1}=u\mid A_t=a)\\
    % \PP(L_{t+1}=l,U_{t+1}=u,Z_{t+1}=z\mid A_t=a,\an(L_{t+1},U_{t+1},Z_{t+1}))\\
                              % &=\PP(Z_{t+1}=z)\PP(L_{t+1}=l,U_{t+1}=u\mid A_t=a)\\
    &=(1/4)\PP(A_t=a\mid L_t=l,U_t=u)\\
    \PP(Z_t=z)&=1/2.\label{eg:markov:kernel:IV}
  \end{aligned}
\end{gather}
The initial state $(L_1,U_1,Z_1)$ is distributed as three i.i.d.
  symmetric bernoulli variables. It follows that the marginal
  distribution of each of $L_t,U_t,A_t,t=1,\ldots,$ is bernoulli with
  success probability 1/2, as with $Z_t$. A DAG is given in Fig. \ref{eg:markov:fig:DAG:IV}. 
  \begin{figure}
    \centering
    \begin{tikzpicture}
    \def\Ax{2}
    \def\Ay{0}
    \def\offset{2.5}
    \def\Bx{\Ax+5}
    \def\By{\Ay}
    \node[shape=circle,draw=black] (A1) at (\Ax,\Ay) {$A_1$};
    \node[shape=circle,draw=black,fill=lightgray] (U1) at (\Ax-\offset,\Ay+\offset/1.5) {$U_1$};
    \node[shape=circle,draw=black] (L1) at (\Ax-\offset,\Ay-\offset/1.5) {$L_1$};
    \node[shape=circle,draw=black] (Z1) at (\Ax,\Ay-1.4*\offset) {$Z_1$};
    \node[shape=circle,draw=black] (A2) at (\Bx,\By) {$A_2$};
    \node[shape=circle,draw=black,fill=lightgray] (U2) at (\Bx-\offset,\By+\offset/1.5) {$U_2$};
    \node[shape=circle,draw=black] (Z2) at (\Bx,\By-1.4*\offset) {$Z_2$};
    \node[shape=circle,draw=black] (L2) at (\Bx-\offset,\By-\offset/1.5) {$L_2$};
    \node[shape=circle,draw=black] (Y) at (\Bx+3,\By) {$Y$};

    \draw [-latex] (U1) to (A1);
    \draw [-latex] (L1) to (A1);
    \draw [-latex] (U2) to (A2);
    \draw [-latex] (L2) to (A2);
    \draw [-latex] (Z1) to (A1);
    \draw [-latex] (Z2) to (A2);
    % \draw [-latex] (Z1) to (Z2);
    % \draw [-latex] (A1) to (A2);
    \draw [-latex] (A1) to (L2);
    \draw [-latex] (A1) to (U2);

    % \draw [-latex] (L1) to (Z1);
    % \draw [-latex] (L1) to (Z1);
    % \draw [-latex] (L1) to (Z2);
    % \draw [-latex] (L1) to [bend left=0] (L2);
    % \draw [-latex] (L1) to [bend right=15] (U2);
    % \draw [-latex] (L1) to [bend left=0] (A2);
    % \draw [-latex] (U1) to [bend left=0] (U2);
    % \draw [-latex] (U1) to [bend left=15] (L2);
    % \draw [-latex] (U1) to [bend left=0] (A2);

    \draw [-latex] (L1) to [bend right=25] (Y);
    \draw [-latex] (L2) to [bend right=0] (Y);
    \draw [-latex] (U1) to [bend left=25] (Y);
    \draw [-latex] (U2) to [bend left=0] (Y);
    \draw [-latex] (A2) to [bend left=0] (Y);
    \draw [-latex] (A2) to [bend left=0] (Y);

    % \draw [-latex] (L1) to [bend left=0    ] (U2);
    % \draw [-latex] (L1) to [bend left=0] (L2);
    % \draw [-latex] (L1) to [bend right=10] (A2);
    % % \draw [-latex] (L1) to [bend left=10] (L2);
    % \draw [-latex] (A1) to [bend left=10] (U2);
    % \draw [-latex] (A1) to [bend left=0] (Z2);
    % \draw [-latex] (A1) to [bend right=10] (L2);
    % \draw [-latex] (A1) to [bend right=10] (A2);
    % \draw [-latex] (A1) to [bend right=10] (L2);
    % \draw [-latex] (Z1) to [bend left=0] (A1);
    % \draw [-latex] (Z2) to [bend left=0] (A2);
    % % \draw [-latex] (Z1) to [bend right=10] (L2);
    % \draw [-latex] (Z1) to [bend right=0] (A2);
    % % \draw [-latex] (Z0) to [bend right=10] (L2);
    % \draw [-latex] (Z0) to [bend right=0] (Z2);

    % \draw [-latex] (L0) to [bend left=25] (Y);
    % \draw [-latex] (L2) to [bend left=30] (Y);
    % \draw [-latex] (U0) to [bend left=35] (Y);
    % \draw [-latex] (U2) to [bend left=0] (Y);

    % \draw [-latex] (\Bx+2.8,\By) to [bend left=0] (Y);
    % \draw [-latex] (\Bx+2.8,\By+.4) to [bend left=0] (Y);
    % \draw [-latex] (\Bx+2.8,\By-.4) to [bend left=0] (Y);
    % \draw [-latex] (A2) to (\Bx+1.3,\By);

  \end{tikzpicture}
    \caption{DAG for the two-state markov model with unknown
      confounding and IV, with 2 time points.[[links between L and U]]}
    \label{eg:markov:fig:DAG:IV}
  \end{figure}
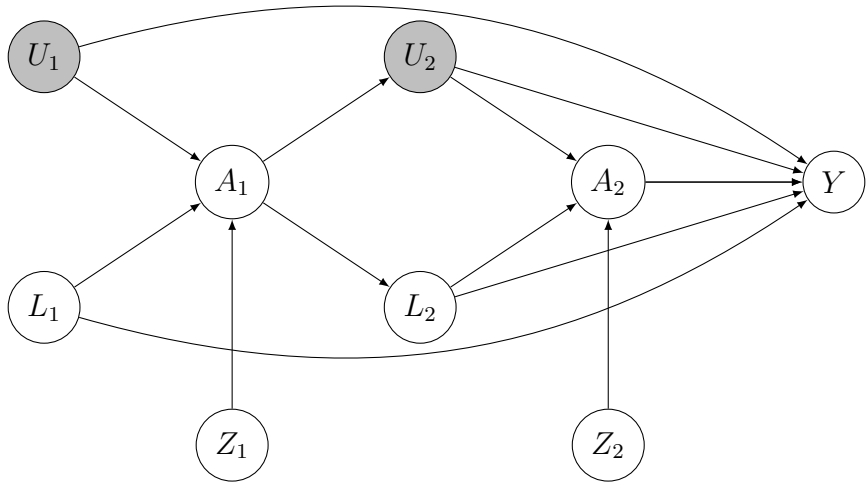
                %     \\
  \begin{figure}
    \centering
    % \begin{tikzpicture}
%   \def\Tx{0}
%   \def\Ty{0}
%   \def\offset{2.5}
%   \def\Ux{\Tx+2*\offset}
%   \def\Uy{\Ty}

%   \node[shape=circle,draw=black] (L1) at (\Tx,\Ty) {$L_1$};
%   \node[shape=circle,draw=black] (AL1) at (\Tx+\offset,\Ty) {$A_1^L$};
%   \node[shape=circle,draw=black] (L2) at (\Ux,\Uy) {$L_2$};
%   \node[shape=circle,draw=black] (AL2) at (\Ux+\offset,\Uy) {$A_2^L$};
%   \draw [-latex] (L1) to [bend left=0] (AL1);
%   \draw [-latex] (AL1) to [bend left=0] (L2);
%   \draw [-latex] (L2) to [bend left=0] (AL2);
%   \draw [-latex] (L1) to [bend left=0] (AL1);
%   \node[shape=circle,draw=black] (Y) at (\Ux+2.2*\offset,0) {$Y$};
%   \draw [-latex] (L1) to [bend left=25] (Y);
%   \draw [-latex] (L2) to [bend left=25] (Y);
%   \draw [-latex] (AL1) to [bend left=25] (Y);
%   \draw [-latex] (AL2) to [bend left=0] (Y);
%   \node[below] at ({\Tx+\offset+(\Ux-\offset)/2},\Ty-.2) {$p_{AL}$};  
%   \node[below] at (\Ux+\offset/2,\Ty-.2) {$p_{LA}$};  

% \end{tikzpicture}

  \begin{tikzpicture}
  \def\Tx{0}
  \def\Ty{0}
  \def\offset{2.5}
  \def\Ux{\Tx+2*\offset}
  \def\Uy{\Ty}
  \node[shape=circle,draw=black,minimum size=10mm,inner sep=0pt] (L1) at (\Tx,\Ty) {$L_t$};
  \node[shape=circle,draw=black,minimum size=10mm,inner sep=0pt] (AL1) at (\Tx+\offset,\Ty) {$A^L_t$};
  \node[shape=circle,draw=black,minimum size=10mm,inner sep=0pt] (L2) at (\Ux,\Uy) {$L_{t+1}$};
  \node[shape=circle,draw=black,minimum size=10mm,inner sep=0pt] (AL2) at (\Ux+\offset,\Uy) {$A^L_{t+1}$};

  \draw [-latex] (L1) to [bend left=0] (AL1);
  \draw [-latex] (AL1) to [bend left=0] (L2);
  \draw [-latex] (L2) to [bend left=0] (AL2);
  \draw [-latex] (L1) to [bend left=0] (AL1);
  \draw [] (AL2) to (\Ux+2*\offset-\offset/3,\Uy);
  \node[right] at (\Ux+2*\offset-\offset/3,\Uy) {$\ldots$};
  \draw [-latex] (\Tx-2/3*\offset,\Ty) to (L1);
  \node[left] at (\Tx-2/3*\offset,\Ty) {$\ldots$};

  \node[below] at ({\Tx+\offset+(\Ux-\offset)/2},\Ty-.2) {$p_{AL}$};  
  \node[below] at (\Ux+\offset/2,\Ty-.2) {$p_{LA}$};  

  % repeat for U nodes
  \def\Ty{-3}
  \node[shape=circle,draw=black,minimum size=10mm,inner sep=0pt,fill=lightgray] (U1) at (\Tx,\Ty) {$U_t$};
  \node[shape=circle,draw=black,minimum size=10mm,inner sep=0pt] (AU1) at (\Tx+\offset,\Ty) {$A^U_t$};
  \node[shape=circle,draw=black,minimum size=10mm,inner sep=0pt,fill=lightgray] (U2) at (\Ux,\Uy) {$U_{t+1}$};
  \node[shape=circle,draw=black,minimum size=10mm,inner sep=0pt] (AU2) at (\Ux+\offset,\Uy) {$A^U_{t+1}$};

  \draw [-latex] (U1) to [bend left=0] (AU1);
  \draw [-latex] (AU1) to [bend left=0] (U2);
  \draw [-latex] (U2) to [bend left=0] (AU2);
  \draw [-latex] (U1) to [bend left=0] (AU1);
  \draw [] (AU2) to (\Ux+2*\offset-\offset/3,\Uy);
  \node[right] at (\Ux+2*\offset-\offset/3,\Uy) {$\ldots$};
  \draw [-latex] (\Tx-2/3*\offset,\Ty) to (U1);
  \node[left] at (\Tx-2/3*\offset,\Ty) {$\ldots$};

  \node[below] at ({\Tx+\offset+(\Ux-\offset)/2},\Ty-.2) {$p_{AU}$};  
  \node[below] at (\Ux+\offset/2,\Ty-.2) {$p_{UA}$};  

  % mixing parameter
  \node (q1) at (\Tx+\offset,\Ty/2) {q};
  \draw[dotted] (q1) to (AL1);
  \draw[dotted] (q1) to (AU1);
  \node (q2) at (\Ux+\offset,\Ty/2) {q};
  \draw[dotted] (q2) to (AL2);
  \draw[dotted] (q2) to (AU2);

\end{tikzpicture}
    \caption{The distribution of the two-state markov model with unknown confounding (\ref{eg:markov:kernel:IV}) may be obtained by combining two chains with no unknown confounders (See Fig. \ref{eg:markov:DAG:SRA}). Corresponding covariate states $L_t$ and $U_t$ are
 concthatenated along with an exogenous IV to give the new covariate
 state $(L_t,U_t,Z_t)$. The new treatment states are obtained by
 mixing with probability $q$, $A_t\coloneqq B_tA_t^L + (1-B_t)A_t^U$,
 with $B_t$ i.i.d. bernoulli with parameter $q$.}
    \label{eg:markov:fig:DAG:IV:2}
 \end{figure}
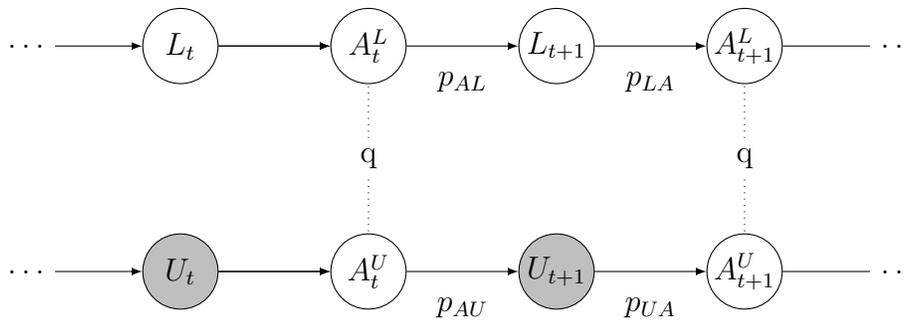
 The model for the conditional density of $A_t$ given $(L_t,U_t,Z_t)$
 may be described by parameters $p_{lu}=\PP(A=0\mid Z=0,L=l,U=u)\in[0,1]$ and $\delta_l$ for
 $l,u\in\{0,1\}$; see Table \ref{eg:markov:table}. Requiring
 $|\delta_l|<1/2-\max(|p_{l0}-1/2|,|p_{l1}-1/2|)$ ensures
 $\PP(a_t\mid l_t,u_t,z_z)>0$. Summing horizontally in Table
 \ref{eg:markov:table} shows $\sum_a\PP(a_t\mid
 l_t,u_t,z_t)=1$. Therefore $\PP(a_t\mid l_t,u_t,z_t)$ is a valid
 density. Moreover, Assumption \ref{assumption:ict} is satisfied since
  \begin{align*}
    \Delta_t(\ol{a}_t,\ol{z}_{t-1},\ol{l}_t,\ol{u}_t)\equiv&\sum_{z_t\in\{0,1\}}(-1)^{1-z_t}\PP(A_t=a_t\mid \ol{L}_t=\ol{l}_t,\ol{U}_t=\ol{u}_t,\ol{Z}_{t-1}=\ol{z}_{t-1},Z_t=z,\ol{A}_{t-1}=\ol{a}_{t-1})\\
    &=\sum_{z_t\in\{0,1\}}(-1)^{1-z_t}\PP(A_t=a_t\mid L_t=l_t,U_t=u_t,Z_t=z_t)\\
    &=(-1)^{1-a_t}\delta_{l_t}
  \end{align*}
  does not
  depend on $\ol{u}_t$. The magnitude
  of $\delta_0$ and $\delta_1$ are interpretable as IV strength. Their difference
  $|\delta_1-\delta_0|$ gives
  the dependence of $\delta$ on $L_t$, which has an analogous role in
  weight stabilization to
  the treatment-confounder dependence $p_{LA}$ parameter in the SRA
  setting.  That is, to the extent
  that this dependence may be approximated by a standardized
  function of $\ol{A}$, an analogue of stabilized weights 
  may be used to decrease the variance of the estimator.

The parameters $q,p_L,p_U$ used to describe the model (\ref{eg:markov:kernel:IV}) are not identified by the
   data $(A_t,Z_t,L_t,U_t)$, nor are the observed parameters $q,p_L$ identified by the observed
   data $(A_t,Z_t,L_t)$. We use them because they allow for easy comparison with the SRA case. For purposes of estimation (e.g., Appendix \ref{appendix:markov}), an
   identifying condition like $p_B=1/2$ or $p_L=p_U$ is needed, or reparameterization.

 The distribution of the resulting markov chain can also be obtained
 by mixing two independent chains of the type described in the ((ref sra section
 above)), say, $\ldots \to L_{t-1}\to A_{t-1}^L\to L_t\to\ldots$ with
 parameters $p_{AL},p_{LA}$, and
 $\ldots\to U_{t-1}\to A_{t-1}^U\to U_t\to\ldots$ with parameters
 $p_{AU},p_{UA}$. See Fig. \ref{eg:markov:fig:DAG:IV:2}. Corresponding covariate states $L_t$ and $U_t$ are
 concatenated along with an exogenous IV to give the new covariate
 state $(L_t,U_t,Z_t)$. The new treatment states are obtained by
 mixing with probability $q$, $A_t=B_tA_t^L + (1-B_t)A_t^U$,
 with $B_t$ i.i.d. bernoulli with parameter $q$. 
  % The distribution
  % of $Z_t$ is independent of the past, $\PP(Z_t=0\mid
  % past)=\PP(Z_t=1\mid past)=1/2$. The conditional density of the
  % treatments $A_t$ is given as
  % $$
  % \PP(A_t=a\mid L_t=l,U_t=u,Z_t=z) =.
  % $$
 The mixing parameter $q$ controls the relative dependence of the
 treatment on known confounding as compared with unknown
 confounding. The IVs are then added as independent, exogenous
 perturbations of the new treatment states $A_t$ in such a way that
 the Assumption \ref{assumption:ict} is satisfied. The parameters
 $p_{LA},p_{AL},p_{UA},$ and $p_{AU}$ have similar interpretations as
 before% , though we set $p_{LA}=p_{AL}=p_L$ and $p_{UA}=p_{AU}=p_U$ to
 % simplify the exposition
 .

  \begin{table}[h]
    \caption{The conditional treatment densities
      $\PP(A=a\mid L=l,U=u,Z=z)$ for the two-state markov model with
      IVs. The densities are the same at all times. There are 6
      parameters, $p_{lu}\in[0,1]$ and $\delta_l$ for
      $l,u\in\{0,1\}$. The magnitude of $\delta_0$ and $\delta_1$ are
      interpretable as IV strength. Their difference
      $|\delta_1-\delta_0|$ gives the dependence of $\delta$ on $L_t$,
      which has an analogous role in weight stabilization to the
      treatment-confounder dependence parameter $p_{LA}$ in the SRA
      setting.}
  \label{eg:markov:table}
    \centering
    \begin{tabular}{l|l|cccc} 
      % \toprule 
      % \multirow{2}{*}{\parbox{4em}{c}} & \multirow{2}{*}{\parbox{6em}{d}}  & \multicolumn{2}{c}{e}\\%\cmidrule{3-4} 
      %                                              &  &  $L=0$ & $L=1$\\ \midrule 
      $U=1$  & $Z=1$ & $p_{01}\pm\delta_0$&   $\ol{p_{01}}\mp\delta_0$& $p_{11}\pm\delta_1$&   $\ol{p_{11}}\mp\delta_1$\\
                                         & $Z=0$ & $p_{01}$&   $\ol{p_{01}}$& $p_{11}$&   $\ol{p_{11}}$\\
      $U=0$ & $Z=1$ & $p_{00}\pm\delta_0$&   $\ol{p_{00}}\mp\delta_0$& $p_{01}\pm\delta_1$&   $\ol{p_{01}}\mp\delta_1$\\
                                         & $Z=0$ & $p_{00}$&   $\ol{p_{00}}$& $p_{01}$&   $\ol{p_{01}}$\\%\midrule 
          \noalign{\hrule height .5pt}
      % \multirow{2}{*}{\parbox{4em}{c}} & \multirow{2}{*}{\parbox{6em}{d}}  & \multicolumn{2}{c}{e}\\%\cmidrule{3-4} 
                                         &  & $A=0$ & $A=1$ & $A=0$ & $A=1$\\ %\midrule
          \noalign{\hrule height .5pt}
                                                   &  &  \multicolumn{2}{c}{$L=0$} & \multicolumn{2}{c}{$L=1$}\\  
    \end{tabular}
  \end{table}

\begin{figure}
    \centering
    \includegraphics[scale=.5]{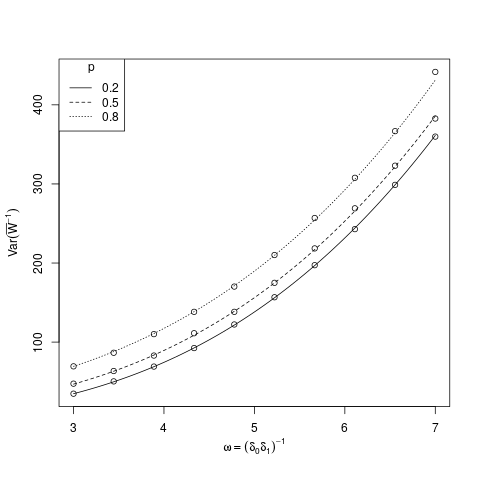}
    \caption{The second moment of the unstabilized IV weights in the two-state markov model. The effect of the known confounding, as measured by $p$, is small relative to the effect of the IV weakness, as measured by $1/(\delta_0\delta_1)$. Another factor, the degree of known confounding of the IV, is fixed in this figure. The lines are the theoretical values and the plotted characters come from a Monte Carlo simulation. For the simulation the number of time points $T$ is 5 and the sample size $n$ is 50.}\label{eg:markov:fig:3}
  \end{figure}

  The conditional density of $Z_t$ is a constant in $(0,1)$ and may be canceled by
  the choice of $h$, so the square of the
  inverse of the IV weights (\ref{defn:weights}) is
  $$
  \ol{W}^{-2}=\prod_{t=1}^T\delta_{L_t}^{-2}.
  $$
  For $t=2,\ldots,T,$ and $l_{t-1}\in\{0,1\}$ define $\phi_t(l_{t-1})=\EE(\prod_{t'=t}^T\delta_{L_{t'}}^{-2}\mid
  L_{t-1}=l_{t-1})$ and $\phi_t=(\phi_t(0),\phi_t(1))$. With this notation,
  $\EE(\ol{W}^{-2})=\sum_{l_1\in\{0,1\}}\EE(\ol{W}^{-2}\mid L_1=l_1)\PP(L_1=l_1)=\delta_0^{-2}\phi_2(0)/2+\delta_1^{-2}\phi_2(1)/2$. Let $p=\PP(L_{t-1}=1\mid
  L_t=1)=\PP(L_{t-1}=0\mid L_t=0)$, which does not in fact depend on $t$ as the chain has been assumed to be started in its stationary distribution. Then $\phi_t$ satisfies the recurrence
  \begin{align}
  \phi_{t-1}=
  \begin{pmatrix}p/\delta_0^2 & (1-p)/\delta_1^2\\
    (1-p)/\delta_0^2 & p/\delta_1^2
  \end{pmatrix}
  \phi_t\label{eg:markov:recurrence}
  \end{align}
  with boundary condition $\phi_{T+1}=(1,1)$.
  % $$
  % \phi_{T}=(\EE(\delta_T^{-2}\mid L_{T-1}=0),\EE(\delta_T^{-2}\mid L_{T-1}=1))=(p\delta_0^{-2}+(1-p)\delta_1^{-2},(1-p)\delta_0^{-2}+p\delta_1^{-2}).
  % $$
  The growth of $\EE(\ol{W}^{-2})$ is determined by the largest eigenvalue of
  the matrix in (\ref{eg:markov:recurrence}),
  \begin{align}
  \lambda_1&=p/2(1/\delta_0^2+1/\delta_1^2)+\sqrt{p^2/4(1/\delta_0^2+1/\delta_1^2)^2-(2p-1)/(\delta_0\delta_1)^2} \label{eg:markov:eigen:unstab}    
    % &=p/2(1/\delta_0^2+1/\delta_1^2)+\sqrt{p^2/4(1/\delta_0^2-1/\delta_1^2)^2+(1-p)^2/(\delta_0\delta_1)^2}.
  \end{align}
  % ((can probably delete second expression, not using any more in what
  % follows))
  The eigenvalue is real for % $\delta_0,\delta_1>0$ and
  $p\in(0,1)$.  To reinterpret this expression, let
  \begin{gather}
  \begin{aligned}
    \omega \coloneqq 1/(\delta_0\delta_1)\\
    \kappa \coloneqq 1/\delta_0^2-1/\delta_1^2\label{eg:markov:omegakappa}
  \end{aligned}
\end{gather}
The product $\omega$ is a measure of IV weakness, and the difference
$\kappa$ is a measure of confounding between the IV and $L$. After
some algebera, it follows that
$1/\delta_0^2+1/\delta_1^2=\pm\sqrt{\kappa^2+4\omega^2}$, and the
principal eigenvalue (\ref{eg:markov:eigen:unstab}) is
  \begin{align*}
    \lambda_1=p/2\sqrt{\kappa^2+4\omega^2}\left( 1 + \sqrt{1-\frac{\omega^2(2p-1)}{\kappa^2+4\omega^2}}\right).
  \end{align*}
  The term in parentheses is at most $2$, so
  $\lambda_1\le p\sqrt{\kappa^2+4\omega^2}$, with equality occurring
  when the transtion probability $p$ is $1/2$. Therefore, $\lambda_1$,
  which determines the exponential growth of the second moment of the
  weights, is approximately linear in the weakness of the IV and the degree
  of IV confounding. In comparison to the case of SRA weights, the
  transition probabilities $p$ have a relatively small effect; see
  Fig. \ref{eg:markov:fig:3}.

  % \comment{ can probably
  %   delete the following inequality now. Using
  %   $0<\delta_0,\delta_1<1$, the largest eigenvalue may be bounded as
  % \begin{align*}
  %   p/2(1/\delta_0^2+1/\delta_1^2)+\sqrt{p^2/4(1/\delta_0^2-1/\delta_1^2)^2+(1-p)^2/(\delta_0\delta_1)^2}\\
  %   \le p/2(1/\delta_0^2+1/\delta_1^2)+p/2|1/\delta_0^2-1/\delta_1^2|+(1-p)/(\delta_0\delta_1)\\
  %   = p/2(1/\delta_0-1/\delta_1)^2+p/2|1/\delta_0^2-1/\delta_1^2|+1/(\delta_0\delta_1)\\
  %   \le p|1/\delta_0^2-1/\delta_1^2|+1/(\delta_0\delta_1),
  % \end{align*}
  % with equality occurring when $\delta_0=\delta_1$. The mean square of $1/\ol{W}^2$ is therefore
  % polynomial in both the product of $1/\delta_0^2,1/\delta_1^2$,
  % a measure of IV weakness, and the difference of
  % $1/\delta_0^2,1/\delta_1^2$, a measure of confounding between the IV and $L$. Since $1/\delta_0^2,1/\delta_1^2>1$ and $p\le1$,
  % the former terms are likely to dominate (\ref{eg:markov:eigen:unstab}) rather
  % than the known confounding, given by $p$. See Fig. \ref{eg:markov:fig:3}.}
  
  % 1. describe delta model, maybe give treatment model here? wont be
  % needed. The squared IV weights are. Letting the density of $Z$ be
  % constant, the weights reduced to the the detla terms. 2
  % additional degrees of freedom for delta. Interpret the magnitude
  % of delta0 and delta1 as the IV strength, and the difference gives
  % the depenence of delta on $L$, which has an analogous effect to
  % the $l->A$ dependence in the SRA setting. That is, to the extent
  % that this dependence may be approximated by a standardized
  % function of $A$, an analogue of stabilized weights ((ref above))
  % may be used to decrease variance of the estimator.\\
  % 2. latent sra weights w/o stabilization behave as ---. This
  % componenet of the weight variance is therefore exponential in
  % both the inverse average size of delta, i.e., weakness of the
  % IV, and the difference of the inverse deltas, i.e., the degree
  % of confounding. \\

  As with SRA weights, the function $h(\ol{A})$ in Theorem \ref{theorem:1} may 
  be chosen to partially stabilize IV weights. A function of
  $\ol{A}$ approximating $\Delta_t$ % =\sum_{z\in\{0,1}(-1)^{1-z}\PP(A_t\ol{L}_t,\ol{A}_{t-1},\ol{Z}_{t-1},Z_t=z)$
  may be used to cancel out the magnitude of the weights and minimize the
  second moment of the weights.  As in the SRA
  case, the influence function (\ref{eqn:influence function}) does not change when $h$ is
  multiplied by a constant scalar, so the minimization is
  well-posed. Analogously to SRA weights, we consider stabilizing a weight term $\delta_{L_t}$ by
  an arbitrary term depending on the treatment previous to $L_t$, say,
  $\gamma_{A_{t-1}}$, with values $\gamma_0,\gamma_1$. For example, analogous to the term $f(A_t\mid \ol{A}_{t-1})=\EE(f(A_t\mid
  \ol{L}_t,\ol{A}_{t-1})\mid \ol{A}_{t-1})$ commonly used to stabilize
  SRA weights, we may take
  \begin{align*}
    \gamma_{ A_{t-1}}&=\EE(\delta_{L_t}\mid \ol{A}_{t-1}) = \EE(\delta_{L_t}\mid
                       {A}_{t-1})=p_L\delta_{A_{t-1}}+(1-p_L)\delta_{1-A_{t-1}},\hspace{.2in} t>1,\\
    \gamma_{ A_0}&=\EE(\delta_{L_1}\mid A_{0})
                   = \EE(\delta_{L_1}).
  \end{align*}
  
  The squared inverse of the weights is 
  $$
  1/\ol{W}^2=\prod_t \gamma_{ A_{t-1}}^2/\delta_{L_t}^2.
  $$
  Proceeding as before, let
  \begin{align*}
  \phi_t(l_{t-1})&\coloneqq \EE(\prod_{t'=t}^T \gamma_{
    A_{t'-1}}^2/\delta_{L_{t'}}^2\mid L_{t-1}=l_{t-1}), \hspace{.2in}l_{t-1}\in\{0,1\},\\
    \phi_t&\coloneqq (\phi_t(0),\phi_t(1)).
  \end{align*}
  Then $\EE(1/\ol{W}^2)=\gamma_0^2(\phi_2(0)/2+\phi_2(1)/2)$,
    $\phi$ satisfies the recurrence
    \begin{gather}
      \begin{aligned}
        \phi_{t-1}=
        \begin{pmatrix}
          p_{LA}p_{AL}\gamma_0^2/\delta_0^2+(1-p_{LA})(1-p_{AL})\gamma_1^2/\delta_0^2 &
          p_{LA}(1-p_{AL})\gamma_0^2/\delta_1^2+p_{AL}(1-p_{LA})\gamma_1^2/\delta_1^2 &\\
          p_{LA}(1-p_{AL})\gamma_1^2/\delta_0^2+p_{AL}(1-p_{LA})\gamma_0^2/\delta_0^2 &
          p_{LA}p_{AL}\gamma_1^2/\delta_1^2+(1-p_{LA})(1-p_{AL})\gamma_0^2/\delta_1^2
        \end{pmatrix}
        \phi_t,
      \end{aligned}\label{eg:markov:matrix:2}
    \end{gather}
    and the growth of $\EE(1/\ol{W}^2)$ is determined by the
    eigenvalues of the matrix $P$ in (\ref{eg:markov:matrix:2}),
    $$
      \tr(P)/2 \pm \sqrt{\tr(P)^2/4-\det(P)}
      $$
    where
    \begin{align*}
      \tr(P)&=p_{LA}p_{AL}(\gamma_0^2/\delta_0^2+\gamma_1^2/\delta_1^2) +
      (1-p_{LA})(1-p_{AL})(\gamma_0^2/\delta_1^2+\gamma_1^2/\delta_0^2)\\
      \det(P)&=(2p_{LA}-1)(2p_{AL}-1)\frac{\gamma_0^2\gamma_1^2}{\delta_0^2\delta_1^2}.
    \end{align*}% ((set p1 equal to p2))

\begin{figure}
      \centering
      \begin{subfigure}{.5\textwidth}
        \centering
        \includegraphics[width=.9\linewidth]{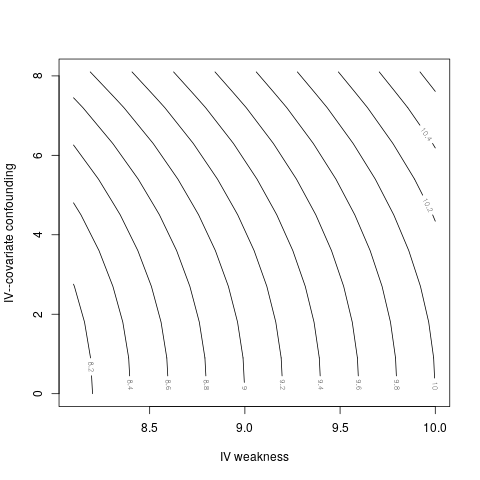}
        \caption{Unstabilized IV weights}
        \label{fig:sub1}
      \end{subfigure}%
      \begin{subfigure}{.5\textwidth}
        \centering
        \includegraphics[width=.9\linewidth]{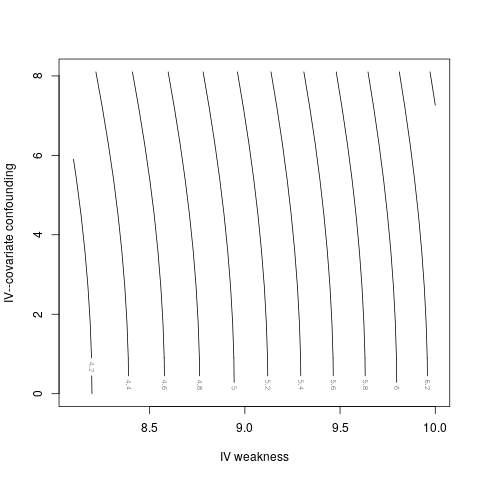}
        \caption{Stabilized IV weights}
        \label{fig:sub2}
      \end{subfigure}
      \caption{The variance of the IV estimator depends on the
        weakness of the IV and the dependence of the IV on the
        covariates. An approximation to the variance is plotted
        against IV weakness and IV confounding using unstabilized and
        stabilized weights. The effect of IV confounding is mitigated
        by stabilization, but the effect of a weak IV remains.}
      \label{fig:eg:markov:IV:weighted}
    \end{figure}

    In terms of the IV weakness and confounding terms (\ref{eg:markov:omegakappa}), the principal eigenvalue may be rewritten as
    \begin{align*}
      \lambda_1=\frac{1}{4}\left(\sqrt{\kappa^2+4\omega^2}(\gamma_0+\gamma_1)(p_{LA}p_{AL}+(1-p_{LA})(1-p_{AL})) + \kappa(\gamma_0-\gamma_1)(1-p_{LA}-p_{AL})\right)\times\\
      \left(1+\sqrt{1-\frac{4\det(P)}{\tr(P)^2}}\right).
    \end{align*}
    The last factor in parentheses has magnitude at most 2. As
    mentioned previously, it may be assumed without loss of generality that $\prod_j\gamma_{A_j}$ has expectation 1 for any law under which $\prod_j\gamma_{A_j}$ has finite expectation. For the variance of the influence function (\ref{eqn:influence function}) does not change on multiplying $h(\ol{A})=h_1(\ol{A})\prod_j\gamma_{A_j}$ by a constant, so that any choice of $\prod_j\gamma_{A_j}$ may be replaced by another with mean 1, i.e., $\prod_j\gamma_{A_j} / \int\prod_j\gamma_{A_j}d\mu$. Letting $\mu$ be counting measure, the assumption becomes
    $$
    1=\int\prod_j\gamma_{A_j}d\mu = \sum_{j=0}^T{T\choose j}\gamma_0^j\gamma_1^{T-j}=(\gamma_0+\gamma_1)^T.
    $$
    Therefore $\gamma_0+\gamma_1=1$ and the principal eigenvalue is
    \begin{align*}
      \lambda_1=\frac{1}{4}\left\{\sqrt{\kappa^2+4\omega^2}(p_{LA}p_{AL}+(1-p_{LA})(1-p_{AL})) + \kappa(\gamma_0-\gamma_1)(1-p_{LA}-p_{AL})\right\}\left(1+\sqrt{1-\frac{4\det(P)}{\tr(P)^2}}\right).
    \end{align*}
    Therefore, the effect of IV confounding $\kappa$ on the variance may be reduced by choosing $\gamma_0-\gamma_1$ close to 0, but no choice of $(\gamma_0,\gamma_1)$ will have an effect on the weakness of the IV, $\omega$, due to the term $\sqrt{\kappa^2+4\omega^2}$. See Fig. \ref{fig:eg:markov:IV:weighted} for a simulation.
    
    % In comparison to the unstabilized IV case, the mean square of the weights is polynomial
    % in $\max_{j,k}\gamma_j^2/\delta_k^2$, rather than in terms of the
    % form $1/\delta_k^2$. In particular, the trace term shows that the
    % growth is  polynomial in a mixture of the like-indexed terms
    % $\gamma_j/\delta_j$ and oppsite-indexed terms $\gamma_j/\delta_k,
    % j\neq k$,
    % with the mixing given by the probability of staying in the same
    % state or switching states in the thinned chain $\ldots
    % L_{t-1}\to L_t\to L_{t+1}\ldots$. ((terminlogy "polynomial in"...actually
    % this is the base for the exponential growth in T)) Since the expression ((expected weight)) is quadratic in
    % $\gamma_0=1-\gamma_1$, the optimum weight for given $\delta_j$ and
    % $p_j$ may be obtained in closed form, but the resulting expression
    % is somewhat complicated.

  % \end{example}

  Given below is a summary of the discussion of the asymptotic variance of the estimator in the four situations considered in this example.
  \begin{enumerate}
  \item SRA weights, unstabilized: The variance is exponential in $T$,
    and for fixed $T$ the variance blows up at a quadratic rate as the
    confounding $p_{LA}=\PP(A\mid L)$ approaches $0$ or $1$.
  \item SRA weights, stabilized: The variance is bounded as long as
    the confounding $p_{LA}=\PP(A\mid L)$ is of the same order as the ``predictiveness'' $p_{AL}=\PP(L\mid A)$.
  \item IV weights, unstabilized: The variance of the weight terms is exponential in $T$, and for fixed $T$ is linear in a terms relating to the weakness of the IV and the degree of dependency between the IV and covariates.
  \item IV weights, stabilized: The variance due to dependency between the IV and covariates may be reduced, but the variance due to the weakness of the IV remains.
  \end{enumerate}

  The difference between the SRA and IV cases seems to be the
  following. In both cases the stabilization terms may be assumed to
  integrate to 1, due to the scale invariance property of the variance
  of the influence function mentioned earlier. In the case of SRA
  weights, the weights themselves also satisfy this type of property,
  being densities. Specifically, the terms $\prod_tf(a_t\mid l_{t-1})$
  cannot be uniformly small across all choices
  $a_t,l_{t-1},t=1,\ldots,T$. One may therefore hope to choose the stabilizing
  terms to match the magnitude of the corresponding weight terms. The
  IV weights do not satisfy this type of property, i.e., $\delta_0$
  and $\delta_1$ may both be arbitrarily small at the same time, and
  no choice of $(\gamma_0,\gamma_1)$, which cannot both be small at the
  same time due to the scale invariance, will control the weights.

  \section{Appendix: Details for the simulation section}\label{appendix:simulation details}
  
\subsection{Data generation}\label{subsection:data generation}
Lemma \ref{lemma:1} gives appropriate conditions on the endogenous noise term $\eta$ for sampling outcomes $Y=m(\ol{A})+\eta+\epsilon$ consistent with
a MSMM model and the assumptions of Theorem \ref{theorem:1}. For example, we may sample
outcomes as
\begin{align*}
  \EE(Y\mid\ol{A},\ol{Z},\ol{L},\ol{U}) &=
  \sum_t\left(f_t(L_t,U_t)-\EE(f_t(L_t,U_t)\mid \ol{A}_{t-1},\ol{L}_{t-1},\ol{U}_{t-1})
  \right) +  m_\beta(\ol{A}),\\
  (L_t,U_t) &\cind \ol{Z}_t \mid \ol{ALU}_{t-1}
\end{align*}
for arbitrary functions $f_t$, once we have chosen a sampling scheme
for $(\ol{A},\ol{Z},\ol{L},\ol{U})$ satisfying the stated conditional indepndence assumption. We choose linear functions, so that
outcome variables $Y$ are sampled as
\begin{align*}
  % Y &= \sum_{t=0}^T\rho_tU_t + \sum_t\beta_ta_t + \epsilon,
        Y &= \sum_{t=0}^T(\tau_t(L_t-\EE(L_t\mid \ol{ALU}_{t-1}))+\rho_t(U_t-\EE(U_t\mid\ol{ALU}_{t-1}))) + m_\beta(\ol{A}) + \epsilon \\
  &= \sum_{t=0}^T(\tau_t(L_t-\EE(L_t\mid \ol{ALU}_{t-1}))+\rho_t(U_t-\EE(U_t\mid\ol{ALU}_{t-1}))) + \beta_0 + \beta_1\sum_tA_t + \epsilon,
\end{align*}
% \comment{why L,U not condionally centered? what happened to L on rhs?}
with $\rho_t,\tau_t\in\mathbb{R}$ and $\epsilon$ standard normal. We
set $\rho_t=\tau_t=1$ in our simulation.

For $1\le t\le T$, $U_t$
is sampled as standard normal and
$Z_t$ is bernoulli with success probability 1/2, all mutually
independent, ensuring the IV assumptions. %  In order to enforce the independent compliance type requirement (\ref{assumption:ict}),
The treatments $A_t$ and covariates $L_t$ are sampled recursively as:
\begin{align}
  \begin{split}
  L_{t+1} &= \lambda_0 + \lambda_1A_{t} + \epsilon_t\\
  \Phi^{-1}(\Delta_{t+1}) &= \Phi^{-1}\left(\PP(A_{t+1}=1\mid \ol{L}_{t+1},\ol{U}_{t+1},\ol{A}_t,\ol{Z}_t,Z_{t+1}=1)\right.\\
          &\qquad \left.-\PP(A_{t+1}=1\mid \ol{L}_{t+1},\ol{U}_{t+1},\ol{A}_t,\ol{Z}_t,Z_{t+1}=0)\right)\\
          &= \alpha_0 + \alpha_1L_{t+1}\\
  \PP(A_{t+1}=1\mid \ol{A}_t,\ol{Z}_{t+1},\ol{L}_{t+1},\ol{U}_{t+1}) &= \Phi(\nu_0 + \nu_1
                                               L_{t+1}+\nu_2
                                              U_{t+1})\times
                                              (1-\Delta_{t+1})+ Z_{t+1}\times \Delta_{t+1}.\label{simulation:algo}
                                            \end{split}
\end{align}
% The $f_1,f_2,\ldots$ are just linear functions. They vary with time $j$
% and would be more properly denoted $f_{1,j},f_{2,j}$, etc. The definitions of
% $\Delta_{j+1}$ and $\PP(A_{j+1}=1)$ are intended to enforce the
% independent compliance type requirement. 
% See Fig. \ref{fig:tikz synthetic}. 
Here, $\Phi$ denotes the standard normal CDF and
$\epsilon_t$ are mutually independent standard normal variables. The
models chosen for $A_t$ and $\Delta_t$ ensure that
Assumption (\ref{assumption:ict})
holds. The parameters
$\lambda_0,\lambda_1\in\mathbb{R}$ control the extent to which the treatment
confounds subsequent covariates, whereas $\nu_1\in\mathbb{R}$ and
$\nu_2\in\mathbb{R}$ control
the extent to which observed and unobserved confounders confounders,
respectively, confound treatment. The reciprocal
arrangement ensures that the confounding is truly longitudinal, so that,
e.g., a series of propensity score analyses would not likely 
estimate the MSMM parameter accurately. The dependence between treatment and a confounder
unavailable for estimation, provided $\nu_2\neq 0$, violates SRA. The
parameters $\alpha_0,\alpha_1,$ bear on the strength of the IV. We set $\lambda_0=\lambda_1=.5,\alpha_0=\alpha_1=.3,\nu_0=-.2,$ and $\nu_1=\nu_2=.2$ in the simulation described below.

\subsection{Estimation}
\label{sec:estimation}
As $f_{Z_t}$ is known under our data generation method (\ref{simulation:algo}), only $\beta, \alpha, $ and $\nu$ require
estimation. We use (\ref{eqn:estimating eqn}) as an estimating equation for $\beta$ and obtain $\frac{\partial s_\beta}{\partial \beta,\alpha}$ from (\ref{eqn:information}) by substituting
$$
% \frac{\partial}{\partial\beta}\mu(\beta)=\frac{\partial}{\partial\beta}\beta^T\ol{A}=\ol{A} \text{ and }
\frac{\partial}{\partial\beta}\mu(\beta)=\frac{\partial}{\partial\beta}\left(\beta_0+\beta_1\sum A_t\right)=(1,\sum A_t) \text{ and }
\frac{\partial}{\partial\alpha}\Delta_t(\alpha)=\frac{\partial}{\partial\alpha}\Phi(\alpha^TL_t)=\phi(\alpha^TL_t)L_t.
$$
We use maximum likelihood to estimate $\alpha$ and $\nu$, pooling over the time points. After integrating out $U_t$, model (\ref{simulation:algo}) implies the observed-data model
$$
\pi_t(\alpha,\nu) = \PP(A_t=1\mid \ol{A}_{t-1},\ol{L}_t,\ol{Z}_{t}) = \Phi(\nu^TL_t)(1-\Phi(\alpha^TL_t)) + Z_t\Phi(\alpha^TL_t),
$$
so that the conditional density of $A_t$ given the observed data is $\pi_t(\alpha,\nu)^{A_t}(1-\pi_t(\alpha,\nu))^{1-A_t}$,
the scores for $\alpha$ and $\nu$ are
$$
s(\alpha,\nu)=\left(\frac{A_t}{\pi_t(\alpha,\nu)}-\frac{1-A_t}{1-\pi_t(\alpha,\nu)}\right)\frac{\partial
                \pi_t(\alpha,\nu)}{\partial \alpha,\nu},
$$
and the information is 
% $$
% \frac{\partial s_{\alpha,\nu}}{\partial\alpha,\nu}=
% $$
% with
% $$
% $$
\begin{align*}  
  &\frac{\partial s(\alpha,\nu)}{\partial\alpha,\nu} =\\
                                                      &\left(-\frac{A_t}{\pi_t(\alpha,\nu)^2}
                                                      - \frac{1-A_t}{(1-\pi_t(\alpha,\nu))^2}\right)\frac{\partial\pi_t(\alpha,\nu)}{\partial\alpha,\nu}\left(\frac{\partial\pi_t(\alpha,\nu)}{\partial\alpha,\nu}\right)^T
                                                      +
                                                      \left(\frac{A_t}{\pi_t(\alpha,\nu)}
                                                      -
                                                      \frac{1-A_t}{1-\pi_t(\alpha,\nu)}\right)\frac{\partial^2\pi_t(\alpha,\nu)}{\partial(\alpha,\nu)^2}
\end{align*}
with % \comment{should $t$ be dropped from $\pi$?}
\begin{align*}
  \frac{\partial\pi_t(\alpha,\nu)}{\partial\alpha,\nu} &= \left((Z_t-\Phi(\nu^TL_t))\phi(\alpha^TL_t)L_t, (1-\Phi(\alpha^TL_t))\phi(\nu^TL_t)L_t\right)\\
  \frac{\partial^2\pi_t(\alpha,\nu)}{\partial(\alpha,\nu)^2}&=
                                                            \begin{pmatrix} -(Z-\Phi(\nu^TL_t))\phi(\alpha^TL_t)(\alpha^TL_t)L_tL_t^T &
                                                              -\phi(\nu^TL_t)\phi(\alpha^TL_t)L_tL_t^T\\
                                                              -\phi(\alpha^TL_t)\phi(\nu^TL_t)L_tL_t^T & -(1-\Phi(\alpha^TL_t)\phi(\nu^TL_t)(\nu^TL_t)L_tL_t^T
                                                            \end{pmatrix}.
  % \frac{\partial s}{\partial\beta,\alpha,\gamma,\nu} &= \begin{pmatrix} 
\end{align*}
The remaining entries of $\partial
s/\partial(\beta,\alpha,\gamma,\nu)$ are 0. The ``sandwich estimator''
for the variance of
$\hat{\beta}$ can then be computed as the empirical covariance matrix
of (\ref{eqn:influence function}).\comment{should these formulas go in
  an appendix?}

A closed-form expression for the estimator $\hat{\beta}$ when the MSMM is linear, as in this example, is given in (\ref{eqn:estimation:closed form}).% The estimator for the parameter of interest, $\beta$, has a closed
% form expression as a weighted OLS regression on the treatments $\ol{a}$. That is, for a sample size of $n$,
% letting $D_n$ be the design matrix
% containing the observed regressors, $Y_n$ the vector of observed
% outcomes, and $\hat{W}_n$ the diagonal matrix formed from the inverted
% weights using the estimates for the nuisance parameters, we estimate $\beta$ as
% $$
% \hat{\beta} = (D_n^T\hat{W}_nD_n^T)^{-1}D_n^T(\hat{W}_nY_n).
% $$
% \comment{use Pn notation, also put earlier so it can be referenced in examples}

\end{document}